EFFECT OF SELECTED FOOD PROCESSING TECHNIQUES ON CONTROL

RELEASE KINETICS OF ENCAPSULATED LYCOPENE IN SIMULATED HUMAN

GASTROINTESTINAL TRACT

by

MOHAMMAD ANWAR UL ALAM

A DISSERTATION

Submitted in partial fulfillment of the requirements

for the degree of Doctor of Philosophy

in the Department of Food and Animal Sciences

in the School of Graduate Studies

Alabama A&M University

Normal, Alabama 35762

August 2022

Submitted by MOHAMMAD ANWAR UL ALAM in partial fulfillment of the requirements for the DEGREE OF DOCTOR OF PHILOSOPHY OF FOOD SCIENCE specializing in Food Safety and Processing.

Accepted on behalf of the Faculty of the Graduate School by the Dissertation Committee:

_______________________________________________   Dr. Lamin Kassama
                                                  Major Advisor

_______________________________________________   Dr. Martha Verghese

_______________________________________________   Dr. Armitra Jackson Devis

_______________________________________________   Dr. Judith Boateng

_______________________________________________   Dr. Aschalew Kassu

_______________________________________________   Dr. John Jones
                                                  Provost and Dean of the
                                                  School of Graduate Studies

_______________________________________________   Date

This dissertation is dedicated to my loving mother and father and my dear wife, Mrs. Mhamuda Khatun, for their love, prayers, encouragement, and support throughout this research.

EFFECT OF DIFFERENT FOOD PROCESSING TECHNIQUES ON CONTROL RELEASE KINETICS OF ENCAPSULATED LYCOPENE IN SIMULATED HUMAN GIT

Anwar Ul Alam, Mohammad, Ph.D., Alabama A&M University, 2022. 248 pp.

Dissertation Advisor: Dr. Mamadou Lamin S. Kassama.

Nanoencapsulation has become a widespread technique to improve the bioavailability of bioactive compounds like lycopene. Many researchers suggest that the consumption of lycopene-rich foods is effective in preventing cancer, diabetes, and cardiovascular diseases due to its strong, oxygen-quenching ability. But the functional activity of lycopene is compromised by light, oxygen, and heat. A biodegradable polymer such as Polylactic acid (PLA) or Polylactic co-glycolic acid (PLGA) is the most effective carrier for encapsulating lycopene due to its excellent biodegradability, biocompatibility, and non-toxigenic effect on the human metabolic system. Hence, the primary objective of this study was to evaluate the effect of pasteurization on bio-accessibility and control release kinetics of encapsulated lycopene nanoparticles in in-vitro human GIT. A three-factor factorial design was used to accomplish all experiments. In the first objective, sonication time, surfactant, and polymer concentration were considered as the three factors to synthesize polymeric lycopene nanoparticles (LNP) whereas in the second objective, type of pasteurization (Conventional, Microwave), encapsulation, and juice concentration (5 and 15° Brix) were three factors. In Objectives 4 and 5, the type of encapsulation and pasteurization, and digestion time were evaluated as the three factors used to evaluate the in-vitro bio-accessibility of encapsulated lycopene NP. The study evidenced that encapsulation improved the lycopene's bio-accessibility by 70% and more than 60% for conventional pasteurized (CP) and microwave pasteurized (MP) nanoemulsions, respectively without compromising the physicochemical properties of both PLA and PLGA-lycopene nanoparticles. The in-vitro bio-accessibility study also showed that CP reduced the functional activity of PLA-LNP by 20% whereas MP had no significant effect ($p < 0.05$) on the bio-accessibility of PLA-LNP. It is unlikely that the PLGA-LNP was more sensitive against MP nanoemulsion (degraded by 47%) than the CP (degraded by 27%). In the case of rumen digestion, it can be shown that MP made conformational changes to both PLA and PLGA which further contributes additional protection for the lycopene in PLA and PLGA. In conclusion, it was shown that the PLA-LNP treated with MP provided the highest bio-accessibility after in-vitro human digestion. The food industries can use novel PLA-LNP to develop functional food for the therapeutic treatment of chronic diseases.

# TABLE OF CONTENT

xii

LIST OF TABLES

LIST OF FIGURES

# LIST OF ABBREVIATION

| | | |
|---|---|---|
| AD | - | After digestion |
| ADME | - | Absorption, digestion, metabolism and excretion |
| A-K AutoGIT | - | Anwar-Kassama Automatic Gastrointestinal Tract |
| ANOVA | - | Analysis of variance |
| BD | - | Before digestion |
| BHT | - | Butylated hydroxy toluene |
| CD | - | Conductivity |
| CTN | - | Control Non-pasteurized |
| CoF | - | Confidence of fitness |
| CP | - | Conventional pasteurized |
| DC-FCCS | - | Dual-color flurescene cross-correlation spectroscopy |
| DLS | - | Dynamic light scattering |
| DLS | - | Dynamic light scattering |
| DPPH | - | 2,2-diphenyl-1-picrylhydrazyl |
| E | - | Encapsulated |
| EE | - | Encapsulation efficiency |
| F | - | Fortified |
| FDA | - | Food and drug administration |
| FT-IR | - | Fourier transform infrared spectroscopy |
| GIT | - | Gastrointestinal tract |
| HD | - | Hydrodynamic diameter |
| IR | - | Infrared |
| KCl | - | Potassium chloride |
| L & D PLA | - | L & D Polylactic acid |
| LNP | - | Lycopene nanoparticles |
| LTLT | - | Low temperature long time |
| M | - | Mobility |
| MP | - | Microwave pasteurized |
| MPS | - | Mononuclear phagocyte system |
| NaCl | - | Sodium chloride |
| NE | - | Non-encapsulated |
| NF | - | Non-fortified |
| NLC | - | Nanostructured lipid carriers |
| PBS | - | Phosphate buffer solution |
| PDI | - | Polydispersity index |
| PDLA | - | Poly-D-lactic acid |
| PGA | - | Poly glycolic acid |
| PLA | - | Polylactic acid |
| PLGA | - | Polylactic co-glycolic acid |
| PLLA | - | Poly-L-lactic acid |
| RPM | - | Revolution per minutes |
| SEM | - | Scanning electron microscopy |
| SLN | - | Solid lipid nanoparticles |
| $T_g$ | - | Glass transition temperature |
| $T_m$ | - | Melting temperature |
| TSS | - | Total soluble solids |
| USDA | - | United states department of agriculture |
| UVS | - | Ultra Violate Spectroscopy |
| VFA | - | Volatile free fatty acids |
| ZP | - | Zeta potential |
| ABTS | - | 2,2'-azino-bis(3-ethylbenzothiazoline-6-sulfonic acid) |

ACKNOWLEDGMENTS

My deepest gratitude first goes to the Almighty Allah for giving me the strength, grace, and wisdom to carry out my research work. This work was supported by a grant from the USDA National Institute of Food and Agriculture, [USDA Evan-Allen Accession No. 1013057: Project Title: Modeling in Vitro Control Release and Diffusion of Loaded Nanoparticles (LNP) in the GI tract - Impact of processing], and the Agriculture Experimental Research Station, Alabama A and M University for providing financial assistance to support this research project.

My heartfelt thanks go to my supervisor and chairman of my committee, Dr. Lamin S. Kassama, for his encouragement, patience, hard work, and contributions towards my research. I appreciate all he did for me and for always encouraging me to do better. I am also extremely grateful to Dr. Armitra Jackson Davis, for all her encouragements and support during my research. I am incredibly grateful for the support and advice rendered to my advisory committee members: Dr. Martha Verghese, Dr. Judith Boateng and Dr. Aschalew Kassu for their support during my dissertation program.

Special thanks to my parents, the Late Mr. Shamsul Alam and Mrs. Shahida Alam, and my entire family for all their prayers, love, and support. My warmest thanks and appreciation go to my ever loving and supportive wife, Mahmuda Khatun, and my beautiful, amazing kids for their love, patience, care, and support throughout my research.

Finally, my gratitude goes to my food engineering and microbiology lab mates, especially to Edwin Ochieng, Dalais Bailey, Aaron Dudley, Sai Vinay Kumar Madala, and Dr. Bio Bruno Bamba during the time of my research and the FAS faculty and staff for their support and encouragement during my study at AAMU. Many special thanks also go to Dr. Mohammad Arif Ul Alam, Assistant Professor of the Department of Computer Science at UMass Lowell for his great help in automation of in-vitro system.

# CHAPTER 1

## INTRODUCTION

### 1.1.  Background

Degenerative diseases like cancer, diabetes, and cardiovascular diseases are the abnormalities as a result of the free radical generation and cellular redox imbalance (Parveen et al., 2015). The free radical and cellular redox imbalance originates from the metabolic degradation of food components in the human digestive system (Labat-Robert & Robert, 2014). Natural antioxidants have a blocking role of auto-oxidation of fats, oils, and different food products (Hussain et al., 2017). In advanced research, it is already acknowledged that regular intake of antioxidants can offset the waves of cancer development at various anatomical sites like the stomach, lung, and prostate gland (Viuda-Martos et al., 2014). Henceforth, there are many ongoing research efforts in developing unique methods of introducing natural antioxidants from fruits, vegetables, herbs, and spices in the food system. For example, tomatoes are one of the vegetables with a rich source of antioxidants. The search for antioxidants from tomatoes and their byproducts has revealed that lycopene, a carotenoid with antioxidant potentials, can control free radical activity which further suppresses cancer and degenerative diseases (Labat-Robert & Robert, 2014; Nasir et al., 2014). However, lycopene is highly susceptible to certain intrinsic and extrinsic environmental factors, such as light, heat, and oxygen. Therefore, encapsulating lycopene in biodegradable polymer attracts considerable interest, as the polymer can breakdown into biological metabolites during digestion without producing any toxicity (Liarou et al., 2018; Nguyen et al., 2018; Huang et al., 2017; Ye et al., 2018) while providing protection of the bioactive components, enzymes, polyphenol, and nutrients against the adverse processing effects and enhance controlled release on targeted sites (Lee et al., 2013). Nano-encapsulation is a platform that is applied in control release delivery of pharmaceutical drugs, and with great potential applications in food systems.

PLGA (Polylactic co-glycolic acid) is a biodegradable polymer and is considered an effective carrier (Kassama & Misir, 2017) as its metabolites are digestible into the human body through the controlled delivery bioactive compounds to the target cells. Polylactic acid (PLA) is also widely used for encapsulating many bioactive compounds (e.g. Quercetin) due to its biodegradability, biocompatibility, high hydrophobicity, strong mechanical strength, and control release properties (Sambandam et al., 2015).

Different food processing methods are known to influence the bio-accessibility of antioxidants in fruit juice (Stinco et al., 2013; Aschoff et al., 2015). The processing techniques, especially the pasteurization techniques such as thermal (conventional and microwave heating) or non-thermal processes (pulse electric field, pulse UV light and etc.), have some impacts on the total soluble solids (TSS), pH, color, particle size, antioxidant content, total phenolic compounds, and titratable acidity of fruit juices fortified with nanoparticles or other nutrients (Mennah-Govela & Bornhorst, 2017).

The hydrodynamic diameter, zeta potential, polydispersity index, rheological properties, thermal properties, and particle size distribution have a definite effect on the control release activity of encapsulated nanoparticle (Kassama & Misir, 2017). Again, understanding the human digestion system is a crucial factor for the validation of the control release potential of bioactive compounds from the biodegradable polymer core. Many scientists have developed a dynamic in-vitro model (Dynamic Gastric Model, DGM) for simulating the human digestion system to observe the bio-accessibility of different nutrients present in foods (Mennah-Govela & Bornhorst, 2017; Mennah-Govela & Bornhorst, 2016; Bornhorst et al., 2016). Enzymes catalyze chemical reactions in the human digestion system and thus contributes extensive impact on the particle size, antioxidant activity, phenolic content, and the rate of biodegradation of the polymer (Sambandam et al., 2015).

## 1.2. Justification of the Study

Lycopene has twice free radical scavenging potential than beta carotene (Vitamin-A) and ten times more than alpha-tocopherol (Vitamin-E) (Suwanaruang et al., 2016). Its antioxidant activities are impeded by oxidation, isomerization, and degradation processes (Calvo & Santagapita, 2017); hence, encapsulation could provide a potential remedy to preserve the integrity of lycopene. Polymer-based encapsulation has some potential to enhance the controlled release and targeted delivery of bioactive compounds with improved efficiency. Biodegradable polymer-based encapsulation at the nano level has become popular due to its non-toxic targeted delivery of natural antioxidants (Akagi et al., 2010). However, the effect of processing on the controlled release of bioactive components in the gastrointestinal tract (GIT) is not well known. Likewise, the adverse effects of processing on the integrity of biodegradable polymer are not known as well (Han et al., 2009). Therefore, the in-vitro control release of encapsulated lycopene undergoing different processing methods will be the scope of this research and therefore contribute new insights in this research area. Hence, it is hypothesized that food processing methods such as conventional heat pasteurization and microwave pasteurization will not significantly change the control release of lycopene encapsulated in biodegradable polymer (PLGA and PLA).

## 1.2.    Objectives

The overall objective of the study is to evaluate the effect of processing methods on the controlled release kinetics of encapsulated lycopene in the gastrointestinal tract using an in-vitro simulation model. Hence, the specific objectives of the study are as follows:

1)  To synthesize lycopene nanoparticles in biodegradable polymers (PLA and PLGA) and determine its encapsulation efficiency.

2)  To assess the effect of conventional and microwave treatment on physicochemical, rheological and antioxidant properties of nanoparticles in food emulsion.

3)  To design and develop an automated simulator of human digestive system to evaluate bioaccessibility of lycopene loaded nanoparticles.

4)  To determine the in-vitro digestion effect on bio-accessibility of lycopene nanoemulsion.

5) To evaluate rumen digestion effect on bio-accessibility of lycopene nanoemulsion.

# CHAPTER 2

# LITERATURE REVIEW

## 2.1.    Background of the Study

### Carotenoids and Lycopene

Carotenoids are lipophilic compounds with a structural backbone of poly-isoprenoid. Most carotenoids contain a series of conjugated double bonds which are sensitive to oxidative modification and cis-trans isomerization. There are six major carotenoids (β-carotene, α-carotene, lycopene, β-cryptoxanthin, lutein, and zeaxanthin) found in the human body fluid, and their characteristics have been observed in the human plasma and tissues. Among them, β-carotene has been the most extensively studied. Most recently, lycopene has drawn significant attention due to its association with reduced risk of chronic diseases, including cancers. Extensive efforts have been spent to identify its biological and physicochemical properties. Relative to β-carotene, lycopene has the same molecular mass and chemical formula, but instead lycopene is an open-polyene chain lacking the β-ionone ring structure as shown in Figure 2.1. While the metabolism of β-carotene has been studied broadly, the metabolism of lycopene remains to be well understood (Mein et al., 2008).

Absorption of lycopene is not only affected by the adverse processing methods, but also the presence of lipids and other lipid-soluble compounds (Rao, 2006). However, it is very alarming that only 10-30% of the dietary lycopene is absorbed through the intestinal lumen due to unavailability of fats and oils, interaction of dietary fiber and other carotenoids (Rao & Agarwal, 1999; Gartner, 1997).

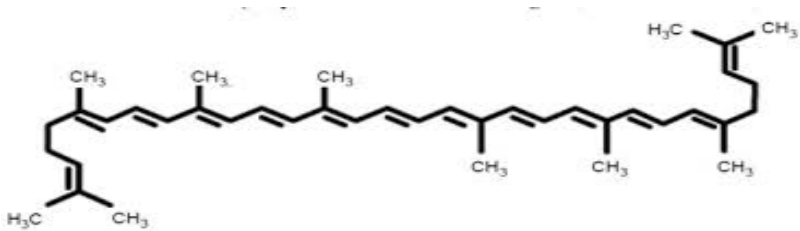

Figure 2.1. Structure of Lycopene (Gupta et al., 2018).

## Lycopene Properties and Processing Effects

Jiang et al. (2019) reported that lycopene, when consumed, may induce an anti-inflammatory effect to inhibit prostate cancer progression. According to Jiang et al., with lycopene treatment, inflammatory factors levels, such as interleukin1 (IL1), IL6, IL8, and tumor necrosis factor-α (TNF-α), were reduced significantly. The survival of mice bearing xenografts prostate cancer was significantly improved (P < 0.01) by the higher dose of lycopene; the burden of the tumor was significantly reduced (P < 0.01) due to lycopene treatment. Lycopene treatment enhances cytoprotective enzymes to be synthesized (Richard et al., 2009). Cytoprotective enzymes could further inhibit adenosine deaminase, an essential component for tumor generation. Parveen et at. (2015), found that consuming tomatoes can reduce the risk of obesity, hypercholesterolemia, cardiovascular disorder, and different types of cancers.

Multi-dimensional factors are responsible to modify the physicochemical properties of lycopene nanoparticles (LNP). Ha et al. (2015) found that high pressure homogenization method facilitates the LNP formulation where higher pressure lowers the hydrodynamic diameter and stability or zeta potential value of the LNP. Unlikely, encapsulation efficiency of lycopene in the same study decreased with increased homogenization pressure. Li et al. (2017) and Ha et al. (2015) used nanoprecipitation technique to encapsulate lycopene and got a relationship with increasing the lycopene concentration increased the hydrodynamic diameter but lower the zeta potential and encapsulation efficiency. Rotary evaporation film ultrasonication method giving a small LNP size ranges from 58 to 105 nm in diameter and high zeta potential value ranging from -37 to -32.5 mV. This LNP also contributes better absorption properties as well as control delivery to the targeted cells (Zhao et al., 2018). Hot homogenization methods providing more encapsulation efficiency for lycopene ranging from 86.6±0.06 to 98.4±0.5% (Nazemiyeh et al., 2016) which further helps to improve absorption of lycopene. Interfacial deposition method also providing high encapsulation efficiency (95.12±0.42%) for lycopene NPs synthesized in PCL (Polycaprolactone) (dos santos et al., 2015). High pressure homogenization method providing highest stability for encapsulated lycopene NPs (ranges from -74.2 to -74.6 mV) and lowest polydispersity index (0.13±0.02 to 0.15±0.05) value (Okonogi & Riangjanapatee, 2015).

Impact of Processing Methods on Food Quality:

Consumer are changing their habits to fresh-like products. As a result, the fruit-juice industries have directed their focus in search for alternative processing technologies that will induce minimum quality changes and hence increase the wholesomeness of their products (Velazquez-Estrada et al., 2012). It is well established that traditional thermal techniques extend the shelf life of juices, thus ensuring their safety and thus maximizes the performance in fruit-juice processing. However, the application of heating has significant impact on the rheological, physicochemical, and organoleptic properties while causing nutritional losses (Gomez et al., 2011). Hence, with growing consumer interest in a healthy diet and a better quality of life, people are liking food rich with phytochemical to minimizing risk of chronic diseases such as coronary heart disease, cancer, and others (Mullen et al., 2007). Similarly, physicochemical attributes of fruit juices such as pH, TSS (total soluble solids), or viscosity determine their stability and storability, hence highlighting the importance of producing a stable product that meets the requirements for increasing market demand. It is generally believed that texture and rheological properties of foods are major determinants for consumer preference and acceptance. Hence, the rheological characterization of food is also important for the design of unit operations, process optimization, and high-quality of liquid and semi-liquid food products (Ibarz, 2003; Rao, 1999). However, the preservation of the organoleptic attributes after unit processes such as pasteurization is crucial in not altering nutrient content and consumer palate.

According to Vikram et al. (2005), at 455W for 180s of microwave pasteurization (MP) had the highest degradation effect on vitamin C content compared to ohmic heating, infrared heating, and conventional heating. However, other authors reported that MP treatment at 900 W power level for 30 s to 10 min preserved vitamin C compared to conventional thermal heating (Igual et al., 2010b; Geczi et al., 2013).

*Microwave processing.* Microwave is an electromagnetic radiation bearing the spectrum between frequencies of 300 MHz and 300 GHz. The Federal Communications Commission (FCC) has approved two frequencies (915±13 MHz, and 2450±50 MHz) within the microwave range for industrial, scientific, and medical (ISM) apparatus application. Microwaves transmit radiation through the food based on their dielectric properties. In contrast to conventional heating systems, microwaves penetrate food, and the heating extends within the entire food material rapidly and more uniformly. It is a nonionizing radiation where heat is generated due to its interactions with the food materials. The two most important mechanisms that explain heat generation in a material placed in a microwave field are ionic polarization and dipole rotation (Benjamin & Gamrasni, 2020).

*Ionic polarization and dipole rotation.* Due to the applied electrical field on food solutions containing ions, the ions move at an accelerated pace due to attraction to the opposite charges. The resulting collisions between the ions cause the conversion of kinetic energy of the moving ions into thermal energy. A solution with a high concentration of ions would have more frequent ionic collisions and, therefore, exhibit an increase in temperature. On the other hand, in dipol rotation, the mechanism of heating is somehow different compared to ionic polarization. As food materials contain water, water is a polar molecule, and as a result, the molecules orient themselves according to the polarity of the electric field applied. In a microwave field with a frequency of 2450 MHz, the polarity changes at 2.45 x $10^9$ cycles per second which causes the rotation of polar molecules based on the rapid changing polarity of the electromagnetic field of the microwave radiation. This rapid rotation of molecules leads to increased friction with system's molecules and hence generates heat. The size, shape, dielectric properties of the materials are major factors that influence the microwave heating of the food items.

## Increase Encapsulation Efficiency of Lycopene Increase Bioavailability or Bio-Accessibility

According to Sharma et al. (2016), the encapsulation of lycopene in biodegradable polymer revealed a potential rise in blood plasma level. The relative bioavailability was found to increase by 297.2% for encapsulated lycopene formula compared with nonencapsulated lycopene-based products. Furthermore, the study also gave a conclusion of better controlled release, high encapsulation efficiency and greater permeability of loaded lycopene into the human intestinal lumen. The zeta potential values or the stability of the nanoparticles did not change at different pH (4 or 9) and temperature (4°C or 25°C) and the encapsulation efficiency was found to be 62.8 ± 2%.

Sambandam et al. (2015) successfully encapsulated quercetin in polylactic acid (PLA) with the hydrodynamic diameter and zeta potential value of 220 ± 30 nm and -21.5 ± 2.2 mV, respectively. The encapsulation efficiency was found to be 73.3% in most optimized composition which contributes to the cumulative release of > 99.7% of Quercetin. A better scavenging effect were found in encapsulated Quercetin than that of non-encapsulated one. Sanna et al. (2012) found PLGA (Poly Lactic-Co-glycolic Acid) to be less efficient as a coating for resveratrol in the presence of chitosan (32%) and alginates (23%). Kassama & Misir (2017) found greater encapsulation efficiency for PLGA to encapsulate freeze-dried Aloe Vera gel (86.3%) and liquid (67%) with zeta potential values of -28 and -21.9 mV, respectively which had better antioxidant properties compared to nonencapsulated one after in-vitro digestion.

Nanoencapsulation and Nanoparticles

Currently, nanotechnology received great attention and has emerged as one of the most promising technologies in food science research which is dealing with manufacturing and application of materials smaller than 1000 nm in the food industry (Sanguansri & Augustin, 2006). Nanotechnology has the potential to revolutionize the food industry from the manufacture of food products to improving the application of ingredient functionality in foods and enhancing the sensory attributes (texture, color, and taste). Additionally, it improves the processability, storage stability and shelf-life of foods, and more exclusively, the development of a great number of new and innovative products with increased functionality and efficacy.

Nanotechnology provides a platform to encapsulate the bioactive compounds at the nano-level to improve its bio-accessibility and bioavailability through human diet. When particles size is reduced to nanoscale range the surface to volume ratio also increases, and subsequently increases the reactivity of the coating materials and the encapsulated ingredients by many folds due to their substantial change in the optical, electrical, and mechanical properties (Neethirajan & Jayas, 2011). Nanotechnology can significantly improve the aqueous solubility, thermal stability, and bioavailability of the bioactive ingredients (Huang et al., 2010; Jafari et al., 2008b; Rashidinejad et al., 2014). An extensive number of studies have been done in nanotechnology for the encapsulation of bioactive compounds to improve its bio-accessibility or bioavailability in the human GIT (Arpagaus et al., 2018; Assadpour & Jafari, 2019a, 2019b; Bao et al., 2019; Capriotti et al., 2019; Fathi et al., 2014; Rashidinejad et al., 2016; Shen et al., 2019; Shishir et al., 2018; Zou et al., 2019). However, there is a need for a comprehensive understanding on nanoencapsulation of bioactive compounds and their incorporation and functionality in novel functional foods.

Release Kinetics of Encapsulated Bioactive Compounds

Release mechanisms of encapsulated bioactive compounds depend on the composition of the particle's stability or the charge of the nanoparticles, preparation method and the release media. Release kinetics can be explained by a mathematical model based on a phenomenon that improves the predictive efficiency (Siepmann et al., 2001). A single mathematical model is not capable of explaining the drug delivery system most of the time, however multiple models can be used more efficiently and precisely. Factors that influence mathematical prediction of the release kinetic of encapsulated bioactive compounds are polymer solubility, dose of the bioactive, molecular weight and size of the polymer, size and shape of the nanoparticles (Kassama & Misir, 2017).

*Higuchi and Baker & Lonsdale model.* The Higuchi mathematical model describes the bioactive ingredients released from an encapsulated core matrix (Higuchi, 1963). The Higuchi model is valid to identify the release of water soluble and low soluble drugs incorporated in gel and solid matrices. The model expression is given by the equation:

$$Q = A \, [D \, (2C - C_s) \, C_s \, t]^{1/2} \qquad (2.1)$$

where Q is the amount of drug released in time t per unit area A, C = initial concentration of drug, $C_s$ = solubility of drug in the media and D = diffusion coefficient of drug. Simplified Higuchi model describes the release of drugs from an insoluble matrix as a square root of the time dependent process based on the Fickian diffusion equation.

$$Q = KH \, t^{1/2} \qquad (2.2)$$

The equation can be explored by plotting the cumulative percentage drug release value against the square root of time. The slope of the plot is the Higuchi dissolution constant (KH). The major benefit of this equation is that it can include the possibility to facilitate device optimization and explain the underlying drug release mechanisms.

Baker and Lonsdale (1974) developed a model with the improvement of the Higuchi model to describe drug release from spherical shaped materials which can be explained by following equation:

$$f = 3/2 \; [1 - (1 - Mt/M\alpha)^{2/3}] - Mt/M\alpha = Kt \hspace{2cm} (2.3)$$

where Mt / Mα is the fraction of drug released at time t and for simplified appearance it can be denoted as Q and the equation will be f = 3/2 [1- (1-Q) 2/3] – Q = Kt. The slope of the plot will be the release constant (K) (Paarakh et al., 2018).

*Hixson-Crowell cube root law.* Hixson and Crowell (1931) first proposed a cube root law to represent the dissolution or decreasing rate of solid surface area as a function of time. This model is mostly applicable for a system where the erosion of surface area and hydrodynamic diameter of particles happens with time. This cube root law can be written as

$$Q_t^{1/3} = Q_0^{1/3} - K_{HC}t$$

$$Q_0^{1/3} - Q_t^{1/3} = K_{HC}t \hspace{2cm} (2.4)$$

where $Q_t$ = remaining weight of solid at time t, $Q_0$ = initial weight of solid at time t = 0, and $K_{HC}$ represents the dissolution rate constant.

*Korsmeyer-Peppas model.* Korsmeyer et al. (1983) formulated a simple relationship between the releases of bioactive compound from the polymeric core-based matrices. Korsmeyer and Peppas (1984) and Ritger and Peppas (1987a, 1987b) established a 15 empirical equation to differentiate between Fickian and non-Fickian release of bioactive compounds from swelling as well as non-swelling polymeric delivery systems. The Korsmeyer Peppas model is explained by following equation:

$$M_t/M_\alpha = kt^n \hspace{3cm} (2.5)$$

where $M_t$ = amount of drug released, $M_\alpha$ = initial drug load, $M_t/M_\alpha$ = fraction of drug released at time $t$, $k$ = Korsmeyer-Peppas constant, $t$ = time, $n$: release exponent. The release mechanism of the lycopene can be determined based on the release exponent ($n$) value, (Ravi and Mandal, 2015).

*Hopfenberg model.* Hopfenberg (1976) developed a mathematical model which explained the release kinetics of bioactive compounds or chemical compounds by surface erosion of polymers until the surface area remains constant during the erosion process. The cumulative fraction of the drug released at time $t$ was described as

$$M_t/M_\alpha = 1 - [1 - K_0 t/CL\ a]^n \hspace{2cm} (2.6)$$

where $K_0$ is the zero-order rate constant describing the polymer degradation (surface erosion) process, CL = initial drug loading throughout the system, a = system's half thickness, and n = exponential value that varies with geometry n = 1, 2 and 3 for slab (flat), cylindrical and spherical geometry, respectively.

*Gallagher Corrigan model.* Bioactive compound releases from its biodegradable polymeric carrier by the combination of diffusion and degradation mechanism is also known as a biodegradable polymeric drug delivery system. Polymer degradation profiles are usually sigmoidal shape best described by the Gallagher and Corrigan mathematical model. The kinetic profile of polymer degradation described by the Gallagher-Corrigan equation consisting of the initial "burst effect" contributed by bioactive adsorbed on the surface of nanoparticles matrix and follow by subsequent slowing down of the release profile due to the polymer matrix erosion (Gallagher & Corrigan, 2000). The total fraction of drug released ($f_t$) at time (t) is

$$f_t = f_{tmax} \left(1 - e^{-K_1 . t}\right) + \left(f_{t\,max} - f_B\right) \left(\frac{e^{K_2 . t \, - \, K_2 . t_{2\,max}}}{1 + e^{K_2 . t \, - \, K_2 . t_{2max}}}\right) \qquad (2.7)$$

where $f_t$ = fraction of drug released in time (t); $f_{tmax}$ = maximum fraction of drug released during process; $f_B$ = fraction of drug released during 1st stage – the burst effect; $k_1$ = first order kinetic constant (1st stage of release); $k_2$ = kinetic constant for 2nd stage of release process–matrix degradation; $t_{2max}$ = time to maximum drug release rate. This calculated $f_t$ is plotted against the time, and the correlation coefficient and coefficient of determination can be calculated to understand the suitability of the model.

## 2.2.    Food Grade Nanometric Delivery Systems

The nutritional value of foods can be improved by the addition of natural bioactive compounds like antioxidants, probiotics, and polyunsaturated fatty acids. However, food bioactives prevent diseases like cancer and heart disease and, in addition, contribute to the overall health and wellbeing (Jimenez-Colmenero, 2013). Nanoencapsulation can help to improve stability (Trombino et al., 2009) , functionality (Swarnakar et al., 2011), solubility (Fathi et al., 2011), cellular uptake (Hu et al., 2012; Hamdy et al., 2009; Harush-Frenkel et al., 2007; Gaumet et al., 2010; Sahoo et al., 2002; Wang et al., 2012; Win & Feng, 2005; Zhang et al., 2008; Li et al., 2010; Yoo & Mitragotri, 2010), and bioavailability (Anand et al., 2010) and may also provide improved controlled release efficacy of the bioactive compounds (Kumari et al., 2010). It is therefore widely accepted that nanoparticles may offer distinct advantages for the delivery of bioactives and improve its stability and the control release to localized targeted region. Due to the increasing interest in the use of nanoparticles for food and oral drug delivery, consumer safety has become a major concern (Vega-Villa et al., 2008). Translocation of the nanoparticles into the tissue due to their small size and higher physiological dose could cause potential safety concerns and challenges to the future of adopting nanotechnology. The type of nanoparticle and its relative chemical, physical, and morphological properties affect their interaction with living cells (phagocytic cells) which further determine the path of excretion from the gastrointestinal (GI) system and possible toxicity effects into the human metabolism.

Type of Delivery Systems

Nano-delivery systems are divided into two main groups, liquid or solids. The liquid delivery system consists of three types: nano-emulsions, nanoliposomes, and nano-polymersomes. The three types of solid nano-delivery systems are lipid nanoparticles, polymeric nanoparticles, and nanocrystals. The lipid particles can also be divided into nanostructured lipid carriers (NLC) and solid lipid nanoparticles (SLN) while the polymeric nanoparticles are classified into nanospheres and nano-encapsulates.

Emulsion and Emulsion Stability

Emulsions are systems that consist of a mixture of two immiscible liquids, generally oil and water. While nano-emulsion droplets deals with particles ranging between 10 and 100 nm in diameter are characteristically looks like transparent as sizes are smaller than the ultraviolet-visible light range. Some of the common surfactants are phospholipids, amphiphilic proteins, or polysaccharides which contribute to the stability of emulsions (Guzey & McClements, 2006). Surfactants enhance the controlled release while improving the entrapment efficiency and likewise minimizing the rate of degradation (McClements & Li, 2010; Carrillo-Navas et al., 2012). For instance, nanodelivery in hydrophobic in nature where the bioactive compounds are dissolved in internal organic phase of an oil-in-water emulsion. But double emulsions are employed for delivery of hydrophilic molecules (Sapei et al., 2012).

There are two methods (mechanical or chemical) that are commonly used for synthesizing nanoemulsion. Mechanical processes are considered a high energy process which employs sonicators and or microfluidizers to breakdown larger emulsion droplets into smaller ones. The spontaneous formation of emulsion droplets due to their hydrophobic effect of lipophilic molecules are considered as low-energy or chemical methods (McClements & Li, 2010).

## Liposomes

Liposomes are self-assembled nanoparticles with phospholipids bilayers that provide major effect in the nanoparticle formulation. Liposomes are ideal for carrying hydrophilic molecules due to their polarity of the vesicle (Tamjidi et al., 2013), but the hydrophobic molecules can be delivered using liposomes to entrap the component in the bilayer. The lipid bilayer protects the encapsulated materials from external conditions of the food component. The most potential benefit of the liposomes is it can be formulated without the use of organic solvents (Madrigal-Carballo et al., 2010). Liposomes can be either small unilamellar vesicles (SUV) or multilamellar large vesicles (MLV). The size of SUV are usually up to 100 nm; whereas MLVs are commonly within the range between 500 nm and 5 μm. Gentle hydration (Wang et al., 2010) and the layer-by-layer electrostatic deposition (Madrigal-Carballo et al., 2010) are the two most common methods to synthesize liposomes. The most common applications of liposomes are creating iron-enriched milk (Xia & Xu, 2005), antioxidant delivery (Madrigal-Carballo et al., 2010), and codelivery of vitamins E and C with orange juice (Marsanasco et al., 2011).

## Polymersomes

Like liposomes, polymersomes are vesicles formulated from amphiphilic copolymer bilayers with bioactives entrapped to form cavities (Bouwmeester et al., 2009). Polymersomes can be used to encapsulate both hydrophilic and hydrophobic bioactive compounds like liposomes. Hence, their improved controlled release and increased stability and versatility are the major advantages for polymer carriers than that of lipid carriers (Rastogi et al., 2009).

<p style="text-align:center">Nanocrystals</p>

Nanocrystals improve the solubility of poor water-soluble drugs by increasing the surface-area-to-volume ratio which increases the dissolution rate of the bioactive in vivo (Sun & Yeo, 2012). The unique advantage of nanocrystals is that they have 100% bioactive loading efficiency and can be prepared without organic solvents (Borel & Sabliov, 2014). Nanocrystals enhanced the solubility and cellular interaction with the bioactive compounds (Hunter et al., 2012). Two methods, chemical or mechanical, popularly used to formulate nanocrystals (Sun & Yeo, 2012). Starch nanocrystals are commonly prepared using stirring ultrasonication, and acid hydrolysis, to prevent aggregation of nanocrystals (Xu et al., 2010; Kristo & Biliaderis, 2007). Wet milling is applied to decrease the size but increase the uniformity of the nanocrystals (Sun & Yeo, 2012). Another technique of forming nanocrystals is nanoprecipitation (Sun & Yeo, 2012). Food grade starch and protein nanocrystals are developed for various applications in the food and biomedical therapies (Tzoumaki et al., 2011; Neto et al., 2013; de Mesquita et al., 2012). Lutein nanocrystals for oral delivery are in the process of development; however, the significant challenges are to maintain crystals sizes within the 100 nm range (Mitri et al., 2011).

Lipid Nanoparticles

Solid lipid nanoparticles (SLN) are similar to emulsion, but lipids are used to interface solids instead. However, it was noted that solidified lipids are digested slowly than liquid interphase (McClements & Li, 2010). The advantage of SLN is that they provide a more stable platform for entrapping and control release of lipophilic molecules without the use of organic solvents, and it also decreases the mobility of the bioactive in the solid matrix (Mehnert & Mader, 2012). Despite its low loading efficiency, it has a burst release behavior from crystal structure at storage condition (Tamjidi et al., 2013).

Researchers are commonly used high energy methods to synthesize SLN for food applications (Tamjidi et al., 2013; McClements & Li, 2010) such as including microfluidization and ultrasonication (Mehnert & Mader, 2012). Hot and cold homogenization techniques are used to form SLNs (Tamjidi et al., 2013). Solid lipid nanoparticles often vary in size depending on the type of surfactant and concentration used; too high or low concentrations often result in particle aggregation. However, use of ionic surfactants often results in smaller particle sizes, hence more than one surfactant is used to increase the stability of emulsions (Mehnert & Mader, 2012). SLNs were developed and used to improve the stability of quercetin (Li et al., 2009), beta carotene, and alpha tocopherol (Trombino et al., 2009) and the shelf life of guava (Zambrano-Zaragoza et al., 2013),

Polymeric Nanoparticles

Polymeric nanoparticles formulated by entrapping bioactive compounds into its matrices with the help of an emulsifier or surfactant (des Rieux et al., 2006; Hunter et al., 2012; Plapied et al., 2011). Hence, particles, nanospheres, and nanocapsules is to protect the entrapped bioactive from degradation (Mishra et al., 2010). The polymeric nanoparticles can be engineered to have mucoadhesive to enhanced intestinal permeability (Chen et al., 2011).

Solvent displacement or nanoprecipitation or desolvation, salting out, and emulsion evaporation methods are popular methods to synthesize polymerics nanoparticles (Sabliov & Astete, 2008). Natural carbohydrates such as gum arabic and maltodextrin are engineered to improve the stability and bioavailability of bioactive compounds (Peres et al., 2011). Protein polymer (zein) (Wu et al., 2012) and gelatin (Shutava et al., 2009) have been successfully used to encapsulate essential oils and polyphenol, with the potential for controlled release of the bioactives.

## 2.3. Production of Nanometric Delivery System

Factors for Nano-delivery System

The physicochemical properties of nanoparticles affect their absorption capacity, size distribution, metabolism, and excretion (ADME) which is essential in modulating it in the *in-vivo* delivery of bioactive compounds. Properties of nanoparticles can influence *in vivo* interactions are size, charge, hydrophobicity, and targeting properties.

*Size of the nanoparticles.* The mechanism of cellular uptake, immune cell stimulation, and particle clearance particularly depend on the size of the particles or nutrient density (Naahidi et al., 2013). Nanoparticles or small molecules enter the cells through endocytosis, and the specific type of endocytosis through which the nanoparticle enters the cell determines the translocation of the entrapped molecule inside the endosome (Sahay et al., 2010). The particle size between 60 and 1,000 nm commonly passes through the endosome while the enterocytes prevent the endocytosis of the larger (> 500 nm) particles and are excreted through the feces before reaching the bloodstream (Yoo & Mitragotri, 2010). The size of particles may affect the immune response as the particle size grows larger than 200 nm, and is more likely to be eliminated by cells in the mononuclear phagocyte system (MPS) (Choi et al., 2011). Nanoparticles between 1 and 20 nm remain in the vasculature longer than larger particles; hence, particles between 30 and 100 nm evade the MPS and thus avoid clearance by macrophage cells, and allow long circulation times (Faraji & Wipf, 2009).

*Hydrophobicity and hydrophilicity.* Hydrophobicity affects cellular uptake, distribution, interaction with immune cells and plasma proteins, and clearance from the body (Naahidi et al., 2013). Because of their large activity coefficient, hydrophobic nanoparticles diffuse into epithelial cells more easily than hydrophilic particles (Powell et al., 2010; Acosta, 2009). Hydrophobic nanoparticles are identified as foreign bodies by macrophages of the MPS in the blood stream and are often excreted through feces with the help of bile salt (Naahidi et al., 2013; Bertrand & Leroux, 2012).

Hydrophilic nanoparticles are quickly excreted through the kidney (Choi et al., 2011) due to their smaller size (< 10 nm) and are less likely to be identified as foreign materials in the bloodstream (Naahidi et al., 2013). Hydrophobic interaction of nanoparticles prevents the nanoparticles from traversing through mucus at a slower rate than that of the hydrophilic particle (Lai et al., 2009).

*Charges of the nanoparticles.* Entry into cells, immune cell stimulation, plasma proteins, and toxicity are significantly influenced by the charge of the nanoparticles passing through it (Naahidi et al., 2013). Positive charges of the nanoparticles are more likely to improve an immune reaction than neutral and negative charges (Naahidi et al., 2013; Bertrand & Leroux, 2012). Moreover, circulation time for neutral charged particles observed to be longer in the blood stream compared to positively or negatively charged particles. Moreover, negative charges are associated with longer circulation times than positive charges (Bertrand & Leroux, 2012).

Targeted delivery of nanoparticles affects the distribution and immune response in human physiology (Naahidi et al., 2013; Markovsky et al., 2012). The physicochemical properties, size, and surface charge affect the targeting of molecules which further affect immune response in the human metabolic system (Markovsky et al., 2012).

## Method for Synthesis of Nano-delivery System

Emulsion evaporation is the most common method of synthesizing nanoparticles. It is a two-step-process, where the first step, polymers, is dissolved in an organic solvent followed with emulsifying in a water solution in the presence of a surfactant. The mechanism for removing solvents depends on the temperature and nature of the solvent which may further lead to nucleation of the polymer on the water–solvent interface (Anton et al., 2008). After evaporation of the solvent, the dispersions are dialyzed to remove the unwanted or low molecular weight polymer and subsequently freeze-dried.

As encapsulated particles are made from droplets, they are added to emulsions by controlling the concentration of the surfactant used (Staff et al., 2011). The revolution per minute (RPM) (Sansdrap & Moes, 1993) of the evaporator, type of solvent (Mainardes & Evangelista, 2005) and, the sonication time are some influencing factors for particles size and polydispersity index (PDI) of the synthesized particles (Sansdrap & Moes, 1993). Low PDI is well desired for biomedical application which mostly depends on the use of stronger emulsifiers (Sansdrap & Moes, 1993).

It is assumed that high PDI value is due to the coalescence of the particles (Loxley & Vincent, 1998). Dynamic light scattering (DLS) is a popular instrument used to characterize nanoparticles (Desgouilles et al., 2012). Recent studies with a dual-color fluorescence cross-correlation spectroscopy (DC-FCCS) revealed that coalescence does not contribute to any effect on droplets if sufficiently stabilized during emulsion formulation (Schaeffel et al., 2012; Staff et al., 2013).

## 2.4. Biodegradable Polymer

### Polylactic Co-glycolic Acid

Polylactic co-glycolic acid is a polymer of lactic acid and glyolic acid, its molecular weight and the lactide to glycolide ratio of PLGA can be modified to adjust the rate of particle degradation. Lowering the glycolide content and increasing the molecular weight will protect the rate of degradation (Chasin and Langer, 1990; Witschi, & Doelker, 1998). Research has shown that increasing the lactide to glycolide content (50:50, 65:35, or 85:15 ratio of PLGA) decreased the release of bioactive compounds from PLGA nanoparticles (Mittal et al., 2007). In the same study, increasing the molecular weight of 50:50 PLGA from 14,500 to 213,000 Da prevent bioactive compounds' degradation from the encapsulated nanoparticles core (Mittal et al., 2007) which also sustained the release of bioactive compounds into the blood plasma. Brannon-Peppas (1994) identified that 50:50, 65:35, and 75:25 lactide to glycolide ratio of PLGA exhibited quick bioactive compounds release, with greater than 70% of bioactive released within the first 24 h while the encapsulation efficiency of bioactive compounds of 65:35 PLGA was found to be higher (54%) than 50:50 PLGA (30%) (Irmak et al., 2014). Moreover, Zaghloul et al. (2005) observed that an increased lactide to glycolide ratio can increase bioactive compounds encapsulation efficiency. In a study that examined polylactic acid (PLA), 85:15 PLGA, 75:25 PLGA, and a mixture of the polymers, in all cases the encapsulation efficiency of bioactive compounds was greater than 80%.

Poly Lactic Acid

*Physical properties.* Physical properties of PLA are significantly dependent on its optical purity. Polylactic acid with 100% L-unit is partially crystalline (45-70%) with a $T_m$ of around 180-184 °C (Jamshidi et al., 1998; Dorgan et al., 2000). With decreasing purity, the degree of crystallinity and $T_m$ of PLA also decreases. Polylactic acid is amorphous in structure if the optical purity is less than 87.5%. High molecular weight with different optical purity PLA has a $T_g$ value within the range between 55 and 61.5°C. Polylactic acid is strong but brittle and soluble in chlorinated organic solvents.

Linear PLA has higher crystallinity than star-shaped PLA with the same purity. The $T_c$ of PLA with various structures are around 115-125 °C. Stereo-complexation have been observed for L & D-PLA and the complex has a $T_m$ at 220°C (Bendix, 1998; Tsuji et al., 1991; Murdoch & Loomis, 1988; Loomis et al., 1990).

Higher rigidity and strength were observed when different fiber incorporated with high molecular weight PLA (>100,000 Da) compared to low molecular weight PLA (Sinclair et al., 1996; Runt et al., 1998; Ajioka et al., 1998).

*Chemical properties.* Hydrolysis is the most important method for PLA degradation but in its dry pure form, it can last for more than 10 years (Huang, 1994; Vert et al., 1995; Gilding & Reed, 1979; Huang et al., 1981; Pitt et al., 1981; Li et al., 1990; Kohn & Langer et al., 1996; Langer & Vacanti, 1993; Heller et al., 1990; Zhang et al., 1994). The rate of hydrolysis varies with many factors; hence, the hydrolysis of PLA with smaller surface-to-volume ratios is much slower and complicated. Polyglycolic acid copolymers are hydrolyzed much faster than PLA and have become the main biodegradable polymeric materials for biomedical applications such as sutures, implants, tissue engineering, and drug release when fast rates of hydrolysis are desirable. PLGA are hydrolyzed much faster compared to PLA. Polylactic acid provides crystallinity, strength, and a fast rate of hydrolysis. Increasing order of rate of hydrolysis is following a chronology: Poly-L-Lactic acid (PLLA) < Poly -D, L-lactic acid (PDLA) < Polylactic co-glycolic acid (PLGA) (Han et al., 1988).

Biodegradations of PLA only appeared due to the activity of proteinase K (Reeve et al., 1994). Evidence has been found of microbial degradation of PLA (Torres et al., 1996), but none of the studies has yet observed at appreciable rates. Lostocco et al. (1998) observed that human microbiome only degrades PCL but not PLA.

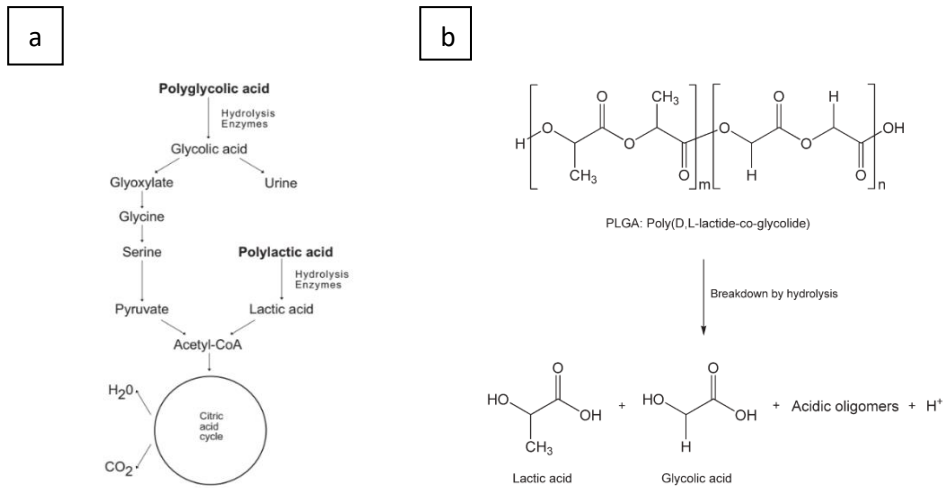

Figure 2.2. Biodegradation of PLA and PLGA in human cell. Breakdown of Polylactic acid to lactic acid and glycolic acid (a) and breakdown of PLGA to lactic and glycolic acid (b).

Polylactic acid and polylactic co-glycolic acid undergo a two-phase degradation process in the body metabolism. The first phase is the physical phase where water molecules hydrolyze long chain polymers into short chain polymers in which molecular weight and strength of the polymeric bond is reduced. Phagocytosis is the second phase of polymeric degradation where degradation is executed by macrophages (Pietrzak et al., 1997). Polyglycolic acid (PGA) undergoes cellular enzymatic hydrolysis to glycolic acid and PLA into lactic acid (Figure 2.2) which further metabolizes to produce ATP (Adenosine tri-phosphate or energy), carbon dioxide and water by citric acid cycle and the residuals are excreted by respiration or urine (Hollinger & Battistone, 1986). The degradation of PGA and PLA is accelerated *in vivo* by cellular enzymes (Vasenius et al., 1990a) and free radicals (Williams, 1992; Ali et al., 1993). The rate of biodegradation depends on shape, size, purity, molecular weight, hydrophobicity and crystallinity of the polymer (Tormala et al., 1998).

### 2.5.  Challenges of Nano-delivery of Bioactive Compound

Nano encapsulated bioactive ingredients can be used to fortify different juices or drinks for delivering bioactive compounds through human diets. Nonetheless, some potential problems should be addressed before using the nanoparticles as a functional ingredient for fruit juices and making them available to consumer markets:

      a.  Though nanoencapsulation provides potential protection against the non-targeted release of bioactive compounds, the pH, and acidity, and

interaction with other nutrients could cause premature release, thus negatively affect the sensory properties of fortified liquid juices.

b. Fortification of beverages can improve the pH and acidity stability, color, viscosity, taste, and mouthfeel of the juices which would impact the overall consumers' behavior toward the products. Even if the encapsulated bioactive has no undesirable effect on the properties of the fortified fruit juices, the producer must deal with some strict regulations which at some point is not even clear to the producer or even not in place yet.

c. The toxicity of the nanoparticles at specific concentrations is still not well understood.

A long-term toxicity study is needed to evaluate the actual effects of new nanoparticles. Therefore, extensive research is required to verify the guidelines around the safety levels of encapsulated bioactive ingredients. Digestion and absorption profiles are crucial for identifying the actual delivery efficiency of bioactive compounds to the target cell. This delivery efficiency data can help to develop new functional foods with improved bio-accessibility.

## 2.6. Bioavailability and Bio-accessibility of Nanoencapsulated Bioactive Compounds

Simulation of Human Digestive System

*The anatomy of the GIT system.* Foods components digested (mechanically or chemically) through the gastrointestinal tract (GIT) to transform it into a form absorbable for the intestine lining (Figure 2.3). The GIT is an 8-9 m length muscular tube which starts at the mouth and ends in the anus (Bornhorst et al., 2014). At the oral cavity (starting point), the food particles are converted from larger to smaller particles by the mechanical action of the teeth and are lubricated by saliva. Additionally, the alpha amylase present in saliva breaks down the starch constituent to certain extend. When the food is sufficiently broken down and lubricated with saliva in the oral cavity, a bolus is formed and ready to be swallowed (Barrett, 2014; Hutchings & Lilliford, 1988). Physical degradation of food particles continues by the peristaltic (contraction of muscle in the gastric antrum) movement and enzyme activity of the stomach during the gastric digestion. The chemical breakdown is facilitated with the gastric secretion (HCl) and enzymes (pepsin and lipase) which hydrolyze the protein and fats, respectively. Once the food particles are degraded to a sufficient degree, it is then emptied from the stomach into the small intestine (Meyer, 1980). Additional chemical breakdown continues when a variety of enzymes (trypsinogen, chymotrypsinogen, amylase, and lipase) are secreted from the pancreas into the small intestine. Hence, the majority of the absorption takes place in the small intestine. The remaining materials are transported to the large intestine, where they are fermented by the anaerobic bacteria prior to its excretion (Bornhorst & Singh, 2013; Gopirajah & Anandharamakrishnan, 2016; Singh et al., 2015). The digestion of food may appear to be purely biochemical, but current research has linked digestion to unit operations that are being applied by many researchers (Boland, 2016; Bornhorst et al., 2016).

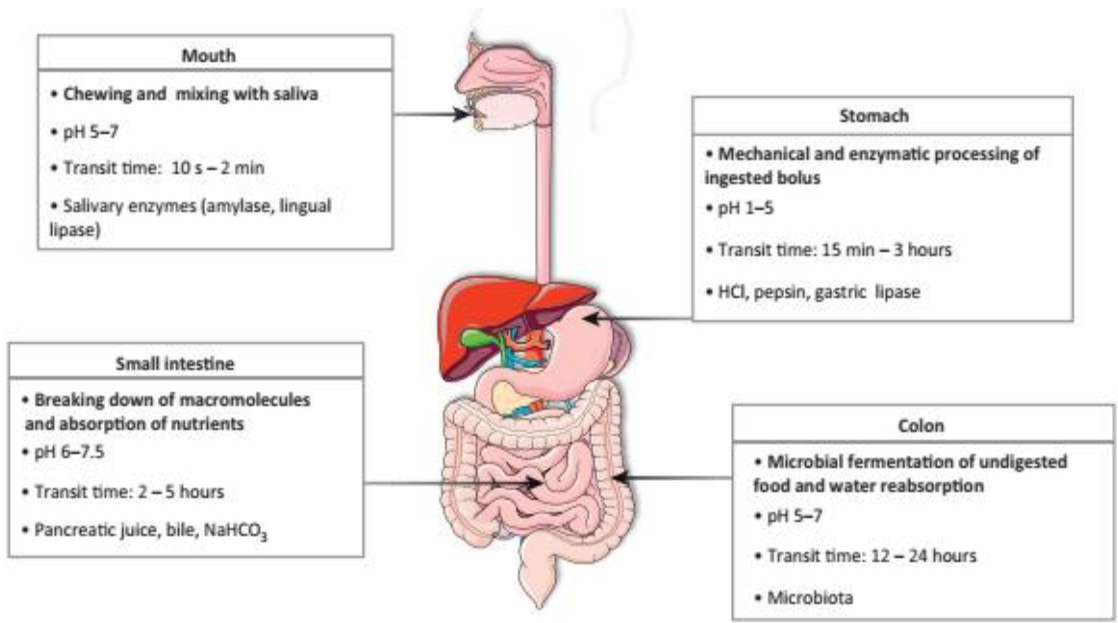

Figure 2.3. Human Digestive System Adopted from (Guerra et al., 2012).

*Simulation and function of the saliva in GIT.* Ptyalin (Salivary amylase) is an enzyme that breaks down carbohydrates into the dextrins after being secreted from the salivary gland of the mouth. The preparation of saliva simulation is also mentioned in some of the research literatures (Grand et al., 2004; Bornhorst & Singh, 2013). About 30% of the total carbohydrate digestion occurs in the mouth (Grand et al., 2004). According to Bornhorst and Singh (2013), to simulate saliva mucin (1 g/L), ptyalinα-amylase (1.18 g/L), NaCl (0.117 g/L), KCl (0.149 g/L) and NaHCO$_3$ (0.21 g/L) should mix in deionized water. The pH of the salivary digestion is adjusted to 7.0 by adding 0.01 N NaOH and the temperature of the solution must be maintained at 37ºC.

*Simulation and function of the Gastric juice in GIT.* In the stomach, the proteins, carbohydrates, and lipids are degraded by the chemical action of pepsin, amylase, lipase, and the hydrochloric acid. The mechanical degradation of the bolus caused by peristaltic movement of the stomach wall, partially degrade the food components into its monomers (Bornhorst et al., 2016). The gastric cells secrete the pepsin where the parietal cell and fundic chief cell secretes hydrochloric acid and lipase. Bornhorst and Singh (2013) also used the following simulated formulation of gastric juice: Mucin (1.5 g/L), NaCl (8.78 g/L), and pepsin (porcine pancreas) (1 g/L). They adjusted the pH to 1.8, using 0.1 N HCl. The maximum activity of the pepsin was at pH 2.0 and then dropped to < 50% at pH 3. Pepsin further is inactivated when the pH > 5 (Kondjoyan et al., 2015; Piper & Fenton, 1965; Pletschke et al., 1995). Gastric lipase is optimum between pH 5 and 6, decreasing to < 40% with pH less than 4 (Hamosh, 1990; Gargouri et al., 1989). Maintaining the peristaltic movement with an automatically agitating water bath at 100-120 rpm could be a significant simulation unit operation for getting similar results of the in-vivo model (Mennah-Govela & Bornhorst, 2016b).

*Simulation and function of intestinal juice in GIT.* After the completion of digestion in the stomach, food particles are transferred to the intestine for further disintegration. There are three types of enzymes (amylase, lipase and protease) responsible for intestinal digestion which are secreted from the pancreas (Bornhorst, 2017). For the intestinal environment, the pH must be adjusted to about 5 by adding 2 mol NaOH/L before 3 mL of a mixture of porcine pancreatin (4g/L) and porcine bile salt (25g/L) which further dissolved in 0.1 NaHCO₃ stabilized with 1% DL-α-tocopherol. The pH will be further increased to 7.5 by adding 2 mol NaOH/L and incubating for 30 min to complete the digestion (Hedren et al., 2002).

*Impact of In-vitro digestion and Bio-accessibility.* The GIT is a long muscular tube, and the digestion starts from the oral cavity passing through the stomach and small intestine and ends in the large intestine. Each step of the human digestion goes through different enzymatic and chemical degradation processes which further impact the bio-accessibility and bioavailability of nutrients. The carbohydrate component of the food started disintegrating in the oral cavity and was completed in the intestine by pancreatin enzymes. The protein and fat digestion begins in the stomach with gastric juice and ends in the intestine with pancreatic enzymic digestion. Each step of the digestion process is regulated through enzymes, which depend on the volume of food and their physicochemical properties of the food particles ingested through the oral cavity (Bornhorst, 2017).

After the digestion process, a marked loss ranging from 37 to 77% of antioxidant capacity, linked to the hydrophilic fraction, was observed. In contrast, the lipophilic and methanolic fractions increased the intensity of antioxidant activity ranging from 9 to 40% after gastric digestion. However, some researchers observed a precipitous loss after completing each digestion (Tommonaro et al., 2017). Encapsulation protects against marked losses of antioxidant potential in the GIT.  Factors such as granular shape, crystallinity, and integrity of corn starch of foods constituent significantly influence bio-accessibility during processing, even with latest non-thermal treatment, such as Pulse

Electric Field processing (Han et al., 2009). Therefore, a significant challenge of encapsulation is the degree of protecting the bioactive compounds from decreasing the rate of bio-accessibility into the human digestive system due to the processing effect.

Therefore, the main purpose of making lycopene emulsion is to protect its antioxidant properties while improving its bio-accessibility (Ha et al., 2015). Higher antioxidant properties are exhibited with droplet sizes between 100 to 200 nm compared to droplet sizes <100 nm. Additionally, smaller droplet (<100 nm) has a slow reaction rate of DPPH and ABTS reactions. According to Ha et al. (2015), droplet sizes larger than 100 nm and lower than 100 nm were found to have a bio-accessibility value of 0.53 and 0.70 which indicates that droplet sizes lower than 100 nm enhance the in-vitro bio-accessibility of lycopene more than droplet sizes higher than100 nm.

Katlijn et al. (2012) found a significant direct effect of olive oil (2%) addition and increase carotenoids uptake with the micellar phase. They also found that the bio-accessibility decreased with increasing particle size which concludes that the cell wall was the main barrier to lycopene bioavailability or bio-accessibility. Hence, the bio-accessibility depends on the integrity of the cell wall as well as its interaction with the different components of the complex food matrix.

*In vitro simulation model of the GIT system.* Many in vitro gastric and small intestinal models (static or dynamic model) have been developed to identify the fate of orally ingested food substances (Guerra et al., 2012). Based on Guerra et al., the static mono compartmental model was found to be the most widely used digestive system. The model simulates the oral, gastric, and intestinal digestion with bioreactors, with fixed pH &

temperatures and specified enzyme activities. The lack of reproducibility, gastric emptying, continuous pH, and secretion flow control are the main shortcomings of this model; hence, the mono and multicompartmental dynamic model were introduced. Even the monocompartmental models do not adequately control the in-vivo peristaltic actions; hence, most researchers adopt the dynamic multicompartmental model as a simulation model to mimic the human digestive system.

## Rumen Digestion

Sheep, cattle and goats popularly known as ruminant livestock. Ruminants are special types of digestive system that is getting energy from fibrous plant other than herbivores. Unlike the monogastric such as swine and poultry, ruminants digestive system can ferment feedstuffs and supply precursors for energy use. To get a better idea about their feed ration understanding about how the digestive system of the ruminant works is mandatory.

*Ruminant digestive anatomy and function.* The ruminant digestive system (Figures 2.3 a & 2.3 b) uniquely qualifies ruminant animals such as cattle to efficiently use high roughage feedstuffs, including forages. The anatomy of the ruminant digestive system includes the mouth, esophagus, four-compartment stomach, small intestine (duodenum, jejunum, and ileum), and large intestine (cecum, colon, and rectum). Four compartments include rumen, reticulum, omasum, and abomasum where mouth consists with teeth and saliva.

Ruminants eat rapidly, and they swallow much of their feedstuffs (< 1.5 inches) without chewing it appropriately. The process of rumination or "chewing the cud" is where forage and other feedstuffs are forced back to the mouth for further disintegration by teeth called chewing and mixing with saliva. Hence, the esophagus functions bidirectionally in ruminants, permitting them to churn out their cud for additional chewing, if necessary. This cud is then engulfed again and passed into the reticulum. Then the solid segment transferred slowly to the rumen for digestion while most of the liquid portion swiftly moves from the reticulorumen into the omasum and then abomasum. The solid portion left behind in the rumen normally stays there for up to 48 hours and forms a concentrated mat in the rumen, where microbes can use the fibrous feedstuffs as a precursor for energy production.

Typically, the rumen and reticulum have similar functions and are separated only by a small muscular fold of tissue. The reticulorumen is the place for microorganisms (rumen bugs) which includes fungi, bacteria and protozoa. These microbes' breakdown the plant cell walls into carbohydrate monomers and hence produce volatile free fatty acids (VFA), such as priopionate (used for glucose synthesis), acetate (used for fat synthesis), and butyrate (used for carbohydrates synthesis). Ruminant later uses the VFA to produce energy.

The rumen acts as a fermentation vat and a host for microbial fermentation. About 50 to 65 percent of starch and soluble sugars in the diet is digested in the rumen. The rumen microorganisms (primarily bacteria) digest cellulose from plant cell walls, complex starch, and synthesize protein from nonprotein nitrogen, and synthesize vitamin B and vitamin K. Rumen pH typically ranges from 6.5 to 6.8. The rumen provides an anaerobic (without oxygen) environment with pH modulating between 6.5 to 5.8, and the gases produced include carbon dioxide, methane, and hydrogen sulfide.

The omasum is spherical in shape and connected to the reticulum by a short tunnel. It is also called the "many piles" or the "butcher's bible" in reference to the many folds or leaves that resemble pages of a book. These folds increase the surface area which increases the area that absorbs nutrients from feed and water, and cattle usually have a highly developed and large omasum.

The abomasum is the "true stomach" of a ruminant, and it is the compartment that is like the stomach in a nonruminant. The abomasum produces hydrochloric acid and digestive enzymes, such as pepsin (breaks down proteins), and receives digestive enzymes like lipase (breaks down fats) secreted from the pancreas with the pH generally ranging from 3.5 to 4.0, and these secretions help prepare proteins for absorption in the intestines. The chief cells in the abomasum secrete mucous to protect the abomasal wall from acid damage.

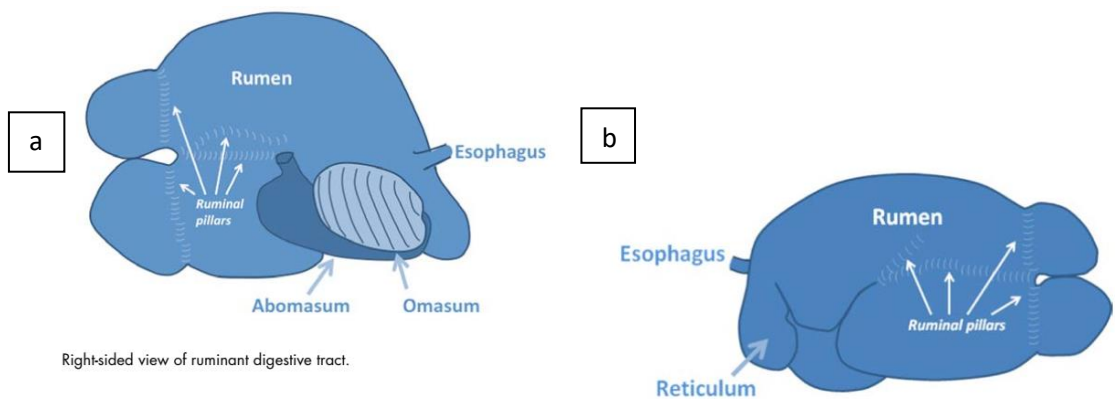

Figure 2.4. Rumen digestive tract from both r                    .

*Nanotechnology in rumen nutrition.* The concept nanotechnology, invented by Richard Feynman in 1959, was first introduced into the research world by Norio Taniguchi in 1974 (Feynman, 1960; Toniguchi et al., 1974; Warad & Dutta, 2005). It is the science of manipulation of atoms, molecules at macromolecular scales (Mannino & Scampicchio, 2007). The FDA (Food and Drug Administration) in 2006, defined nanomaterials as particles smaller than micrometric level. In the agriculture and food industry, nanotechnology was discussed for the first time in the United States Department of Agriculture's (USDA) action plan published in September 2003 (Jospeh & Mornison, 2006). Currently, it is used to provide greater control over the nature of food, such as taste, texture, processing speed, heat tolerance, durability, safety, bioavailability of nutrients and its packaging (Scrinis & Lyons, 2007; Chaudhury et al., 2008; Sekhon, 2010).

Nanotechnology applications are also used in manufacturing feeds for farm animals (Chen & Yada, 2011). Nanomaterials are of sizes ranging from 1 to 100 nm usually used for animal nutrition, because of its extensive stability at high temperature and pressure (Stoimenov et al., 2002), and can easily be incorporated in digestive juices (Feng et al., 2009). The large surface area contributed by smaller size particles exhibit better interface with other biologically active compounds in vivo (Zaboli et al., 2013). Its interface can invade the epithelium of the small intestine and be circulated to the brain, lungs, heart, kidney, liver, and stomach (Hillyer & Albrecht, 2001). Size, shape and inherent stability of the particles ultimately determine its functionalities as nanomaterials (Dickson & Lyon, 2000). Currently, nanomaterials are used to bind and remove toxins and pathogens from the biological metabolism (Fondevila et al., 2009; Fondevila, 2010; Pindela et al., 2012). The use of nanominerals also helps to increase animal production and ensure their healthiness (Rajendran, 2013).

CHAPTER 3

MATERIALS AND METHODS

## 3.1. Synthesis of Lycopene Nanoparticles

Materials

Synthesizing Nanoparticles (PLA or PLGA), the PLA polymer was dissolved in dichloromethane (Alfa Aesar, USA) while PLGA in ethyl acetate (Fisher Scientific, USA) was used to make the organic phase. Tween-80 (Fisher Scientific, USA) and dimethylamine boren (Tokyo Chemicals Industry Co Ltd., Japan) were used as surfactant to formulate PLA and PLGA nanoparticles. Polylactic acid (Sigma-Aldrich, USA), PLGA (PolySciTech, USA) and pure lycopene (Indofine Chemical Company Inc, USA, purity = 95%) bought from fisher scientific, and tomatoe fruits used to extract lycopene by column chromatography were purchased from Walmart supercenter in Huntsville.

In synthesizing the PLA nanoparticles, ethyl acetate was used to dissolve PLA polymer to produce the organic phase. Tween-80 (Fisher Scientific, USA) was used as surfactant to formulate PLA nanoparticles. PLA (Sigma-Aldrich, USA) and pure lycopene (purity = 95%) purchased from fisher scientific where tomatoes were purchased from Walmart supercenter to extract lycopene by column chromatography.

## Extraction of Lycopene from Tomatoes

Roma tomatoes were collected from the local markets (Walmart Supercenter Hunstville, AL) and sliced into small pieces to increase the surface area. A Tray dryer (Honeywell, Proctor & Schwartz Inc, USA) was used for drying the tomatoe samples for 24 h at 45°C. The dried samples were ground, and 8 g was used for extraction. After dissolving the samples in acetone, petroleum, & dichloromethane (1:1), the extracts were filtered with Whatman filter paper. The solvents were released from the solid curd by squeezing the filtrate into a preset column. Next, petroleum ether, or dichloromethane in 1:1 proportion, or 95% petroleum were sequentially used as a mobile phase during each experimental run. Red band was collected into different flasks from the column and evaporated using a rotary evaporator (BM 500, Yamato, Japan). After collecting the dry extract, acetone was used to reconstitute the solution to make the stock solution with desired dilution. The purity of lycopene was verified by using UV-spectrophotometric (UVmini-1240, UV-VIS Spectrophotometer, Shimadzu) method at 505 nm of wavelength.

## Research Design for Synthesis of Lycopene-PLGA and Lycopene-PLA Nanoparticles

In this study, a three-factor factorial experimental design was used to evaluate the effect of sonication time, surfactant, and polymer concentration on physicochemical properties of PLGA-LNP. The factors were the surfactant concentration, sonication time, and polymer concentration and their levels are shown in Figure 3.1 and 3.2. The physicochemical, encapsulation efficiency, and size distribution were the main response variables (Figures 3.1 and 3.2). The ANOVA was applied on the data analysis and the Tukey mean comparison and homogeneity test were done at a 5% level of significance to identify the actual difference point among experimental units. In addition, multiple regression was done to optimize the formulation of nanoparticles.

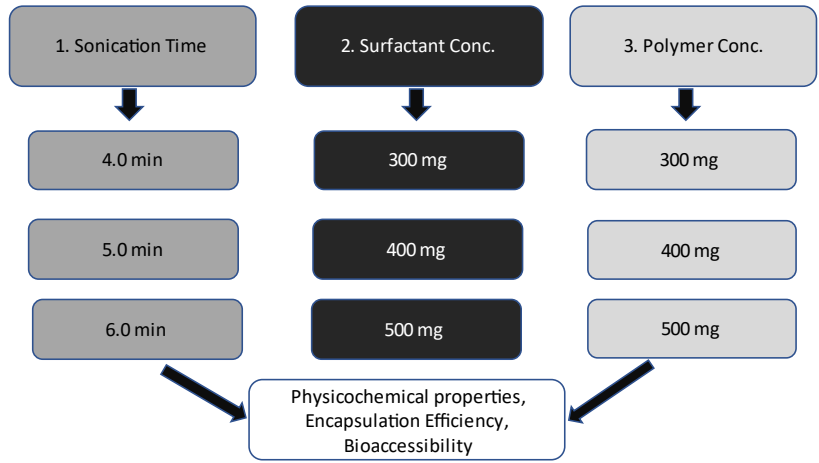

Figure 3.1. Research design for synthesis of PLGA-LNP.

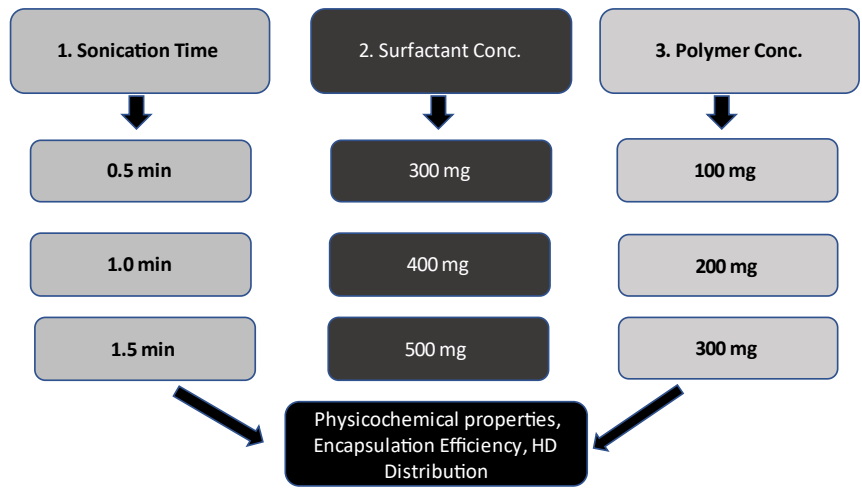

Figure 3.2. Research design for synthesis of PLA-LNP.

## Synthesis of PLGA and PLA-LNP

The emulsion evaporation method was followed to synthesize PLGA-lycopene nanoparticles (Kassama & Misir et al., 2017). Two phases were made with organic and water to formulate the encapsulated lycopene-PLGA/PLA nanoparticles. PLGA and lycopene solution (40 mg/100mL) were dissolved in ethyl acetate (organic phase) for PLGA nanoparticles where for PLA based nanoparticles, lycopene, tween-80 and PLA dissolved in dichloromethane. In the water phase, dimethylamine boren was dissolved in deionized water and transferred into the organic solution at room temperature under magnetic stirring. The organic solution was homogenized in water, and the emulsion was subjected to the Sonicator (QSonica 700, Qsonica LLC, USA) for 4, 5, and 6 min for PLGA and 30 s, 1.0 & 1.5 minutes for PLA-LNP. Next, the emulsion passed through rotary evaporator (RE 301, Yamato Scientific Co. LTD, Tokyo, Japan) to evaporate the organic solvent (ethyl acetate) from the emulsion at $35^{\circ}$C (Kassama & Misir, 2017). After evaporation, nanoparticles were collected by centrifugation for 40 minutes under 10,000 rpm (Farrag et al., 2018) at $4^{\circ}$C, followed by freeze-drying.

## Measurement of Physicochemical Properties of Nanoparticles

A Zeta-sizer (Nano ZS90, Malvern Instrument Ltd, Worcestershire, UK) instrumentation used to determine the average hydrodynamic diameter and polydispersity index of lycopene nanoemulsions at $25^{\circ}$C with a scattering angle of $90^{\circ}$. A plastic cuvette of three milliliters was filled up with lycopene nanoemulsion and placed into the sample chamber of the Zeta-sizer to measure the size and polydispersity index of the nanoparticles.

In addition, Zeta potential or the stability, mobility, and electric conductivity of lycopene nanoemulsions were measured by putting the sample into a "U" shape cuvette and running it through the zeta sizer at 25°C.

## Determination of Percentage Encapsulation Efficiency (EE)

Encapsulation efficiency was measured by the method Farag et al. (2018) developed with some modifications. The 2 mL loaded nanoparticle solution was subjected to high-speed cooling (4°C) centrifugation at 10,000 rpm for 30 min. The supernatant solution was mixed with 2 mL of acetone, followed by a UV-Vis spectrophotometric analysis at 505 nm. Lycopene content was determined by comparing the sample absorption with the standard lycopene curve. Encapsulation efficiency (EE) was calculated by Eq. 3.1.

$$EE\% = \frac{\text{(Total amount of lycopene added − Non bound lycopene)}}{\text{Total amount of lycopene added}} \times 100 \qquad (3.1)$$

## Morphological and Chemical Characteristics

*SEM imaging of PLGA nanoparticles.* The morphology of formulated PLA-LNP and PLGA-LNP were determined by SEM (Zeiss EVO 50, Scanning Electron Microscope, Japan) analysis. The LNP were dispersed in plate and dropped for coating on an aluminium stub using double sided carbon tape. The samples were then coated by gold sputtering unit at 10 Pa vacuum for 10 S (SC7620, gold sputtering unit, Japan). The typical acceleration potential used was 30kV and the image was captured at the desired magnification.

*Spectroscopy.* FT-IR analysis was done with the method developed by Sambandam et al. (2015) with some modification to identify the chemical bond inside the nanoparticles. The spectra were recorded on Perkin Elmer 1 Fourier Transform Infrared Spectrophotometer. The samples were scanned from 400 to 4000 $cm^{-1}$ and the spectra for pure PLA, PLGA, Lycopene, and PLA/PLGA-LNP data was recorded, and graphs were generated for further explanation.

<div align="center">Optimization of Lycopene Nanoparticles Formulation</div>

The ANOVA of the data was done to evaluate the effectiveness of the different treatments at 5% level of significance. The Tukey and homogeneity test were conducted at 5% level of significance to identify the actual causal effect among experimental variables. In addition, multiple regression was done to identify the most contributing factors to the nanoparticle formulation.

### 3.2. Processing Effect on Physicochemical Properties of Lycopene Nanoemulsion

<div align="center">Chemicals</div>

Juices model was developed with table sugar (sucrose, bought from Walmart supercenter) to adjust the total soluble solids (brix). Citric acid (Sigma-Aldrich, USA) and sodium citrate (Sigma-Aldrich, USA) were used to adjust the acidity and pH of the juices. Pectin (Tokyo Chemicals Industry Co. Ltd, Japan) and lecithin (MP Biomedical, USA) were used as thickening and emulsifying agent, respectively. An organic color was put into the juices to enhance its sensory appeal.

## Experimental Design

In this study, a three-factor factorial experimental design was used to evaluate the effect of pasteurization on bio-accessibility of lycopene nano-emulsion. Hence, the factors and levels are as follow: fortification at two levels (non-encapsulated and encapsulated), pasteurization at three levels (CNP = Control non-pasteurized, CP and MP), and juice concentration at two levels (5 and 15ºBrix). Hence, the experiments consisted of 12 experimental units shown in Figure-3.3. Thus, the following dependent variables were determined: color, pH and TSS, physicochemical and rheological properties of the samples. Data analysis was performed with ANOVA of the experimental data at $P < 0.05$ level of significance.   A Tukey HSD mean comparison conducted on experimental variables that were significant and a homogeneity test was done to identify the simple and individual effects of juice concentration, pasteurization, and encapsulation at their different levels.

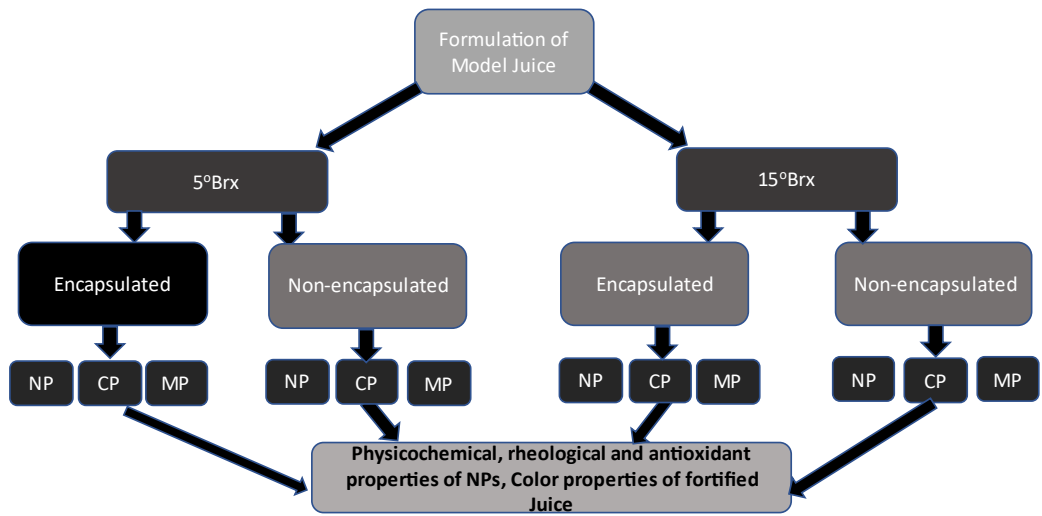

Figure 3.3. The experimental design for evaluating the pasteurization effect on physicochemical, rheological and antioxidant properties of lycopene nanoemulasion.

## Formulation of Model Juices

Table sucrose was used to adjust the juices concentration at 5 and 15° Brix. Citric acid and sodium citrate were used to adjust the pH between 4.10 - 4.40 based on the Henderson-Hasselbalch equation (equation-3.2) (Po & Senozan, 2001). Pectin (0.01%) was used as a thickening agent while the lecithin was used as an emulsifier. Laboratory grade homogenizer (Hamilton Beach 12 Speed homogenizer) was used to homogenize all the ingredients for making the artificial fruit flavored juices for the experiment.

$$pH = pKa + \log([A^-]/[HA]) \tag{3.2}$$

where pH = Acidity of a buffer; pKa = Negative logarithm of Ka; Ka = Acid dissociation constant (For citric acid = $1x10^{-4}$); [HA] = Concentration of an acid; and [A⁻] = Concentration of conjugate base

## Fortification of Juices with Lycopene Nanoparticles

Dried powdered nanoparticles were used to fortify the two different concentrated juices.  The powdered nanoparticles were directly dissolved in the juice before adding the thickening agent and subsequently blending into an emulsion. After fortifying the juices with encapsulated lycopene, they were subjected to pasteurization (CP and MP) treatment followed by colorimetric, rheological and DPPH inhibition analysis.

## Conventional Pasteurization and Microwave Pasteurization Treatments

Low Temperature Long Time (LTLT) thermal pasteurization was applied on 100 mL juice by putting into a beaker and heated in a hot air oven (30GC Lab Oven; Quincy Lab, Inc; USA) at $60°C$ for 30 minutes to obtain a 5-log reduction of microbial population a method established by Wang et al. (2018). On the other hand, the microwave pasteurization treatment was conducted with a Microwave oven (NN-H764WF, Shanghai Matsushita Microwave Oven Co., LTD, Shanghai, China) and the power level was controlled with FISCO workstation software (FISCO V-1.10.9 software, Montreal, Quebec). The temperature profile of the samples was monitored with fiberoptic temperature sensors controlled by FISCO V-1.10.9 software.  The method developed by Hashemi et al. (2019) was adopted with slight modification to run the microwave pasteurization. Microwave power of 840 W and 960 W were used for $5^o$ Brix $15^o$ Brix juice, respectively where total treatment time was 120 s and contact time at $100^oC$ was 40 s. All the samples were instantly cooled down before any subsequent measurements.

## Physicochemical Properties of Lycopene Nanoemulsion

Average hydrodynamic diameter and polydispersity index of lycopene nanoemulsions were determined with a zeta-potential and particle size analyzer (Nano ZS90, Malvern Instrument Ltd, Worcestershire, UK). All the measurements were performed triplicate.

## Antioxidant Properties of Lycopene Nano-emulsion

Antioxidant properties of lycopene nanoemulsions were identified by the DPPH method developed by Moreno et al. (1998) with slight modification. In brief, a DPPH concentration of 1 mg/50 mL of ethyl alcohol (purity = 99%) was used as a control. Samples were prepared by dissolving the nanoparticles with dichloromethane, evaporating the solvent by rotary evaporator, centrifugation, filter & re-dilution with acetone to get the desired sample solution. For the actual analysis, 200 µL of DPPH and 40 µL of the sample (and blank) were placed in a 96 well microplate reader with triplicate and deionized water used as blank. The samples were run through the UV-VIS spectrophotometer (UVmini-1240, UV-VIS Spectrophotometer, Shimadzu) at 517 nm, and data were recorded after 90 minutes of contact time.

Determination of Color and Soluble Solid Content of Fruit Juices Model

*Color.* The *L*∗, *a*∗, and *b*∗ values of the sample were measured based on colorimetric (Chroma Meter CR-400, Konica Minolta Sensing, Inc, Japan) method developed by Wang et al. (2018) with slight modification. The total color difference ($\Delta E$) and hue angle were calculated using the following equations 3.3 and 3.4

$$\Delta E = [(L - L_0)^2 + (a - a_0)^2 + (b - b_0)^2]^{1/2} \qquad (3.3)$$

$$\text{Hue angle} = \text{Tan}^{-1}(b/a) \qquad (3.4)$$

where $\Delta E$ is the total color difference between treatments and control; $L$ and $L_0$ are the lightness of sample and control, $a$ and $a_0$ are the redness to greenness of sample and control, $b$ and $b_0$ are the yellowness to blueness of sample and control, respectively. The TSS (Total Soluble Solid) was measured with a pocket refractometer (Pal-$\alpha$, ATAGO, pocket refractometer, Japan).

## 3.3. Processing Effect on Rheological Properties of Lycopene Nanoemulsion

The rheological properties were measured using a temperature-controlled and programmable Brookfield Viscometer (HBDV-III U, Brookfield Engineering Laboratories, Middleboro, MA). The samples were characterized at 4ºC using the spindle 21, hence the rheological properties of juice samples were determined at shear rate ranging between 50 to 230 s$^{-1}$, based on the preliminary study. Most common rheological models are Power law and Casson model. From this study Casson and Power law model will be carried on evaluating the rheological behavior of fortified or non-fortified juices.

Mathematical models tested are shown in the following Eq. 3.5 and 3.6:

$$\tau^{1/2} = \tau_o^{1/2} + (\eta\gamma)^{1/2} \qquad (3.5)$$

$$\tau = K. \gamma^{n.} \qquad (3.6)$$

where $\tau_o$, $\eta$, n, K represent the yield stress, plastic viscosity, flow behavior index, and the consistency index, respectively.

## 3.4. In-vitro Control Release Kinetics of Lycopene Nanoparticles

The freeze-dried sample (100 mg for PLA and 300 mg for PLGA) loaded lycopene-NPs was placed into a dialysis bag (MWCO 10 kDa) containing 10 mL phosphate buffer solution (PBS) (Swackhamer et al., 2019). An additional 10 mL of PBS was also placed in a centrifuge tube to submerge the nanoparticles containing dialysis tube. The PSB solution outside the dialysis tube was the receiving medium for lycopene that diffused out of nanoparticles inside the dialysis tube. Dimethylamine boren was used as an emulsifier; whereas butylated hydroxy toluene (BHT) was added to protect the lycopene from oxidation with atmospheric oxygen (Rao & Murthy, 2000; Swackhamer et al., 2019). A mechanical agitator was used at 120 rpm and 37°C to simulate in-vitro extracellular fluid movement through bloodstream (Rao & Murthy, 2000). A 2 mL sample was taken every day from the receiving PBS medium to evaluate the content of lycopene, and the content was quantified with a UV/Visible spectrophotometric method (UV 1800, Shimadzu, Japan) at 505 nm (Jeevitha & Amarnath, 2013).

## 3.5.    Development of Simulation Model for Human GIT

Simulation of Chemicals Components in Human GIT

*Preparation of saliva.* Reagent-a) 100 mg mL mucin (Alfa Aesar, USA), 11.7 mg NaCl (Fisher Scientific, USA), 210 mg NaHCO$_3$ (Fisher Scientific, USA), 14.9 mg of KCl (Fisher Scientific, USA), and 118 mg of α-amylase (MP Biomedical, USA) were added in 100 mL of deionized water at pH 7 (Drechsler & Bornhorst, 2018).

*Preparation of gastric juices.* Reagent-b) Porcine pepsin (0.5%) solution (100 mL):

Reagent b was made with Porcine pepsin (0.5%) (Sigma-Aldrich, USA); 3.6 mmol of CaCl$_2$.2H$_2$O (Fisher Scientific, USA); 1.5 mmol of MgCl$_2$.6H$_2$O (Fisher Scientific, USA); 49 mmol of NaCl (Fisher Scientific, USA); 12 mmol of KCl (Fisher Scientific, USA) and 6.4 mmol of KH$_2$PO$_4$ (Merck, Germany) were added into 100 mL of deionized water (Hedren et al., 2002) and stirred until homogenization appeared.

Reagent-c) HCl/L (1 N) solution:

Thirty (30) mL of 6N HCl and made the volume up to 180 mL with distilled water (Hedren et al., 2002). The prepared solution was mixed gently for making it homogenized.

Reagent-d) 0.1 N of NaOH:

Reagent d was made with 1.6 g of NaOH (Fisher Scientific, USA) which was poured into 400 mL volumetric flask and diluted up to the mark with deionized water (Hedren et al., 2002).

*Preparation of intestinal Juice.* Reagent-e) Intestinal juice

The reagent e was made with 0.45 g pancreatin (Alfa Aesar, USA), and 2.8 g bile salt (Oxoid, UK) was added into a 100 mL volumetric flask and diluted up to the mark with deionized water (Tumuhimbise, 2009).

Automation of Human Digestive System

Digestion is the process by which components of food and drinks are broken down into smallest particle sizes, so that the body can use them to build and nourish cells by providing energy. Ingestion of food and its transport flow from one chamber to the other are accurately controlled by a well-established sensing and feedback mechanisms. These mechanisms break the complex molecular structure of food components to its monomer and thus help to release the functional nutrients from the food to the body (Hedren et al., 2002). In this section, the automatic simulation model of human digestion system component shown in Figure 3.4

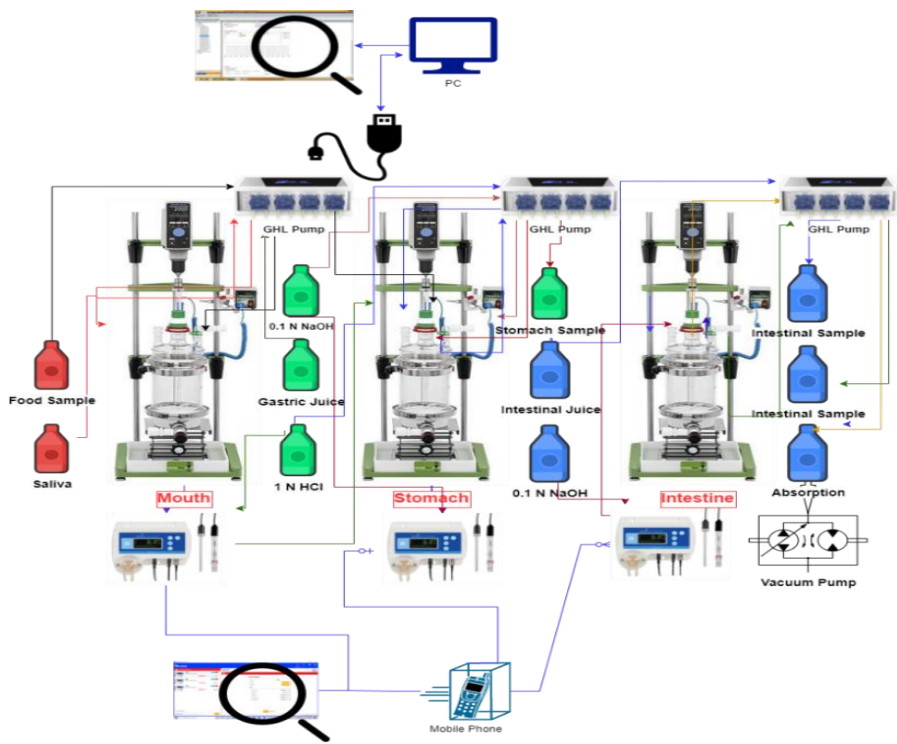

Figure 3.4. Design of Automatic in-vitro digestive system for evaluating bio-accessibility and release kinetics of lycopene-PLGA/PLA nanoparticles.

*Automation of mouth digestion.* The mechanical disintegration of food particles starts in the mouth with the help of teeth called mastication (chewing). In mastication, the food is crushed and mixed with saliva to form a bolus for swallowing. It is a complex mechanism involving opening and closing of the jaw, secretion of saliva, and mixing of food with the tongue. To simulate the mixing of salivary fluid with a non-Newtonian juice model fortified with lycopene nanoparticles. Automatic stirrer (OHS 100 Advance, Chemglass Life Science, USA) was used at 120 RPM to simulate the fluid dynamic action in mouth. The pH was controlled with a mobile app. The solid, and semi-solid foods when ingested are mixed with saliva (Hajishafiee et al., 2019) to break down the complex carbohydrate like starch to its partially disintegrated form of dextrin by the action of α-amylase enzymes (Okada et al., 2007). The saliva (Reagent-a) simulation was done by using the method developed by Assad-Bustillos et al. (2020) with slight modifications. The saliva also helps to lubricate the food and facilitate the subsequent digestion by transferring partially digested foods to the stomach through oesophagus (Okada et al., 2007). The pump (GHL-doser-2.1, USA LLC, Wilmington, NC) has been used to simulate secretion and transfer of salivary components to the mouth and transference of the bolus from mouth to the stomach.

*Automation of stomach digestion.* The stomach stores the foods that are transferred from the mouth, and it controls the transfer of food components to the small intestine. The pH of the stomach is usually extremely low about 1.5 - 2.0 in the beginning of the digestion, and it converts the pepsinogen to its active form pepsin (Piper & Fenton, 1965; Guyton & Hall, 2006). The pH was adjusted to the preset level with a GHL Doser 2.1 pump. The pH of the digesta was controlled by the "Blue lab pH controller" using automatic dosing of 1 N HCl solution into the stomach. The secretion and transfer of the gastric juice was done with a Doser 2.1 pump using its automatic command scheduling. The salivary enzyme and α-amylase coming from the mouth compartment are inactivated when the pH drops to less than 3.8 (Fried et al., 1987). After one-hour of digestion, the Doser pump starts to transfer 0.1 N NaOH solution automatically into the stomach to adjusts the pH to the range of 5.5 - 6.0 and the "Blue lab pH controller" (Software version: PHC_v1 1.0.35) helps to maintain the the pH constant for the next hour. Optimum activation pH for the gastric lipase is 5–5.4 (Sams et al., 2016). Finally, the lipase starts the hydrolytic activities and finishes in an hour. To maintain the peristaltic condition in the stomach, an automatic stirrer was employed to agitate the gastric content at a speed of 120 rpm. The gastric emptying occurs at the rate at which chyme passes from the stomach to the duodenum which is a function of its caloric density and viscosity (Marciani et al., 2001). Gastric emptying is mathematically modelled as follows in Eq. 3.7 (Elashoff et al., 1982):

$$y(t) = 2^{-\left(\frac{t}{T/2}\right)^{\beta}} \qquad\qquad (3.7)$$

where *y(t)* refers to the fraction of the chyme remaining in the stomach at time (*t*); *β* refers to the coefficient describing the shape of the curve; and T/2 refers to the gastric emptying half time calculated using Eq 3.8 (Hunt & Stubbs, 1975)

$$T/2 = V_o \times (0.1797 - 0.1670^{-K}) \qquad\qquad (3.8)$$

The parameter T/2 depends on $V_0$ = the volume of food going through digestion, and *K* = the caloric density of the foods.

*Automation of intestinal digestion.* The chyme passes from the stomach to the small intestine that comprises duodenum, jejunum and ileum where gastric contents blend with the bile and pancreatic enzymes, and the pH (7.0) was maintained at neutral for duodenal digestion. The pancreatic enzymes, i.e. lipases, proteases, and amylase, are secreted into the duodenum. In this system, intestinal juices (reagent-e) are automatically controlled through the command of GHL Doser 2.1. The volumetric flow rate of the intestinal juice was 10 mL/min. Here, high enzyme activity in the intestine induces a hydrolysis of the carbohydrates, proteins, and fats. The peristaltic movement in the intestine was simulated by an automatic stirrer at 120 rpm.

*Automation of cellular uptake of nanoparticles.* After completion of intestinal enzymatic activities, nanoparticles are absorbed through duodenal lining. In the meanwhile, before absorption of the nanoparticles, they pass through a protective mucus layer to the underlying epithelium for further hydrolysis by a brush border enzyme (Bornhorst et al., 2016). The absorption specifically happens through the duodenum for polymeric nanoparticles. The average pore size of enterocytes present in the duodenum is 200 nm (Borel & Sabliov, 2014). Hence, to simulate the absorption, a membrane filter was placed in between sample collector chamber and the receiving chamber for absorption after digestion. The membrane filtration has been done in vacuum (0.7 - 7 bar) and flow rate was constant (Singh & Heldman, 2014). The pore size of the membrane filter is about 200 nm and samples can be received under vacuum conditions.

**3.6.  In-vitro Bio-accessibility, Release Kinetics and Cellular Uptake of Lycopene Nanoparticles**

Human Digestion

*Research design.*

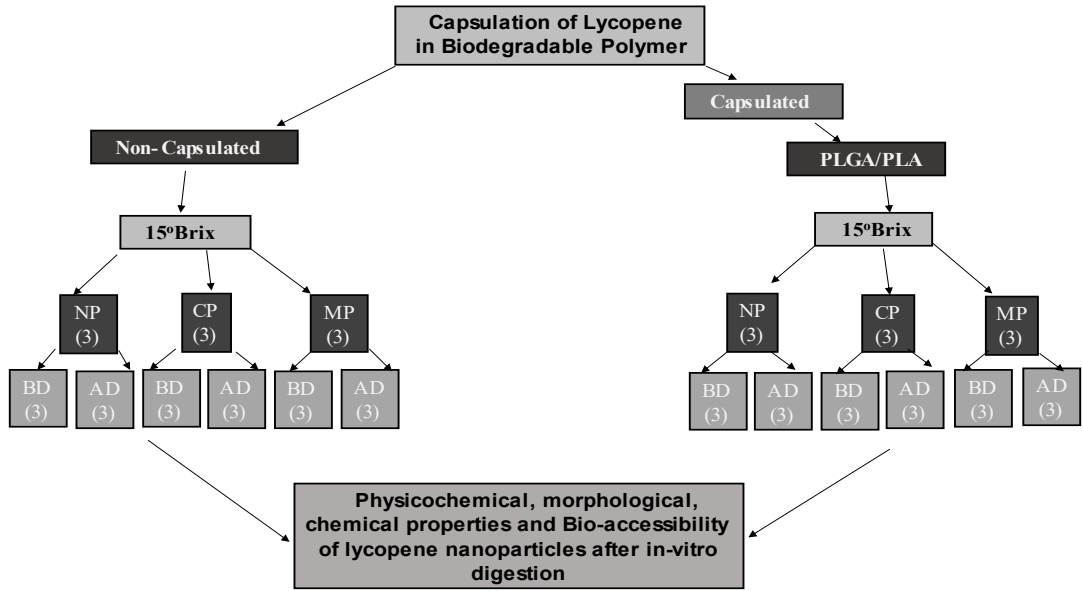

Figure 3.5. Research design for in-vitro bioavailability of lycopene nanoparticles.[1]

Three-factor factorial design (Figure 3.4) was used to evaluate the effect of pasteurization and digestion on the physicochemical, morphological, chemical and bio-accessibility of lycopene nanoparticles. Hence, the type of fortification, pasteurization and digestion time referred to be three factors for lycopene nanoemulsion.

---

[1] NP = Control Non-pasteurized, CP = Conventional Pasteurization, MP = Microwave Pasteurization, BD = Before in-vitro digestion and AD = After digestion.

*Physicochemical properties of lycopene nanoparticles.* The stomach digestion effect was studied, and samples were collected at every 15 minutes for the first hour and then 30 minutes interval for the next hour of digestion. Samples were collected every hour form the intestinal chamber to study the intestinal effect. After collecting all the samples, they are filtered through 450 nm membrane filter to separate the nanoparticles from the chyme. Physicochemical properties were determined by method developed previously in section-3.1.4.

*Bio-accessibility of lycopene nanoparticles.* The salting-out technique was used to deactivate the enzymes present in samples before the bio-accessibility study. The PLGA samples were dissolved in ethyl acetate while dimethyl amine boren for the PLA nanoparticles to break and free the lycopene from NPs core. The DPPH activity (Section-3.2.6) was used to evaluate the bio-accessibility of lycopene nanoparticles.

*In-vitro release kinetics.* The mechanism of controlled release of bioactives from the core of biodegradable polymeric nanoparticles are a combination of diffusion and degradative or erosion of the polymeric systems. Such release profiles usually are sigmoidal shaped. The Gallagher-Corrigan model is a mathematical model used to explain a single fraction of bio-actives released from a biodegradable polymeric particle. The kinetic profile shows an initial burst release due to bioactives that are not tightly bound to the polymetric monolayers followed by slow release due to the surface erosion of the nanoparticles (Gallagher & Corrigan, 2000). The bioactive release profiles usually are sigmoidal in shape, hence shows regions of instantaneous release describe as the burst release.

The total fraction of drug released ($f_t$) at time ($t$) is described by the following equation:

$$f_t = f_{tmax}\left(1 - e^{-K_1.t}\right) + (f_{t\,max} - f_B)\left(\frac{e^{K_2.t - K_2.t_{2\,max}}}{1 + e^{K_2.t - K_2.t_{2max}}}\right) \quad (3.9)$$

where, $f_t$ = fraction of drug released in time ($t$); $f_{tmax}$ = the maximum fraction of drug released during process; $f_B$ = fraction of drug released during 1st phase (the burst effect); $k_1$ = first order kinetic constant (1st phase of release); $k_2$ = kinetic constant for 2nd phase of release process–matrix degradation; $t_{2max}$ = time for maximum drug release rate. The release time ($f_t$) is plotted against the time; and the correlation coefficient and coefficient of determination determined the suitability of the model (Balcerzak & Mucha, 2010; Arezou et al., 2015).

*Measurement of in-vitro absorption profile of lycopene nanoparticles.* Absorption of encapsulated lycopene nanoparticles is described by the Korsmeyer Peppas Model as shown in equation (3.10).

$$M_t/M_\alpha = kt^n \qquad\qquad (3.10)$$

where $M_t/M_\alpha$ = the fraction of drug released at time $t$, $k$ = Korsmeyer-Peppas is a constant, $t$ = time, $n$ = mechanism of transport.

The best fit for each model, the correlation coefficient ($R^2$) values were used as a determining indicator. The release mechanism of the lycopene has been determined based on the "$n$" value (Ravi & Mandal, 2015).

Table3.1. Identification of drug transport mechanism based on release exponent of Korsmeyer Peppas model equation (equation 3.10).

| Release exponent (n) | Drug transport mechanism |
| --- | --- |
| n < 0.5 | Fickian diffusion |
| 0.45 < n < 0.89 | Non-Fickian transport |
| 0.89 | Cass II transport |
| n > 0.89 | Super case II transport |

## Rumen Digestion

*Collection of rumen fluids.* Esophageal intubation method was used to collect rumen fluid of cows (ID: 289, 212 and 202) from Winfred Thomas Agricultural Research Station (Alabama A & M University). The equipment used for that method named RumenMate. The cows were restrained in a chute before collecting the sample and the esophageal directed through the mouth so that it could pass through the esophagus to the rumen. About 600 mL of rumen fluid been collected from each cow in the early morning of the collection day and preserved the fluid at -20 to -80ºC before used for digestion (Martinez-Fernandez et al., 2019). All the rumen fluid used up within for digestion in the same day of its collection to minimize the microbial death.

*Research design for rumen digestion.* Two by three (2 x 3) factor-factorial experimental design was followed to evaluate the effect of rumen digestion on lycopene nanoparticles. Biodegradable polymers and pasteurization methods were observed to be experimental independent variables where physicochemical properties, and bio-accessibility were response variables. The ANOVA were tested to evuluate the effect of the experimental variables. The Tukey mean comparison and homogeneity test were conducted to evaluate main effect among factors of consideration.

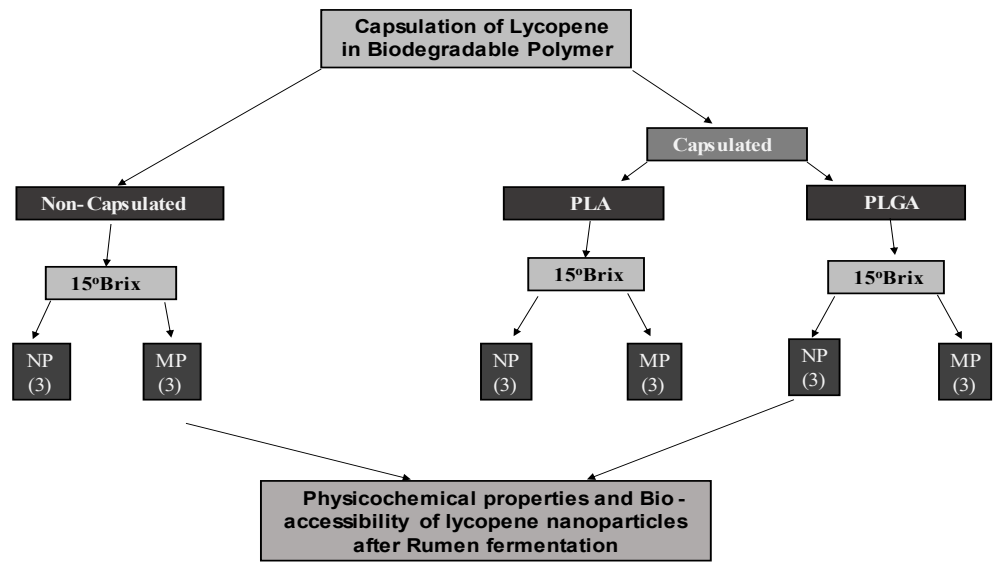

Figure 3.6. Research design for evaluating the effect of rumen digestion on physicochemical properties and bio-accessibility of lycopene nanoparticles[2].

---

[2] CTR = Control, CP = Conventional Pasteurization, MP = Microwave Pasteurization.

*Physicochemical properties of lycopene nanoparticles.* The freeze dried PLGA/PLA-NPs samples were used to fortify the $15^o$ brix juices. The fortified juices were divided into two groups based on application of MP treatment before the rumen fermentation. Two hundred milliliters of rumen fluid were added into the automatic digestive system, and digestion continued for 2 hours. The temperature was maintained at $39^o$C to simulate the rumen environment where the system was agitated at 100 rpm and the pH controlled between 5 and 5.8 to facilitate the rumen digestion (Arce-Cordero et al., 2022). The sample for analysis were collected in every 15 minutes interval for the $1^{st}$ hour, and then every 30 minutes for $2^{nd}$ hour. Physicochemical and antioxidant properties of the nanoparticles for each samples collected were conducted by using zeta sizer, and DPPH activity based on the methods describe in Sections 3.2.6, respectively. The enzyme activities were inhibited by salting out method before measured the physicochemical properties of lycopene nanoparticles. To break the bond between rumen fluid and nanoparticles, rumen digesta were hydrolyzed by 0.1N HCl which further filtered through 200 nm pore size of membrane filter and filtrate were preserved for future analysis.

*Bio-accessibility of lycopene nanoparticles.* The samples (10 mL) were poured into a conical flask and diluted to 20 mL by adding 10 mL of ethyl acetate into the PLGA based digesta and 10 mL of dichloromethane for PLA based digesta. After 15 minutes of stirring, samples were collected from the bottom of the conical flask to run the DPPH inhibition activity analysis with a UV-spectrophotometric method as described in section 3.2.6.

*In-vitro release kinetics.* The DPPH inhibition activity was observed for each sample collected and compared to the initial DPPH inhibition activity of lycopene nanoparticles.

The ANOVA test was conducted to evaluate the effect of the treatment on the experimental variables (pasteurization and digestion time on bio-accessibility of lycopene nanoparticles). The Tukey mean comparison test was conducted at 5% level of significance to determine the main effect of the independent variables to bio-accessibility of lycopene nanoparticles.

CHAPTER 4

RESULTS AND DISCUSSION

The development of biodegradable polymeric materials in biomedical applications has markedly advanced during the last half-century. More so, the polymeric biomaterials sector is expected to growth over 22%, due to its promising potential in biomedical applications particularly, in the global implantable biomaterial market (Taylor et al., 2015). Biodegradable polymeric materials mostly facilitate the development of therapeutic devices concerning transient implants and three-dimensional platforms for tissue engineering. Further advancements have occurred in the controlled delivery of bioactive compounds like lycopene. These applications require analysis of the polymeric materials' physicochemical, biological, and degradation properties which can help identify the therapeutic agent's delivery efficiency to the target site. As a result, researchers have designed a wide range of natural or synthetic polymers which undergoes through controlled hydrolytic or enzymatic degradation and play a potential role in biomedical applications (Song et al., 2018).

Polylactic co-glycolic acid (PLGA) is an improtant biodegradable polymer in biomedical applications and can be modified with varying lactide/glycolide ratios to achieve transitional degradation rates between polylactic acid (PLA) and poly-glycolic acid (PGA). Generally, the copolymer PLGA offers more incredible control release properties than its co-constituent polymer by varying its monomer ratio (Felix et al., 2013). Similarly, Shi et al. (2009) reported that PLGA/hydroxyapatite microsphere composites significantly improve osteoblast proliferation by upregulating an essential osteogenic enzyme, alkaline phosphatase.

Lycopene belongs to the tetraterpene carotenoid family, found in red fruit and vegetables. Eleven conjugated double bonds provide its exceptional potential to scavenge lipid peroxyl radicals, reactive oxygen species, and nitric oxide. However, there is a poor statistical association between dietary intake and serum lycopene levels if ingested orally. Hence, it is very improbable that nutritional intervention alone could be instrumental in correcting lycopene deficiency or chronic diseases prevention. Therefore, new nutraceutical formulations of lycopene or other carotenoids with enhanced bioavailability are urgently needed to support the prevention of chronic diseases (Petyaev, 2016). Furthermore, natural antioxidants have tremendous loss (37 to 77%) during the digestion process if it is hydrophilic. Researchers proved that encapsulating in biodegradable polymer at nano-level might be an attractive platform to improve the bio-accessibility, control release, and targeted delivery of bioactive compounds without producing any toxicity (Liarou et al., 2018; Nguyen et al., 2018; Ye et al., 2018; Lee et al., 2013).

## 4.1.    Extraction of Lycopene by Column Chromatography

The purity of the lycopene extract was measured by comparing it with the standard curve of pure lycopene concentration. Figure 4.1 shows the standard curve for pure lycopene when it was dissolved with 50 mL of distilled water in presence of dimethyl amin boren. The purity of the lycopene extracted by column chromatography was 85%.

The standard curve of lycopene had an $R^2$ value of 0.9968. The relation between absorption and lycopene concentration explained by the equation y = 0.0123x - 0.0057 where x = concentration of lycopene and y = absorption of lycopene determines at 505 nm wavelength with a UV-Spectrophotometric method. Column of column chromatography method separate different carotenoids into different layer based on their characteristics color. Red is the characteristics color for lycopene which appeared just above the yellow band shown in Figure 4.2.

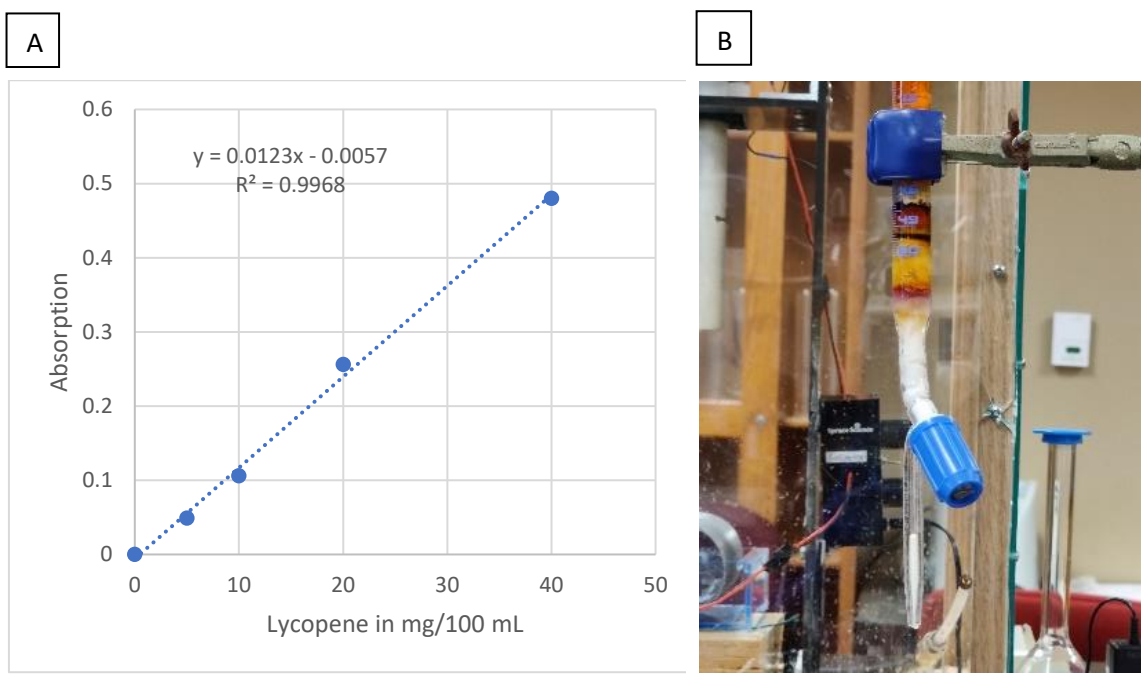

Figure 4.1. Extraction of lycopene by column chromatography (B) and determin lycopene purity by establishing standard curve of pure lycopene.

## 4.2. Synthesis of Lycopene-PLGA-LNP

### Physicochemical Properties

*Hydrodynamic diameter.* Figure 4.2 identified that the lowest hydrodynamic diameter (x nm) was observed when the samples were sonicated for 6 minutes, with 500 mg of surfactant, and 300 mg of polymer concentration. Lai et al. (2007) observed that nanoparticles between the sizes of 100 and 200 nm had a moderate absorption rate through the intestinal lining. Based on the literature and homogeneity test results of the present study, 5 minutes of sonication time, 300 mg of surfactant, and 300 mg of polymer concentration was the optimum condition for producing nanoparticles ranged between 100 and 200 nm to get optimum absorption. On the other hand, the other treatments need higher surfactant, polymer, or sonication time to obtain the same hydrodynamic profile (100 - 200 nm).

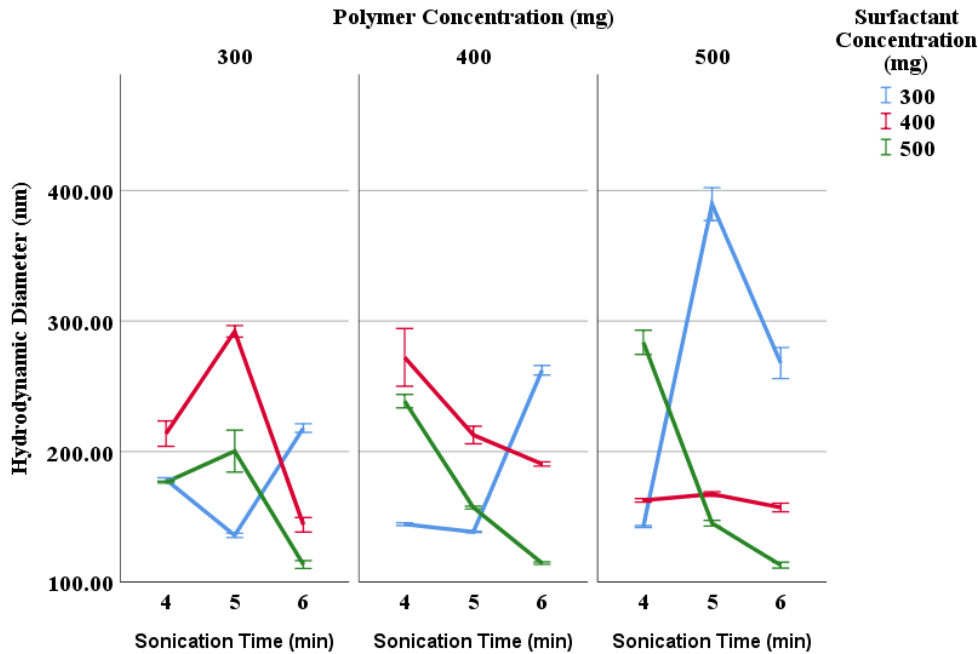

Figure 4.2. Effect of sonication time, surfactant and polymer concentration on hydrodynamic diameter of lycopene nanoparticles.

The outcome of research has revealed that high surfactant concentration (Joung et al., 2016; Chuacharoen & Sabliov, 2016) compared to the generally used concentrations for nanoparticle synthesis (Chuacharoen & Sabliov, 2017) might cause a detrimental effect on the delivery of the entrapped bioactives (Barzegar & Moosavi-Movahedi, 2011). The excessive use of surfactant may also induce undesirable characteristics in foods fortified with nanoaprticles (Barzegar & Moosavi-Movahedi, 2011). For example, the excessive surfactant used during fabrication of milk based nanoemulsion may cause undesirable off-flavored lipid oxidation during the processing and storage (Haidar et al., 2017). Chuacharoen et al. (2019) observed that the lowest concentration of surfactant uses in the synthesis of nanoparticles contributed to the largest size of nanoparticles for the curcumin nanoparticles. Chuacharoen and Sabliov (2016) observed 193.93 nm curcumin nanopartices at lowest surfactant concentration.

*Polydispersity index.* The probability of intestinal absorption is related to the polydispersity index (PDI) The lower the PDI (Figure-4.3), the higher the absorption through the intestinal lumen, and the higher the probability of obtaining highefficiency targeted delivery of encapsulated bioactive compounds. The lowest PDI value was found when the nanoparticle was synthesized at 6 minutes of sonication time, with 400 and 500 mg of surfactant and polymer concentration. The lower PDI value (̃0.2) was also obtained with lower surfactant concentration (300 mg), 5 minutes of sonication time, and 300 mg of polymer concentration. Hence, the surfactant concentration (300 mg) is considered an optimum value  for moderate absorption. In addition, Haidar et al. (2017) observed that the lower the surfactant concentration the higher the stability of fortified media used for delivering nutrients through diet.

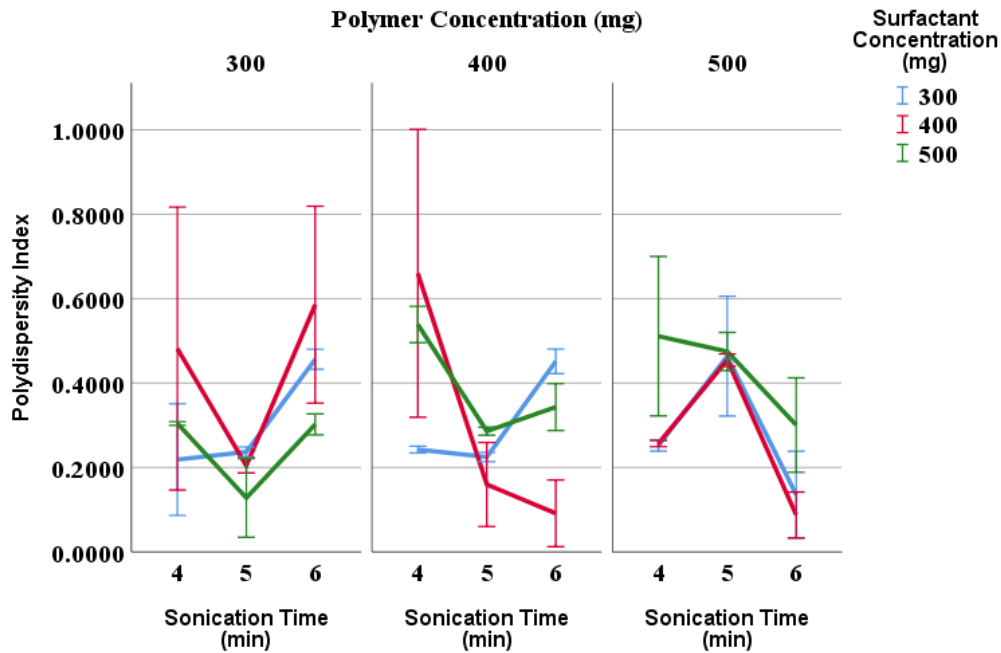

Figure 4.3. Effect of sonication time, surfactant and polymer concentration on PDI of encapsulated lycopene nanoparticles.

*Zeta potential value.* According to Wissinga et al. (2004) and Jacobs et al. (2000) the zeta potential values (ZP) of greater than 30 mV provides a good stability, whereas 60 mV is excellent stability. Faster aggregation of nanoparticle can be occurred if the ZP value are as low as -5 mV to +5 mV, but greater than 20 mV will provide only short-term stability.

The higher the zeta potential indicates the higher stability of the nanoparticles. According to Figure4.4, the ZP of all the treatments were greater than 60 mV, where the nanoparticles with the highest ZP were synthesized for 4 minutes of sonication time, with 400 mg of surfactant, and 400 mg of polymer concentration. Chuacharoen and Sabliov (2016) reported zeta potential values (-54.27 and -48 mV) at increasing lecithin because it induced a stronger anionic bond resulting in a decreased negative surface charge.

Nanoparticles with slight negative charge with a diameter of 150 nm were more efficient to accumulate in tumor cells than the positive charged with significant large diameters (Honary & Zahir, 2013). In addition, if both positive and negative charges become equal, the positively charged NPs will have greater phagocytic uptake than negatively charged (Honary & Zahir, 2013).

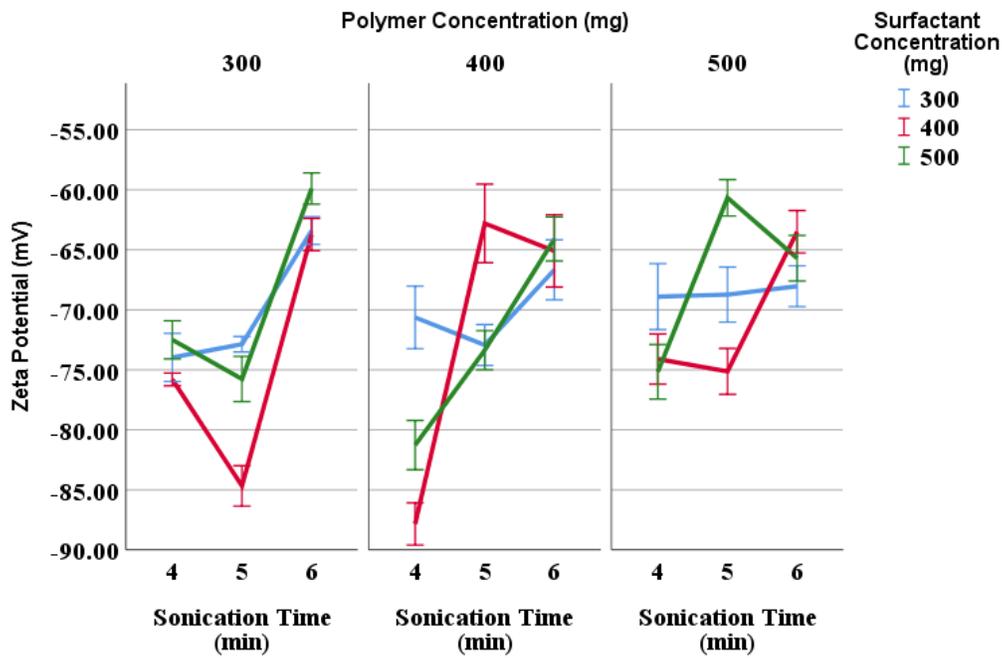

Figure 4.4. Effect of sonication time, surfactant, and polymer concentration on zeta potential of lycopene-PLGA nanoparticles.

*Conductivity of Lycopene Nanoparticles.* The higher the conductivity the more significant the damage due to heat treatment such as microwave, pulse electric field, or ohmic heating, as these treatments depend on the dielectric properties of the food materials and its electron flow behavior. Figure 4.5 shows the lowest electric conductivity was observed from the nanoparticle synthesized at 5 minutes of sonication time, 400 mg of polymer, and 300 mg of surfactant concentration. The electric conductivity of the blend increases with an increase in sonication time (Mohsin et al., 2019).

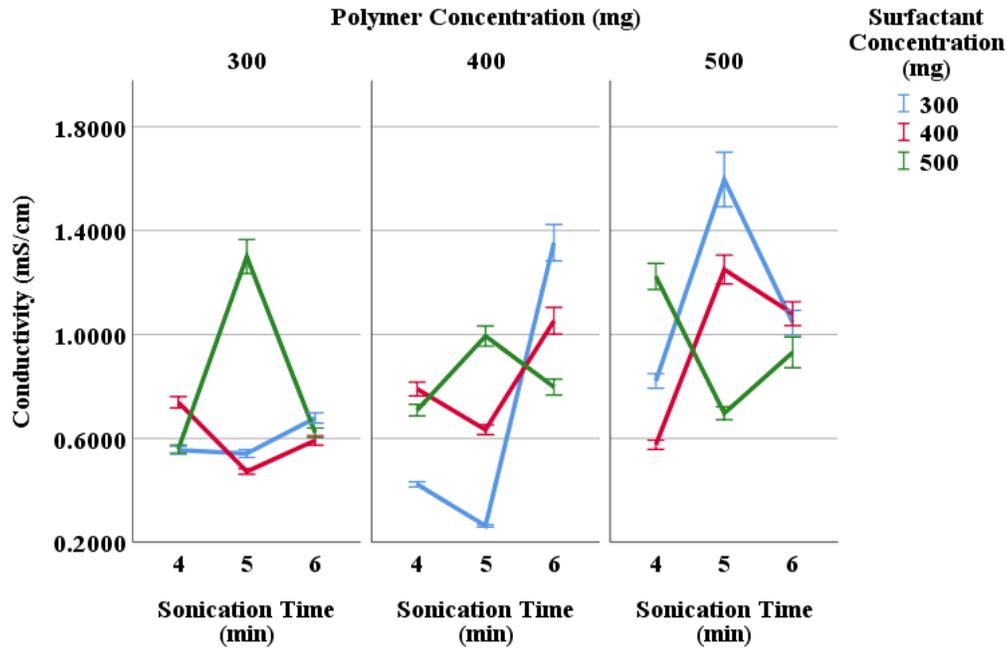

Figure 4.5. Effect of sonication time, surfactant, and polymer concentration on conductivity of lycopene-PLGA nanoparticles

*Encapsulation efficiency of lycopene nanoparticles.* ANOVA test results showed (Figure 4.6) that the interaction effect of surfactant, polymer concentration, and sonication time significantly affects lycopene's encapsulation efficiency in polymeric nanoparticles. Figure 4.6 also explained the highest encapsulation efficiency was observed on the following synthesis treatments: 6 minutes of sonication time irrespective of polymer or surfactant concentrations. When the sonication time of 5 minutes was combined with 300 mg of surfactant and 300 mg of polymer concentration contributed to more than 90 percent encapsulation efficiency of lycopene observed in polymeric nanoparticles, excellent stability (> 60 mV) and low PDI ($\geq$ 0.2). Encapsulation efficiencies was observed at 78.05% at sonication time of 4 minutes and it further increases to 97.28% when sonication was increased to 6 minutes.

Wu et al. (2017) observed that the addition of surfactant could improve the encapsulation efficiency of curcumin into the polymer. Therefore, it can be concluded that the droplet size, distribution, surface charge, and encapsulation efficiency of bioactive compounds in polymeric solution depended on the ratio of surfactant to the total composite of that system (Wu et al., 2017).

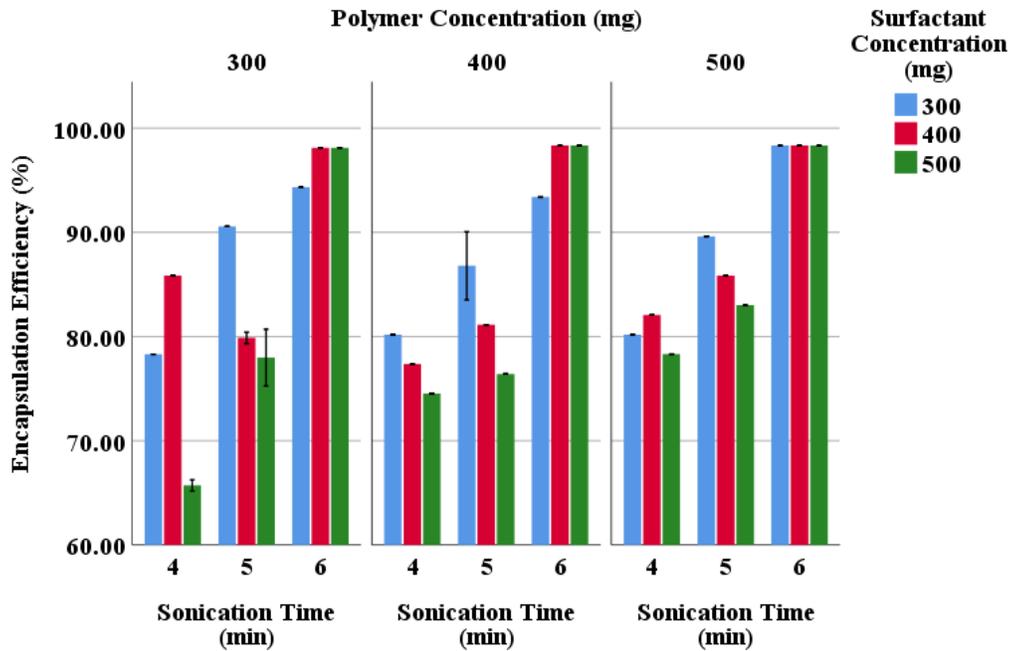

Figure 4.6. Effect of surfactant, polymer concentration and sonication time on encapsulation efficiency of lycopene nanoparticles.

## Optimization of Nanoparticles

The ANOVA table explained how the interactive effect of the independent variables contribute to the significant ($p < 0.05$) changes of the physicochemical properties of formulated nanoparticles. Tukey HSD and homogeneity test depicted that sonication time did not affect ($p > 0.05$) the polydispersity index of the nanoparticles, where increasing sonication time decreases stability and mobility but improved encapsulation efficiency and conductivity of the nanoparticles significantly ($P<0.05$). Sonication time had a different effect on the hydrodynamic diameter of the nanoparticles. Increasing the sonication time from 4 to 5 minutes did not affect ($p > 0.05$) hydrodynamic diameter but had a significant ($p < 0.05$) inverse effect at 6 minutes of sonication time. Variability of the PDI was very high for both the 4- and 6-minutes sonication time treatment, thus creating a problem for the reproducibility of nanoparticles synthesis. Mohsin et al. (2019) inferred increased sonication time leads to a decreased size of the polymeric nanoparticles. Sari et al. (2015) also observed that surfactants significantly ($p < 0.05$) contributed changes on encapsulation efficiency of bioactive compounds in polymeric nanoparticles.

Increasing the surfactant concentration had an inverse effect on the hydrodynamic diameter of the nanoparticles; hence, it means increasing surfactant concentration in the composite nanoparticles causes deterioration of the hydrodynamic diameter. A mixed effect was observed on other physicochemical properties of nanoparticles due to the change in the surfactant concentration. Surfactant concentration did not significantly affect ($p > 0.05$) the PDI; whereas, increasing the surfactant concentration from 300 to 400 mg improved the nanoparticles' stability (zeta potential), but further increment to 500 mg caused significant surface erosion. The encapsulation efficiency was constant due to the increment of surfactant concentration from 300 to 400 mg, but a significant reduction was observed when 500 mg of surfactant was used to formulate nanoparticles. Chuacharoen and Sabliov (2016) identified a low PDI (0.18) with the lowest surfactant concentration for formulating curcumin nanoemulsions. Chuacharoen et al. (2019) also observed an increased PDI (0.19 to 0.23) with increased surfactant concentration about 10 to 30 times. Hence, the lowest surfactant concentration of this study was observed at 300 mg to get optimum PDI of LNP.

Table 4.1. Homogeneity test results to evaluate the effect of sonication time, surfactant and polymer concentration on physicochemical properties of lycopene nanoparticles.

| Independent Variables | Physicochemical Properties of Lycopene Nanoparticles | | | | | |
|---|---|---|---|---|---|---|
| | HD (nm) | PDI | ZP (mV) | Mobility (µm cm/cv) | CD (mS/cm) | EE (%) |
| Sonication Time (min) | | | | | | |
| 4 | 201.43[a] | 0.38[a] | -75.57[a] | -5.92[a] | 0.71[a] | 78.06[a] |
| 5 | 204.28[a] | 0.29[c] | -71.88[b] | -5.63[b] | 0.86[b] | 83.47[b] |
| 6 | 175.62[b] | 0.30[ac] | -64.45[c] | -5.05[c] | 0.91[c] | 97.29[c] |
| Surfactant Concentration (mg) | | | | | | |
| 300 | 208.58[a] | 0.29[a] | -72.52[a] | -5.45[a] | 0.80[a] | 87.96[a] |
| 400 | 201.36[b] | 0.33[a] | -69.82[b] | -5.67[b] | 0.81[a] | 87.43[a] |
| 500 | 171.38[c] | 0.35[a] | -69.57[b] | -5.47[a] | 0.87[b] | 83.41[b] |
| Polymer Concentration (mg) | | | | | | |
| 300 | 185.81[a] | 0.32[a] | -71.63[a] | -5.59[a] | 0.67[a] | 85.43[a] |
| 400 | 192.29[b] | 0.33[a] | -71.40[a] | -5.62[a] | 0.78[b] | 85.16[a] |

| | 500 | 203.21[c] | 0.33[a] | -68.88[b] | -5.40[b] | 1.02[c] | 88.23[b] |

Different superscript letter within the column means significant differences ($P<0.05$). All the value is a mean value of triplicates.

Increasing the polymer concentration also increases the hydrodynamic diameter and the electrical conductivity of the nanoparticles which is undesirable for better bio-accessibility. Moreover, a lower hydrodynamic diameter has higher absorption and higher electrical conductivity, thus facilitating a higher degradation effect against the heat treatments like microwave, ohmic, and pulse electric fields. However, PDI does not change due to change in polymer concentration No effect was observed on the stability and encapsulation efficiency of the nanoparticles when polymer concentration varies from 300 to 400 mg. However, nanoparticles synthesized with 500 mg of polymer concentration had a deterioration effect and yet improves the stability and encapsulation efficiency of the encapsulated nanoparticles. Therefore, the polymeric concentration should not be too high, just enough to achieve high efficiency without compromising the stability of nanoparticles.

Multivariate regression analysis in Table 4.2 shows the surfactant concentration of t nanoparticles had an inverse relation ($p < 0.05$) with the hydrodynamic diameter of the nanoparticles. However, the sonication time during the manufacturing of the nanoparticles may led the change of zeta potential ($P < 0.05$) and encapsulation efficiency ($P < 0.05$) by 63.8% and 85.7%, respectively. The electrical conductivity of the nanoparticles can be modified (44.9%) by changes in polymer concentration in the composite formula. So, it can be inferred that sonication time strongly influenced the zeta potential and encapsulation efficiency of the nanoparticles, where surfactant and polymer concentration strongly influence the hydrodynamic diameter and electrical conductivity of the encapsulated lycopene nanoparticles, respectively.

Table 4.2. Multiple Regression analysis for optimizing the synthesizing parameters for PLGA-LNP.

| Response Variables | Factors | Beta | t | Sig. | Correlations | | |
|---|---|---|---|---|---|---|---|
| | | | | | Zero-order | Partial | Part |
| Hydrodynamic Diameter | (Constant) | | 4.473 | 0.000 | | | |
| | Surfactant | -0.231 | -2.121 | 0.037 | -0.231 | -0.235 | -0.231 |
| Zeta Potential | (Constant) | | -18.037 | 0.000 | | | |
| | Sonication Time | 0.638 | 7.394 | 0.000 | 0.638 | 0.644 | 0.638 |
| Conductivity | (Constant) | | -1.689 | 0.095 | | | |
| | Sonication Time | 0.250 | 2.564 | 0.012 | 0.250 | 0.281 | 0.250 |
| | Polymer Conc. | 0.449 | 4.610 | 0.000 | 0.449 | 0.465 | 0.449 |
| Encapsulation Efficiency | (Constant) | | 9.378 | 0.000 | | | |
| | Sonication Time | 0.857 | 16.419 | 0.000 | 0.857 | 0.882 | 0.857 |
| | Surfactant Conc. | -0.203 | -3.888 | 0.000 | -0.203 | -0.405 | -0.203 |

The regression model fitness test shows that none of a single model can significantly explain the relationship between independent variables (Sonication time, surfactant, and polymer concentration) and hydrodynamic diameter, PDI, and conductivity of the nanoparticles. Quadratic model best explained (P < 0.05) the relationship between sonication time with zeta potential ($R^2 = 0.422$) and sonication time with encapsulation efficiency ($R^2 = 0.780$), respectively.

Table 4.3. Parametric estimation of dependent variable for independent variables.

| Dependent Variable: Zeta Potential and Independent variable: Sonication Time. | | | | | | | | | |
|---|---|---|---|---|---|---|---|---|---|
| | Model Summary | | | | | Parameter Estimates | | | |
| Equation | $R^2$ | F | $df_1$ | $df_2$ | Sig. | Constant | $b_1$ | $b_2$ | $b_3$ |
| Quadratic | 0.422 | 28.440 | 2 | 78 | 0.000 | -53.05 | -13.09 | 1.87 | |
| Dependent Variable: Encapsulation Efficiency and Independent variable: Sonication Time | | | | | | | | | |
| Quadratic | 0.780 | 138.63 | 2 | 78 | 0.000 | 140.44 | -32.40 | 4.20 | |

Based on these regression results, the proposed model for sonication time and zeta potential value of the nanoparticles is given in Eq 4.1

$$Y = -53.048X^2 - 13.09X + 1.86 \qquad (4.1)$$

where Y = zeta potential value of the nanoparticles, and x is the sonication time.

Based on this regression analysis, the proposed quadratic model for showing the relationship between sonication time and encapsulation efficiency of the nanoparticles is given in Eq. 4.2

$$Y = 140.44X^2 - 32.4X + 4.2 \qquad (4.2)$$

where Y = Encapsulation efficiency lycopene, and x for sonification time.

Physical and Morphological Characteristics of Optimized Nanoparticles

*Raman Spectroscopy Analysis for Optimized Nanoparticles.* The band wave number at 1412 cm[-1] shows a peak intensity for glycolic acid which appears on the nanoparticles, as PLGA has a monomer of glycolic acid (Figure 4.7). The low-intensity peak found at 1774 cm[-1] wave number shows the ester bond in PLGA build between lactic acid and glycolic acid (van Apeldoorn et al., 2004). However, intensity peaks at 1032 cm[-1] and 1580 cm[-1] wave number shows lycopene visible on both lycopene and PLGA loaded lycopene nanoparticle's (Radu et al., 2016; Qin et al., 2012; Nikbakht et al., 2011) which verify the presences of lycopne in encapsulated nanoparticles.

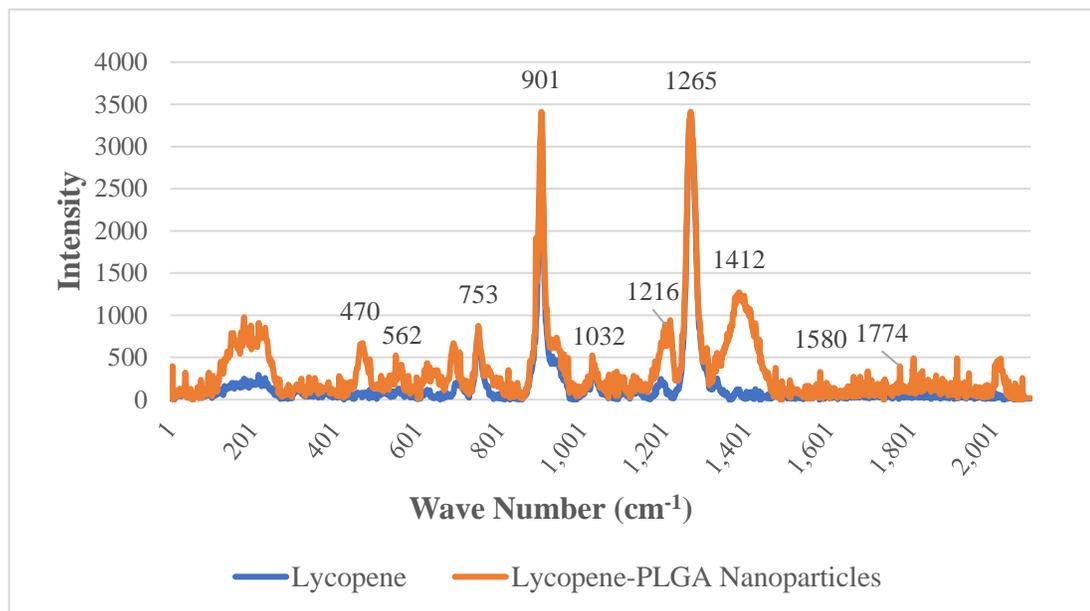

Figure 4.7. Raman spectroscopic analysis of standard lycopene and lycopene-PLGA nanoparticles.

*Morphological characteristics of optimized nanoparticles.* Figure 4.8 (a) shows the hydrodynamic diameter of optimized encapsulated lycopene nanoparticles are normally distributed which is a potential characteristic for targeted delivery of encapsulated bioactive compounds. A sharp peak appeared at the middle of the distribution means uniformity among different particle's diameters. Normal distribution with the highest peak at the middle of the curve is also shown for the zeta potential distribution (Figure 4.8 b) which again means uniform surface stability of LNP.

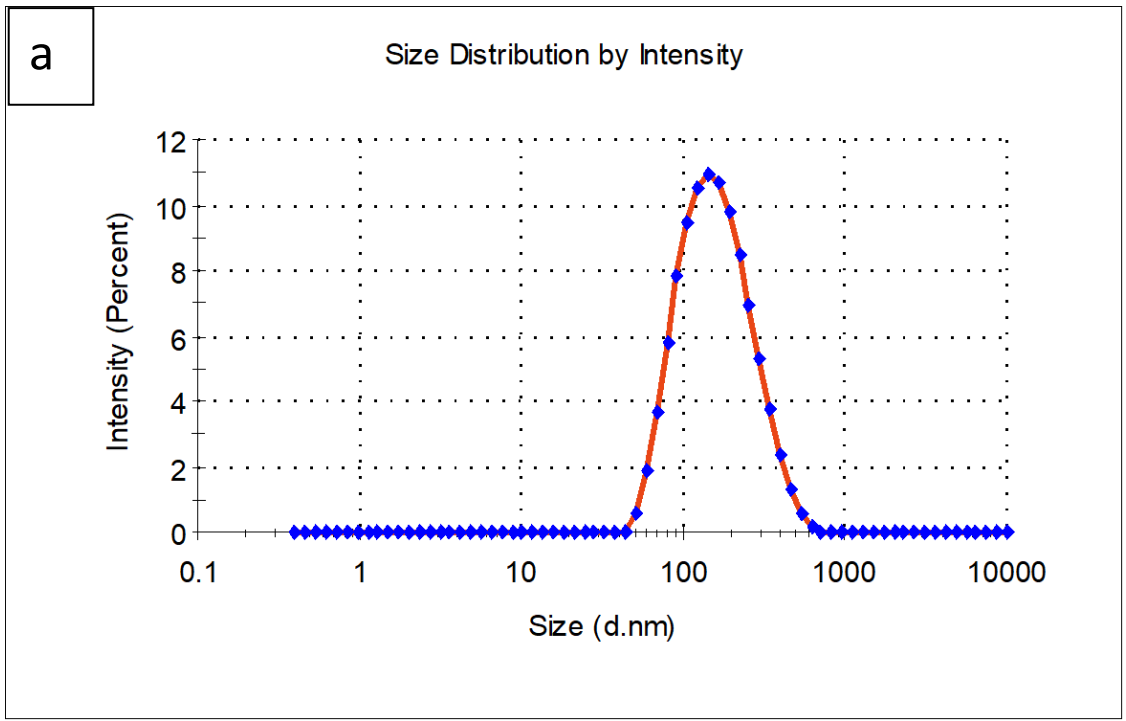

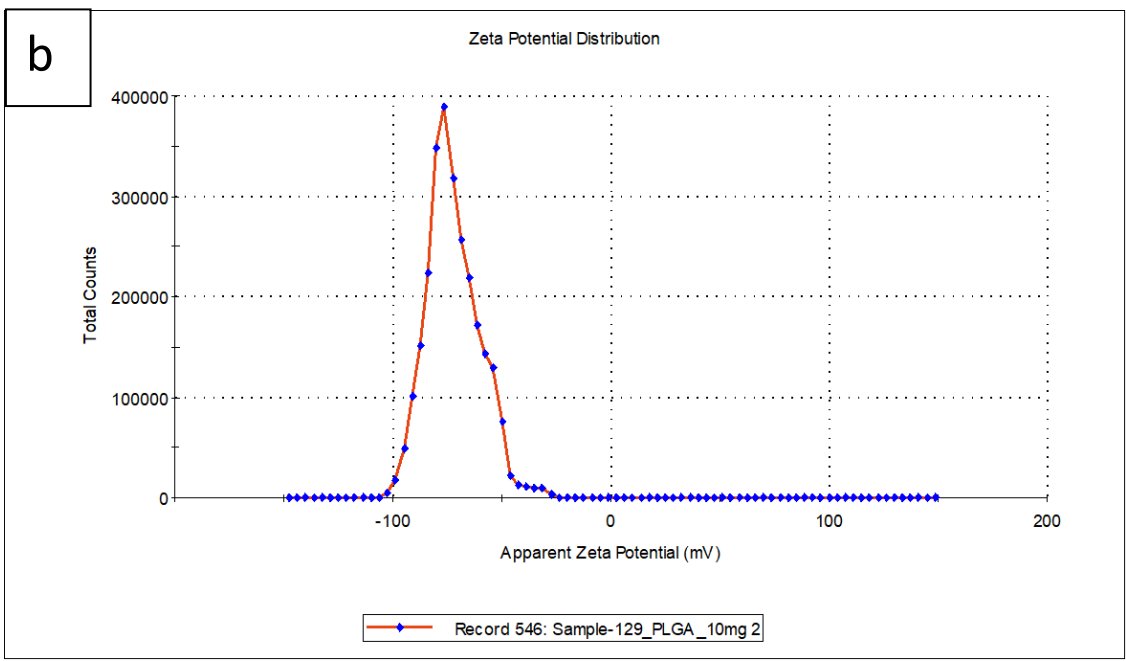

Figure 4.8. Hydrodynamic diameter (a) and zeta potential (b) distribution for optimized lycopene-PLGA nanoparticles.

Scanning electron microscopy (SEM) images shows that encapsulated lycopene nanoparticles were spherical and potential for high absorption through the intestinal lining (Figure 4.9).

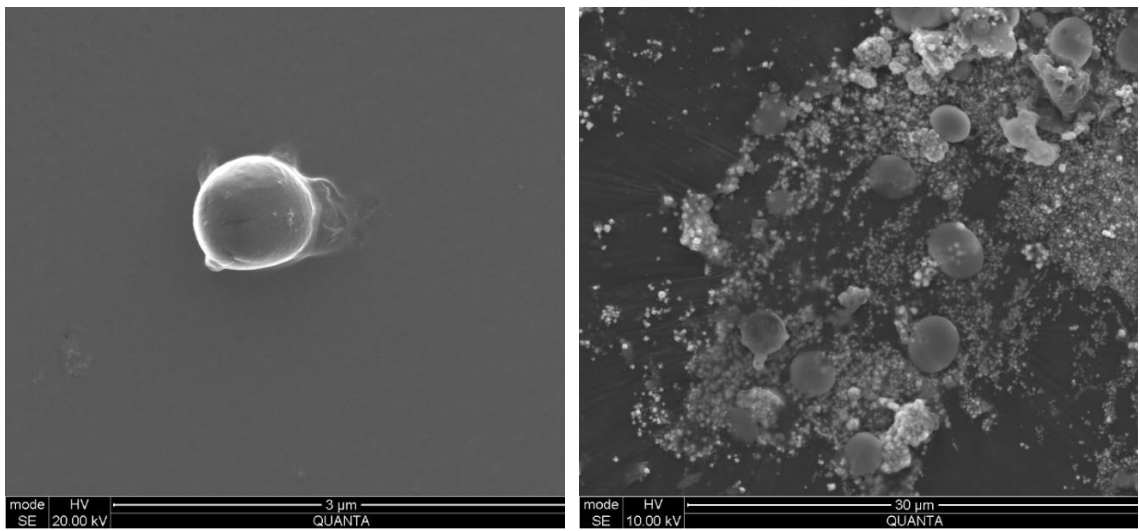

Figure 4.9. SEM imaging of Lycopene-PLGA nanoparticles.

In-vitro Control Release Kinetics of the Optimized PLGA-LNP

PLGA-Lycopene NPs released lycopene (Figure 4.10)) at a reasonably constant rate for the first 12 days of the experiment, it released 53.9 ± 6.7 % of the total content at a constant rate of 4.49 percent/day into the invitro phosphate buffer solution (PBS, pH =7.4) at 37ºC. Burst release was observed on the 13th day, where 41.7 ± 3.8 % of the total content was released within a single day. The biodegradable polymers usually follow three distinct release mechanisms i.e., desorption of the active compound from from the NP's surface, diffusion from the polymeric matrix, and re-adsorption, degradation, and dissolution/erosion of the polymeric network. Bano et al., (2020) also observed a burst release of bioactive compounds through PLGA initially; whereas Khan et al. (2018) found consistent release when chitosan was used to make an additional barrier for PLGA NPs. Shadab et al. (2014) found similar results in PLGA nanoparticles where burst release occurred within 30 min of experimental, and the release was sustained for over 25 days period. According to Korsmeyer – Peppas Model, the equation 4.3 is used to describe the release kinetics of the PLGA-lycopene nanoparticles (Figure 4.10):

$$Y = 3.81X^n \qquad\qquad\qquad (4.3)$$

where Y = Fraction of lycopene released, x is time in the day, and n is the mechanism of transport of drug through the polymer.

The Korsmeyer-Peppas mathematical model fitness test ($R^2$ = 96%) as shown in Figure 4.10. The exponential value (n = 0.99) shows the release kinetics followed the non-Fickian type of diffusion. The non-Fickian diffusion are diffusion where the release of bioactive compounds happened due to the surface erosion of polymeric nanoparticles (Paarakh et al., 2018).

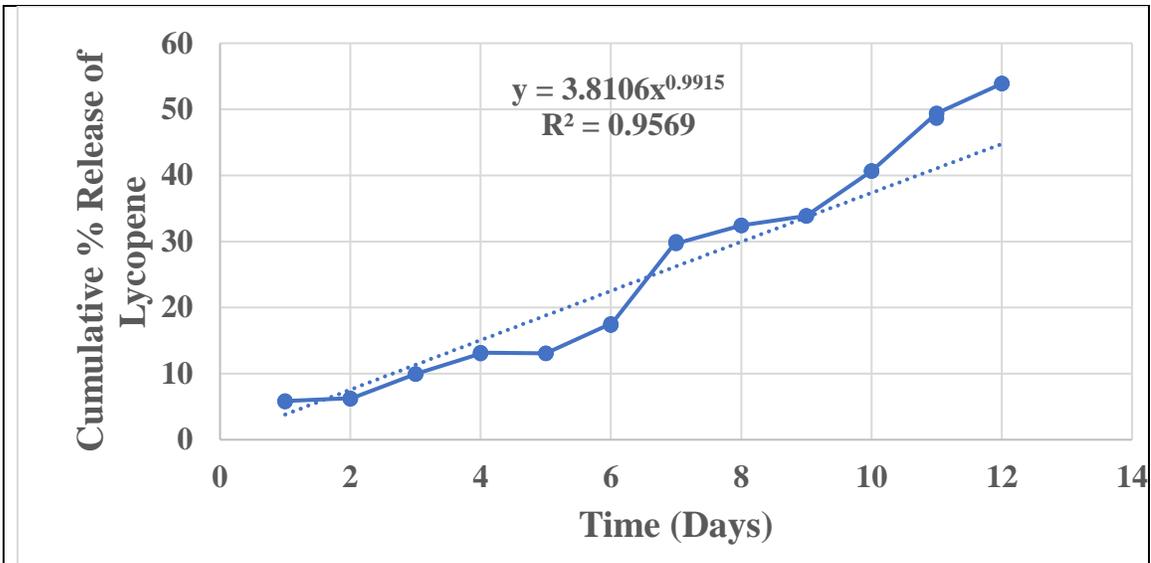

Figure 4.10. Korsmeyer-Peppas model fitness for control release of PLGA-LNP by dialysis method.

## 4.3.    Synthesis of Lycopene-PLA Nanoparticles

Physicochemical Properties

*Hydrodynamic diameter.* Figure 4.11 shows the effect of sonication time on the HD during the synthesis of LNP. The lowest HD was observed at one minute sonication and while the highest after 1.5 minutes.   On the other hand, surfactant concentration at 500 mg contributed to the highest hydrodynamic diameter compared to 300 mg which showed lowest HD.  Figure-4.11 shows for the sonication time of 0.5 minutes, 300 mg of surfactant and 100 mg of biodegradable polymer contributed to be lowest hydrodynamic diameter of the nanoparticles which is desirable for high absorption through the intestinal lining. Li et al. (2018) observed that in combination of modified starch and medium chain triacylglycerol  enhanced the synthesis of nanoparticles between the range 145.1±2.31 and 151.2±2.56 nm with high pressure homogenization method. Zhao et al. (2020) used Sesame oil in combination with 2% of lactoferrin for encapsulating lycopene and managed to produce nanoparticles in between 206.67±5.66 and 216.7±2.61 nm.

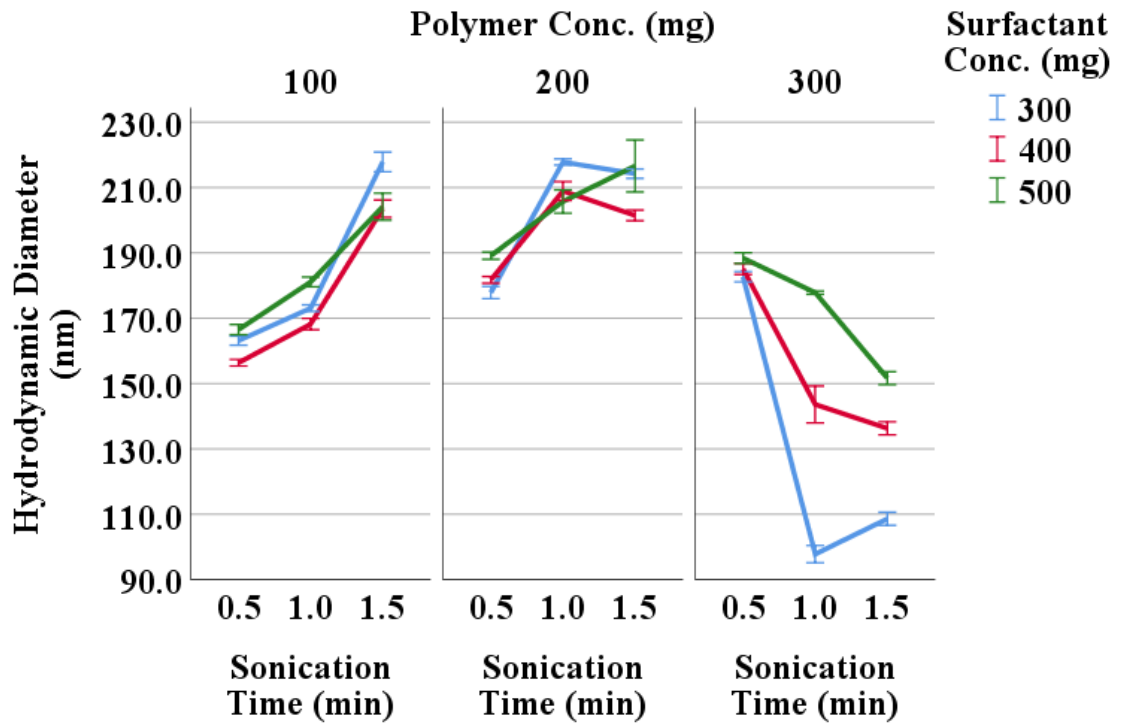

Figure 4.11. Effect on sonication time, surfactant and polymer concentration on hydrodynamic diameter of PLA-LNP.

*Polydispersity index.* Figure 4.12 shows the Sonication time at both 1.0 and 1.5 minute contributed to the lowest PDI; whereas 0.5 minutes increased the PDI value to the highest when combined with 100 mg of polymer used to encapsulate lycopene. On the other hand, surfactant concentration of 500 and 300 mg contributed the lowest and highest PDI value of the nanoparticles, respectively. The reason can be explained as higher the surfactant concentration, reduces the surface tension which enhance the sonication process to synthesize a more homogeneous nanoparticles by cavitation. At higher the polymer concentration (100 to 300 mg), the higher the PDI value could be the result of uneven deposition of extra layer on the surface of nanoparticles. High pressure homogenization method provides better homogeneity for lycopene nanoparticles formulation (Okonogi et al., 2015; Li et al., 2018); whereas the emulsion evaporation method provides slightly lower homogeneity for the encapsulated lycopene nanoaprticles (Jain et al., 2017; Ha et al., 2015).

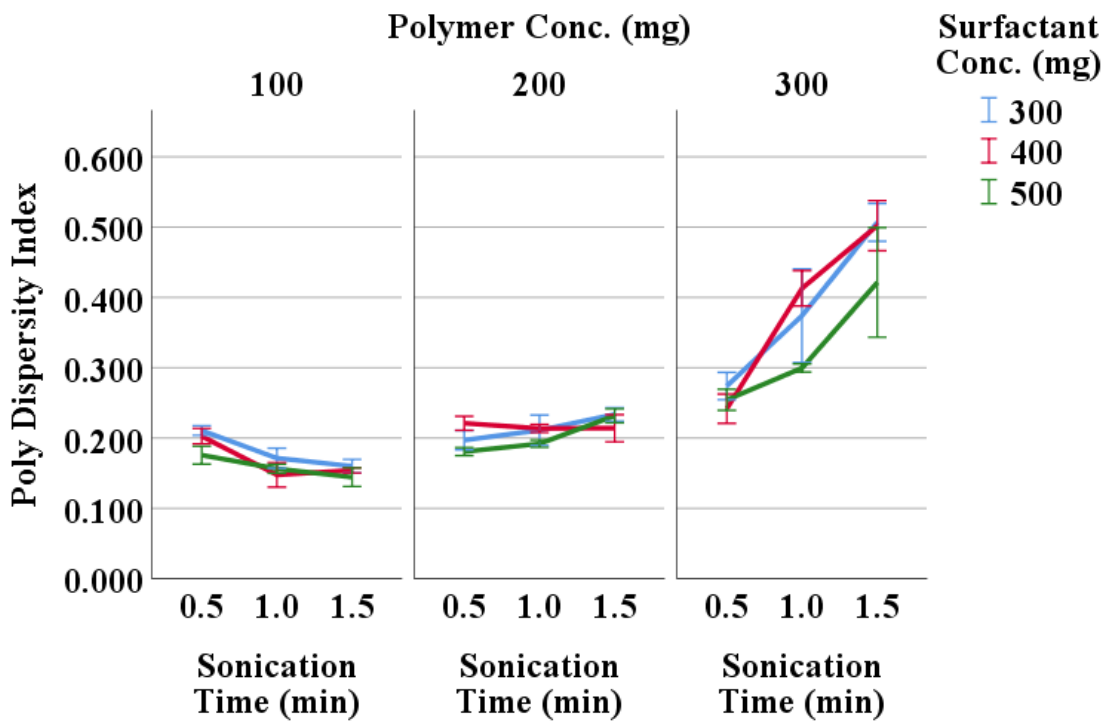

Figure 4.12. Effect on sonication time, surfactant and polymer concentration on polydispersity index of PLA-LNP.

*Zeta potential.* Sonication time at 1.5 the zeta potential value was found to be lowest when the synthesis of NP occurs a sonication time of 1.5 minutes, and at 0.5 minute, it was highest as shown in Figure 4.13. Similarly, at polymeric concentration of 300 and 100 mg the lowest and highest zeta potential value were observed, respectively. That means the lower the polymer concentration, the lower will the deposition on the surface of polymeric nanoparticles, hence making the bond stronger. Unlikely the sonication time and polymer concentration, the surfactant concentration has a different effect on zeta potential of the nanoparticles. At the surfactant concentration of 300 and 500 mg, the stability of the nanoparticles was found to be highest and lowest, respectively. This can be inferred to the lowest sonication time (0.5 minutes), and surfactant (300 mg) and polymer concentrations (100 mg) were identified to be the optimum for providing the highest stability of the nanoparticles. Singh et al. (2017) used ultrasonication in the presence of tween-80 and Ploxamer-188 in ratio of 1:2 and managed to get a lycopene nanoparticle of 121.9 ± 3.66 nm of diameter with 84.5% of encapsulation efficiency and a zeta potential value of -29.0 ± 0.83 mV; whereas, the rotary evaporation film ultrasonication method provided a smaller encapsulated lycopene nanoparticles (58 - 105 nm) but higher zeta potential value (-37 to -32.5 mV) (Zhao et al., 2018) in the presence of soybean phosphatidylcholine as a surfactant.

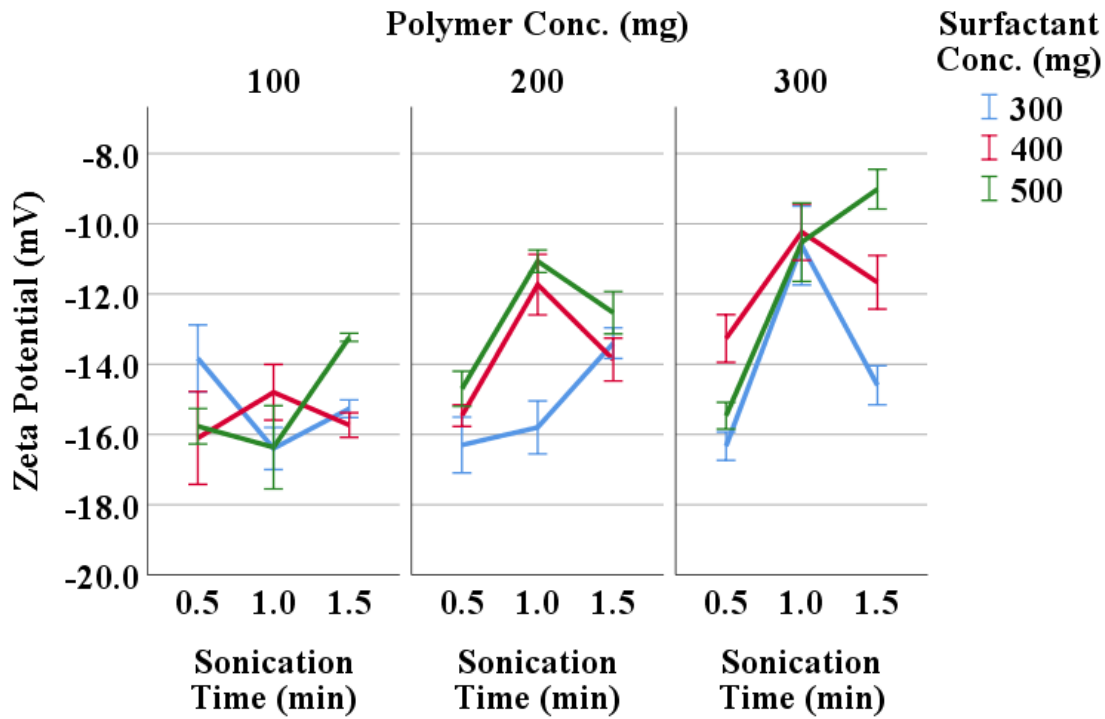

Figure 4.13. Effect on sonication time, surfactant and polymer concentration on zeta potential value of lycopene-PLA-LNP.

*Conductivity of LNP.* Figure 4.14 depicts the effect of sonication time, sufactant concentrations and polymer concentrations on the electrical conductivity of the lycopene-PLA-NPs. When the NPs is sonicated for 1.0 minute, the conductivity observed was the highest, and while at 1.5 minutes, the lowest value was observed for the formulated nanoparticles. Similarly, at 300 mg of polymer concentration, the electrical conductivity was highest, but when the polymer concentration decreased to 200 mg the lowest was observed. The surfactant concentration at 500 mg contributed the highest electrical conductivity, hence inferring that highest conductivity was found at the highest polymer and surfactant concentrations which is undesirable for electromagnetic or UV radiation-based treatment.

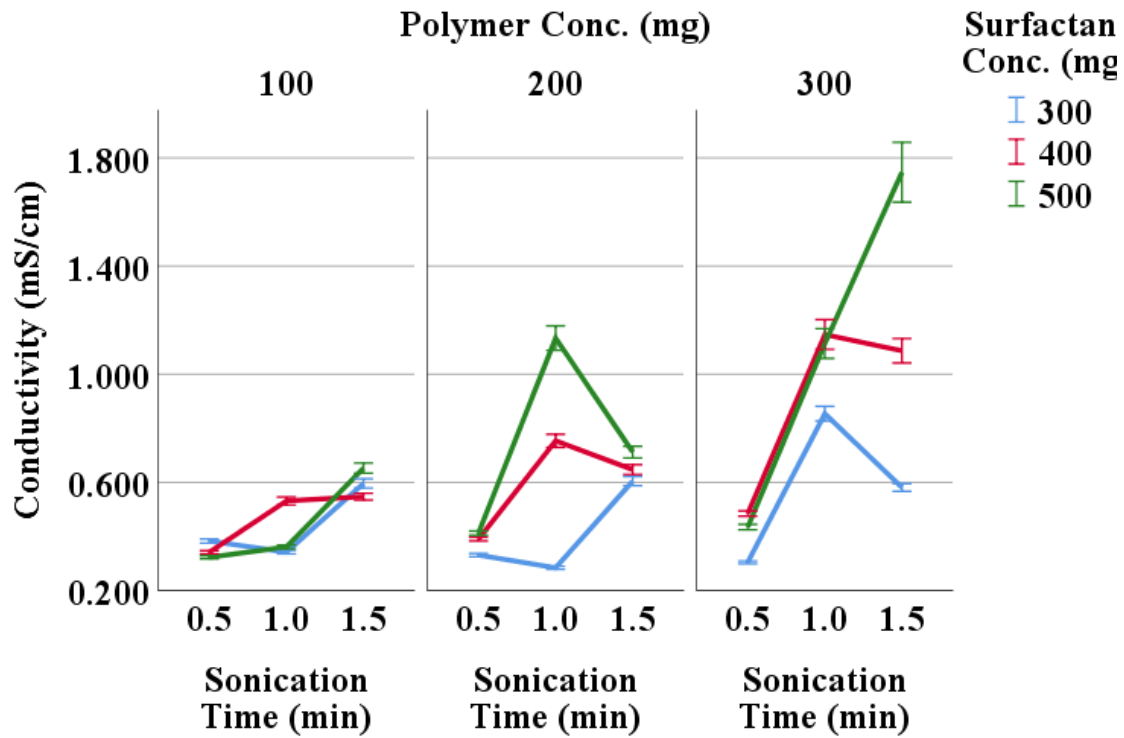

Figure 4.14. Effect of sonication time, surfactant and polymer concentration on electric conductivity of lycopene-PLA-NP.

*Encapsulation efficiency of lycopene in PLA-LNP.* Encapsulation efficiency is an important and most significant response variable for synthesizing nanoparticles. It is the one of the key components of drug discovery. Figure 4.15 shows sonication time of 0.5, 1.0 and 1.5 minutes contributed the same encapsulation efficiency of lycopene when surfactant was 300 mg, and polymer concentrations of 100, 200 and 300 mg respectively. Ha et al. (2015) observed encapsulation efficiency of 51.63 ± 6.06% for lycopene when emulsion evaporation method was used at 140 MPa pressure in the presence of tween-80 as surfactant; whereas Jain et al. (2017) found an encapsulation efficiency to be 79.6 ± 2.9% when compritol-888 ATO and Gelucire at a ratio of 2:1 incorporated with 0.3% phospholipid and 0.1 (w/w) of Pluronic F-68 was used in his emulsion evaporation.

Nazemiyeh et al. (2016) have also shown encapsulation efficiency of lycopene of 98.4±0.5 when glycerol palmito stearate, myristic acid, and poloxamer-407 were used as a surfactant in hot homogenization method.

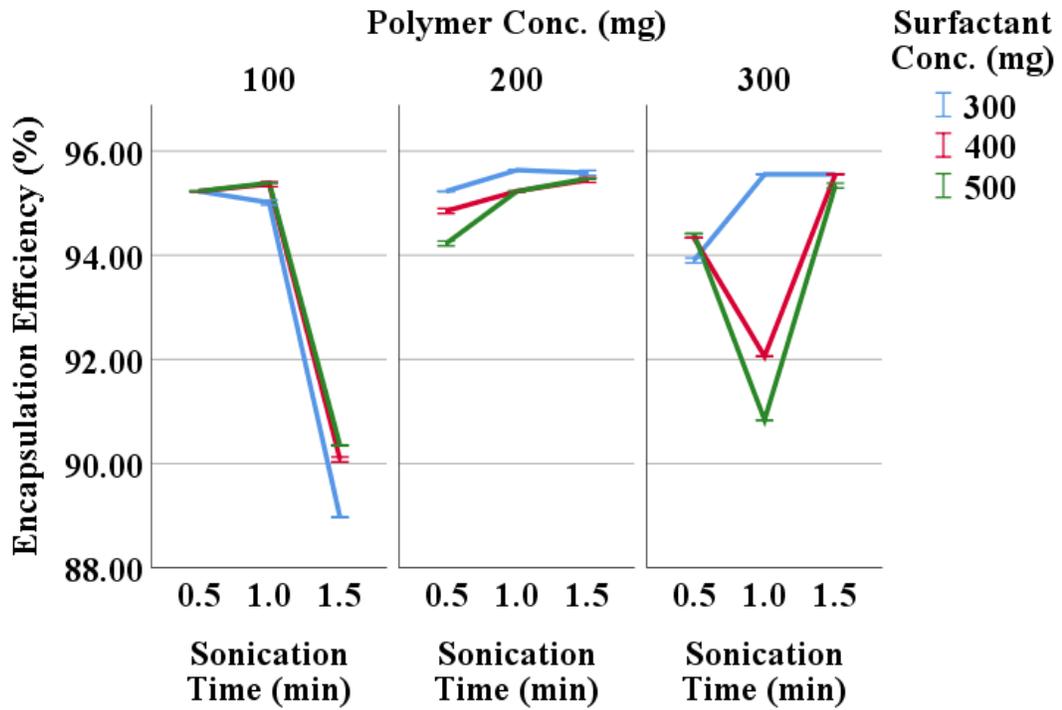

Figure 4.15: Effect of sonication time, surfactant and polymer concentration on encapsulation efficiency of PLA-LNP.

Optimization of Nanoparticles Formulation by Statistical Analysis

The ANOVA shown in Table 4.4 has shown interaction effect ($p < 0.05$) of sonication time, surfactant, and polymer concentration on the hydrodynamic diameter, polydispersity index, conductivity, and encapsulation efficiency, but not on zeta potential of the nanoparticles. The Zeta potential was only affected by the bimodal interaction of sonication time and polymer concentration.

Table 4.4. ANOVA test to identify interaction effect of sonication time, surfactant and polymer concentration on physicochemical properties of PLA-LNP.

| Source | | Type III Sum of Squares | df | Mean Square | F | Sig. |
|---|---|---|---|---|---|---|
| Corrected Model | HD | 77136.81[a] | 26 | 2966.801 | 401.082 | 0.000 |
| | PDI | 0.85[b] | 26 | 0.033 | 52.607 | 0.000 |
| | ZP | 694.38[c] | 26 | 26.707 | 3.580 | 0.000 |
| | Mobility | 2.38[d] | 26 | 0.092 | 31.075 | 0.000 |
| | Conductivity | 9.52[e] | 26 | 0.366 | 371.992 | 0.000 |
| | Encapsulation_Efficiency | 297.62[f] | 26 | 11.447 | 15586.419 | 0.000 |
| Sonication Time | HD | 1196.08 | 2 | 598.044 | 80.850 | 0.000 |
| | PDI | 0.06 | 2 | 0.032 | 51.124 | 0.000 |
| | ZP | 124.43 | 2 | 62.218 | 8.339 | 0.001 |
| | Mobility | 0.50 | 2 | 0.250 | 84.521 | 0.000 |
| | Conductivity | 2.70 | 2 | 1.352 | 1372.628 | 0.000 |
| | Encapsulation_Efficiency | 19.44 | 2 | 9.720 | 13235.111 | 0.000 |
| Surfactant | HD | 32859.68 | 2 | 16429.844 | 2221.152 | 0.000 |
| | PDI | 0.58 | 2 | 0.288 | 463.458 | 0.000 |
| | ZP | 198.22 | 2 | 99.109 | 13.283 | 0.000 |
| | Mobility | 0.70 | 2 | 0.349 | 118.216 | 0.000 |
| | Conductivity | 2.34 | 2 | 1.171 | 1188.882 | 0.000 |
| | Encapsulation_Efficiency | 43.29 | 2 | 21.645 | 29472.112 | 0.000 |
| Polymer | HD | 2957.40 | 2 | 1478.700 | 199.906 | 0.000 |
| | PDI | 0.01 | 2 | 0.008 | 12.943 | 0.000 |
| | ZP | 59.38 | 2 | 29.691 | 3.979 | 0.024 |
| | Mobility | 0.20 | 2 | 0.100 | 33.828 | 0.000 |
| | Conductivity | 1.16 | 2 | 0.580 | 589.244 | 0.000 |
| | Encapsulation_Efficiency | 2.95 | 2 | 1.472 | 2004.778 | 0.000 |
| Sonication Time X Surfactant | HD | 28859.61 | 4 | 7214.90 | 975.383 | 0.000 |
| | PDI | 0.17 | 4 | 0.04 | 67.545 | 0.000 |
| | ZP | 95.51 | 4 | 23.88 | 3.200 | 0.020 |
| | Mobility | 0.40 | 4 | 0.10 | 33.986 | 0.000 |
| | Conductivity | 1.02 | 4 | 0.26 | 259.590 | 0.000 |
| | Encapsulation_Efficiency | 193.01 | 4 | 48.25 | 65700.612 | 0.000 |
| Sonication Time X Polymer | HD | 892.90 | 4 | 223.23 | 30.178 | 0.000 |
| | PDI | 0.002 | 4 | 0.00 | 0.648 | 0.631 |
| | ZP | 23.30 | 4 | 5.82 | 0.781 | 0.543 |
| | Mobility | 0.17 | 4 | 0.04 | 14.692 | 0.000 |
| | Conductivity | 0.50 | 4 | 0.12 | 126.618 | 0.000 |
| | Encapsulation_Efficiency | 10.10 | 4 | 2.52 | 3437.278 | 0.000 |
| Surfactant X Polymer | HD | 6069.14 | 4 | 1517.28 | 205.122 | 0.000 |
| | PDI | 0.01 | 4 | 0.002 | 3.575 | 0.012 |
| | ZP | 76.70 | 4 | 19.17 | 2.570 | 0.048 |
| | Mobility | 0.13 | 4 | 0.03 | 11.238 | 0.000 |
| | Conductivity | 0.62 | 4 | 0.15 | 157.813 | 0.000 |
| | Encapsulation_Efficiency | 10.24 | 4 | 2.56 | 3487.278 | 0.000 |

Table 4.4.(Continued)

| | | | | | | |
|---|---|---|---|---|---|---|
| Sonication Time X Surfactant X Polymer | HD | 4301.99 | 8 | 537.75 | 72.698 | 0.000 |
| | PDI | 0.02 | 8 | 0.002 | 3.207 | 0.005 |
| | ZP | 116.85 | 8 | 14.61 | 1.958 | 0.070 |
| | Mobility | 0.28 | 8 | 0.03 | 11.894 | 0.000 |
| | Conductivity | 1.18 | 8 | 0.15 | 149.276 | 0.000 |
| | Encapsulation_Efficiency | 18.60 | 8 | 2.32 | 3165.278 | 0.000 |
| Total | HD | 2658189.30 | 81 | | | |
| | PDI | 5.88 | 81 | | | |
| | ZP | 15961.77 | 81 | | | |
| | Mobility | 97.93 | 81 | | | |
| | Conductivity | 42.07 | 81 | | | |
| | Encapsulation_Efficiency | 720117.32 | 81 | | | |

Tukey HSD (Honest significance difference) and homogeneity test inferred that the sonication time increment has significant effect on HD, PDI, conductivity and encapsulation efficiency but not for ZP and mobility when sonication time changed from 1 to 1.5 minutes

Table 4.5. Homogeneity test to identify main effect among different levels of independent variables.[3]

| Independent Variables | Physicochemical Properties of Lycopene Nanoparticles | | | | | |
|---|---|---|---|---|---|---|
| | HD (nm) | PDI | ZP (mV) | Mobility (μm cm/cv) | CD (mS/cm) | EE (%) |
| Sonication Time (min) | | | | | | |
| 0.5 | 176.77[c] | 0.22[a] | -15.25[a] | -1.20[a] | 0.38[a] | 93.59[a] |
| 1.0 | 174.89[a] | 0.24[b] | -13.06[b] | -1.02[b] | 0.72[b] | 94.48[b] |
| 1.5 | 183.82[b] | 0.29[c] | -12.33[b] | -1.04[b] | 0.80[c] | 94.74[c] |
| Surfactant Concentration (mg) | | | | | | |
| 300 | 181.54[a] | 0.17[a] | -15.28[a] | -1.20[a] | 0.45[a] | 93.43[a] |

---

[3] Superscript through the column describes the significant difference. Physicochemical properties of LN explain as mean value.

| | | | | | |
|---|---|---|---|---|---|
| 400 | 201.50[b] | 0.21[b] | -13.87[a] | -1.08[b] | 0.58[b] | 94.17[a] |
| 500 | 152.44[c] | 0.36[c] | -11.49[b] | -0.97[c] | 0.86[c] | 95.21[b] |
| Polymer Concentration (mg) | | | | | | |
| 100 | 172.55[a] | 0.26[a] | -14.73[a] | -1.15[a] | 0.48[a] | 94.52[a] |
| 200 | 176.14[b] | 0.26[a] | -13.19[ac] | -1.07[b] | 0.66[b] | 94.24[a] |
| 300 | 186.78[c] | 0.23[b] | -12.73[c] | -1.03[c] | 0.76[c] | 94.05[a] |

Homogeneity test suggested that increasing sonication time significantly (P < 0.05) increases the hydrodynamic diameter, PDI and, electrical conductivity of the PLA-lycopene nanoparticles. On the contrary, increasing sonication time significantly (p < 0.05) decrease the stability, mobility, and encapsulation efficiency of lycopene nanoparticles.

The homogeneity test for surfactant concentration has shown almost similar effect as the sonication time with minor exceptions. Increase surfactant concentration contributes to the increment of the PDI, mobility and conductivity. The hydrodynamic diameter increased from 181.54 to 201.5 nm due to change the concentration from 300 to 400 mg, respectively. Further increase (500 mg) caused a detrimental effect on the HD, as it dropped down to 152.44 nm as higher surfactant prevent agglomeration by caping on nanoparticles and thus provide smaller particles than at lower concentration. Unlikely, increasing the surfactant concentrations increases the encapsulation efficiency significantly from 93.43 to 95.21%. The zeta potential values did not change for the change in concentration from 300 to 400 mg but significantly (p < 0.05) dropped from -13.87 to -11.49 mV when the concentration increased to 500 mg. As higher surfactant prevents polymer agglomeration and thus allows less layer on the surface of nanoparticles that ultimately has given lower stability to nanoparticles.

The biodegradable polymer concentration showed that increasing the concentration of polymer caused an increase of the size of the nanoparticles due to high deposition per particles surface as revealed by the homogeneity test. Similarly, the increasing polymer concentration also increases the PDI, and electrical conductivity, but reduced the mobility and encapsulation efficiency. The Zeta potential of the nanoparticles reacting differently with polymer concentration changes. A significant difference ($p < 0.05$) was found between 100 and 300 mg when the zeta potential value decreased from -14.73 to -12.73 mV.

Table 4.6. Multiple Regression analysis results for optimization of independent variables.

| Model | Standardized Coefficients Beta | t | Sig. | Correlations Zero-order | Partial | Part |
|---|---|---|---|---|---|---|
| Dependent Variable: HD | | | | | | |
| 1      (Constant) | | 8.92 | 0.000 | | | |
|         Surfactant Conc. | -0.38 | -3.75 | 0.000 | -0.38 | -0.39 | -0.38 |
| Dependent Variable: PDI | | | | | | |
| 2      (Constant) | | 1.139 | 0.258 | | | |
|         Sonication Time | 0.26 | 4.06 | 0.000 | 0.26 | 0.42 | 0.26 |
|         Surfactant Conc. | 0.77 | 11.73 | 0.000 | 0.77 | 0.80 | 0.77 |
| Dependent Variable: ZP | | | | | | |
| 3      (Constant) | | -11.02 | 0.000 | | | |
|         Sonication Time | 0.32 | 3.42 | 0.001 | 0.32 | 0.36 | 0.32 |
|         Surfactant Conc. | 0.42 | 4.44 | 0.000 | 0.42 | 0.45 | 0.42 |
| Dependent Variable: Encapsulation Efficiency | | | | | | |
| 4      (Constant) | | 75.59 | 0.000 | | | |
|         Sonication Time | -0.24 | -2.25 | 0.028 | -.244 | -.248 | -0.244 |

The regression analysis shows that the sonication time have contributed the 26 and 32 and -24.4% of the variability of the PDI, zeta potential, and encapsulation efficiency, respectively, whereas the surfactant influenced 77 and 42% of the total variability for PDI, and zeta potential, respectively. The sonication time (-24.4%) and surfactant concentration (-38%) had inverse effect on the encapsulation efficiency, and hydrodynamic diameter, respectively. Hence, at the lower sonication time, surfactant and polymer concentration is the optimum conditions for the highest encapsulation efficiency, stability lowest electrical conductivity and PDI value. Mohsin et al. (2019) reported that an increased sonication time leads to a decrease in size of the polymeric particles. Sari et al. (2015) also reported that surfactants significantly contributed to higher encapsulation efficiency of bioactive compounds in polymeric nanoparticles. Chuacharoen and Sabliov (2016) identified a PDI of 0.18 with the lowest surfactant concentration for formulating the curcumin nanoemulsions.

Table 4.7. Linearity of the dependent variables based on parametric regression analysis.

Dependent Variable: HD, Independent variable: Surfactant.

| Equation | Model Summary | | | | | Parameter Estimates | | | |
|---|---|---|---|---|---|---|---|---|---|
| | R Square | F | $df_1$ | $df_2$ | Sig. | Constant | $b_1$ | $b_2$ | $b_3$ |
| Linear | 0.15 | 13.659 | 1 | 79 | 0.000 | 207.59 | -0.145 | | |
| Quadratic | 0.42 | 28.685 | 2 | 78 | 0.000 | 92.58 | 1.24 | -0.003 | |
| Cubic | 0.42 | 28.685 | 2 | 78 | 0.000 | 159.92 | 0.000 | 0.003 | $-1.12 \times 10^{-5}$ |

Model fitness analysis explained that Quadratic model better fit the relationship between HD and surfactant concentrations ($R^2 = 0.42$). Based on the model fitness, test predicted mathematical model for surfactant and hydrodynamic diameter is given in Eq. 4.4.

$$Y = 92.58X2 + 1.24X - 0.003 \qquad (4.4)$$

Morphological Characteristics

*FT-IR data analysis.* The lycopene absorption curve from FTIR analysis is shown in Figure 4.16, and the peak wave number 1100 cm$^{-1}$ was found to identify the lycopene in the PLA nanoparticles. Another identical characteristic peak found at 1430 cm$^{-1}$ wavenumber is also identified as lycopene in the PLA nanoparticles. According to Irudayaraj et al. (2003) the C-C and C-C-H stretching identifies a different characteristic of lycopene found between 1100-1400 cm$^{-1}$ wavenumber. The wavenumber peak between 1477-1400 cm$^{-1}$ usually shows C-H (trans) bending which is a distinct characteristic of lycopene existing in PLA nanoparticles (Irudayaraj et al., 2003). Both lycopene and encapsulated lycopene PLA nanoparticles appeared on the same distinct peak where the C-C, C-C-H and C-H (trans) bending into their structural characteristics identified by FTIR analysis. So, it is inferred that lycopene was indeed encapsulated in the biodegradable polymeric PLA.

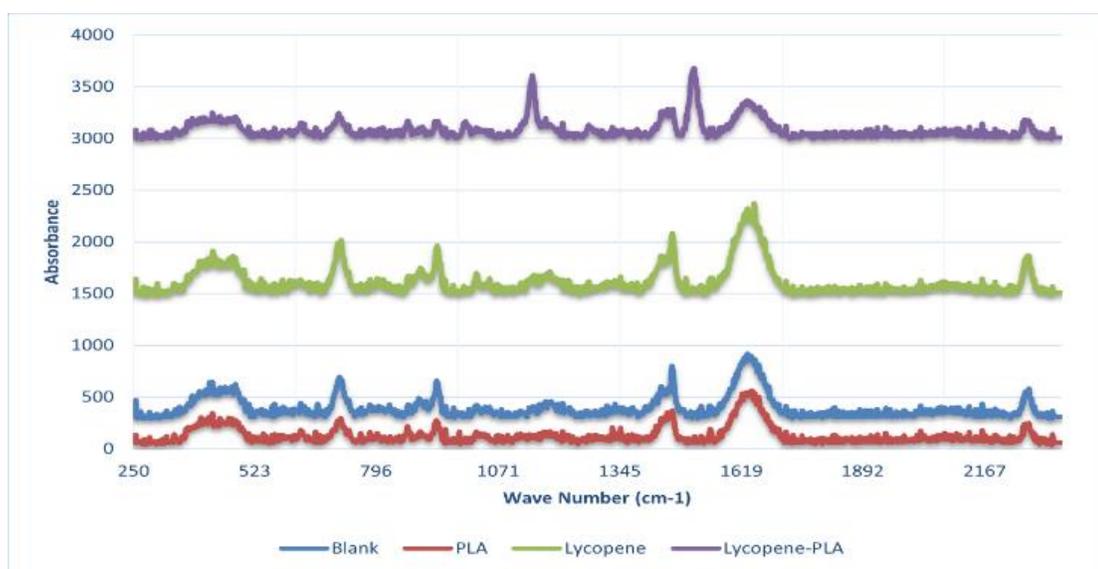

Figure 4.16. FT-IR analysis results for standard lycopene and lycopene-PLA-NP.

*HD and ZP distribution.* The particle size distribution curve of the hydrodynamic diameter shows uniformity of the nanoparticles. The width of the hydrodynamic diameter curve is small, that an indication of spherical shaped nanoparticles. Hence, the zeta potential distribution curve is also an indication of surface stability and control release activity of the nanoparticles, and small width evaluate as better uniformity. Therefore, it can be concluded that the optimum condition provided us better control release kinetics and bio-accessibility for the encapsulated lycopene.

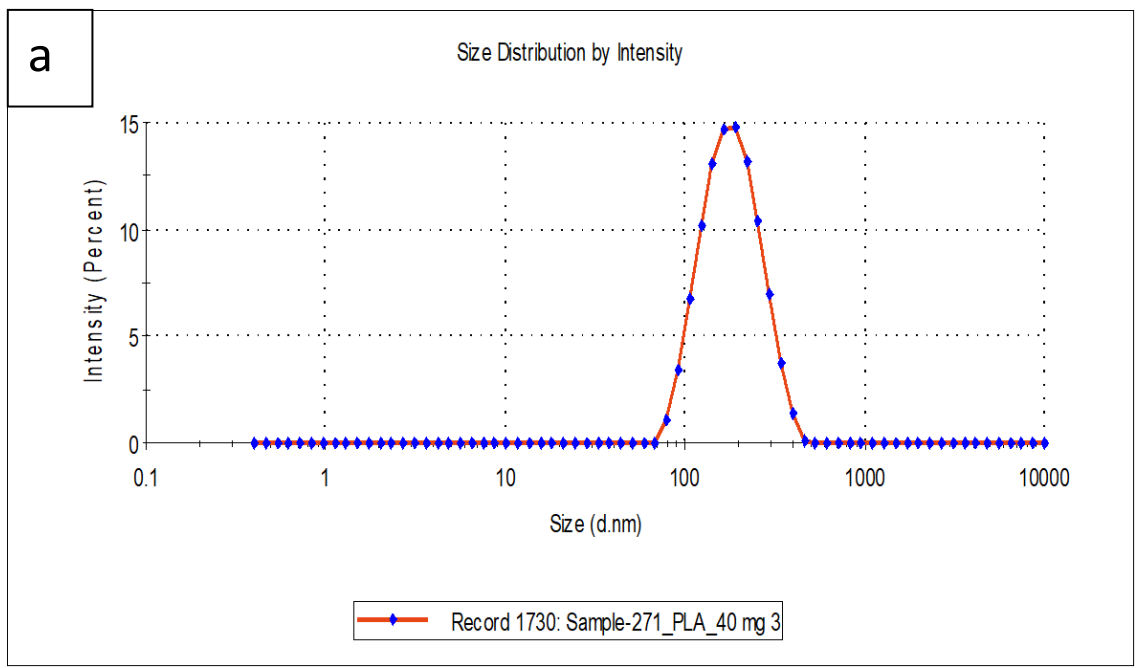

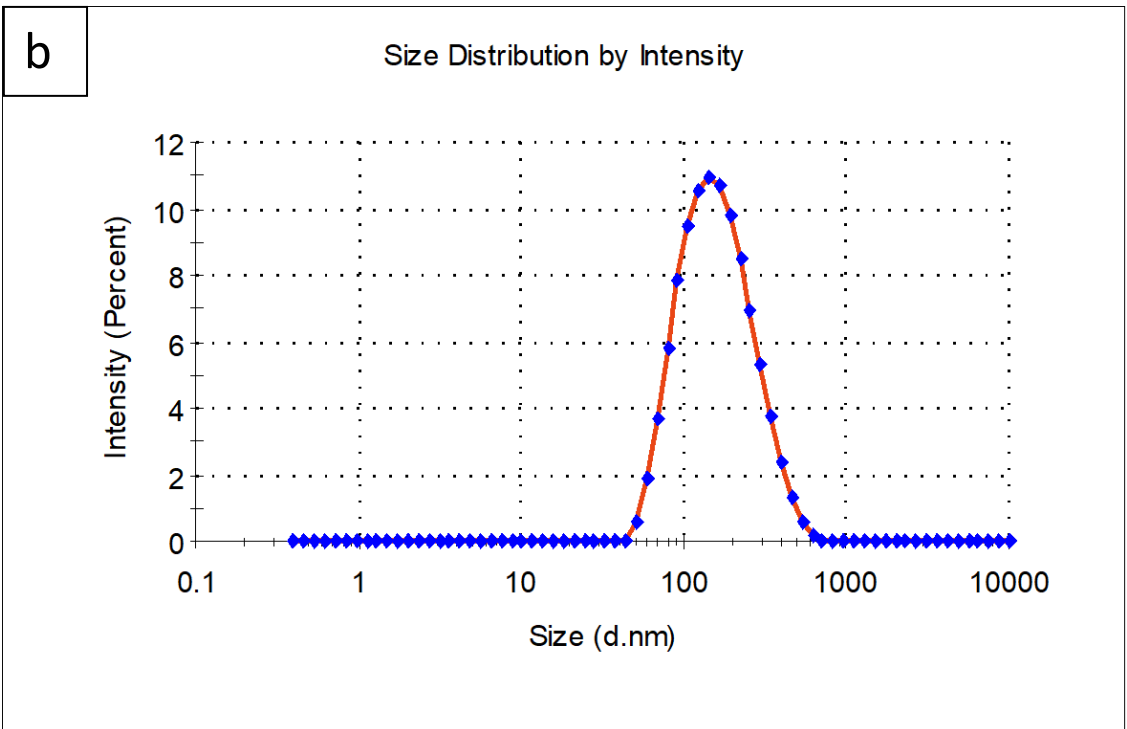

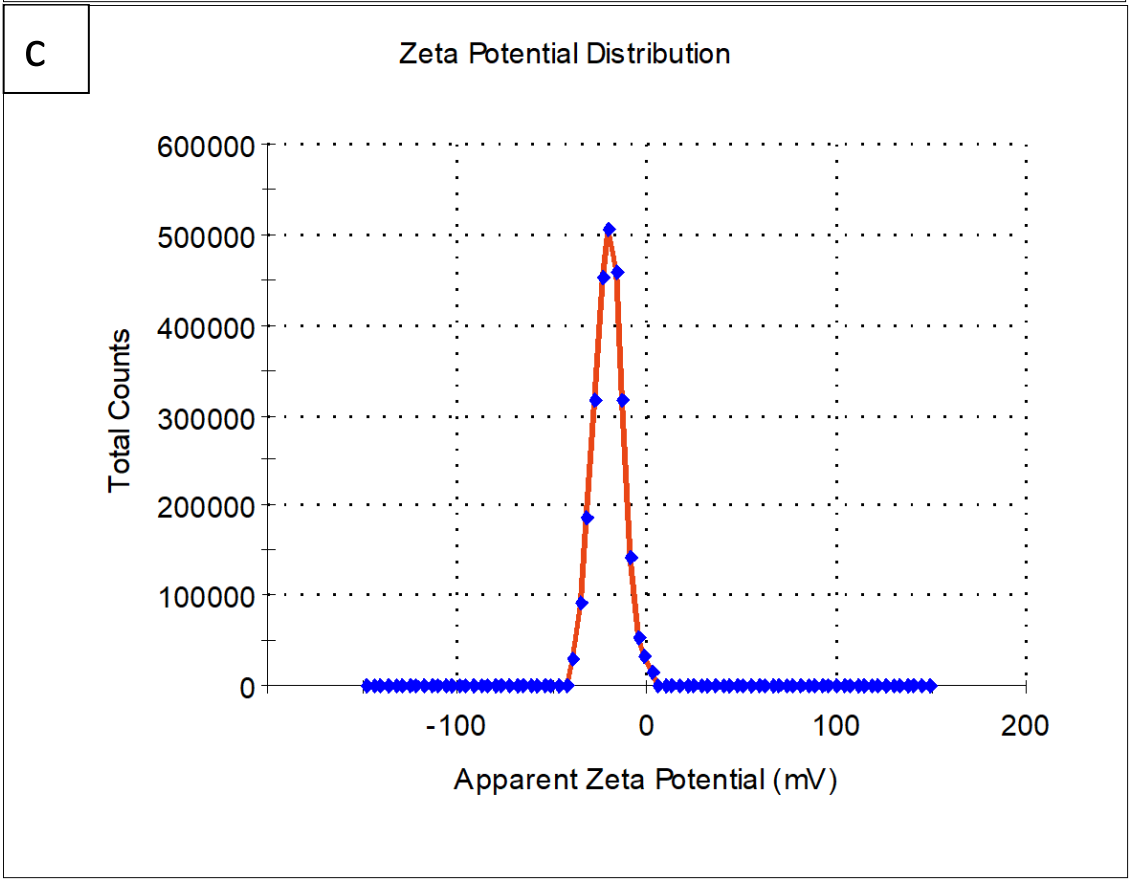

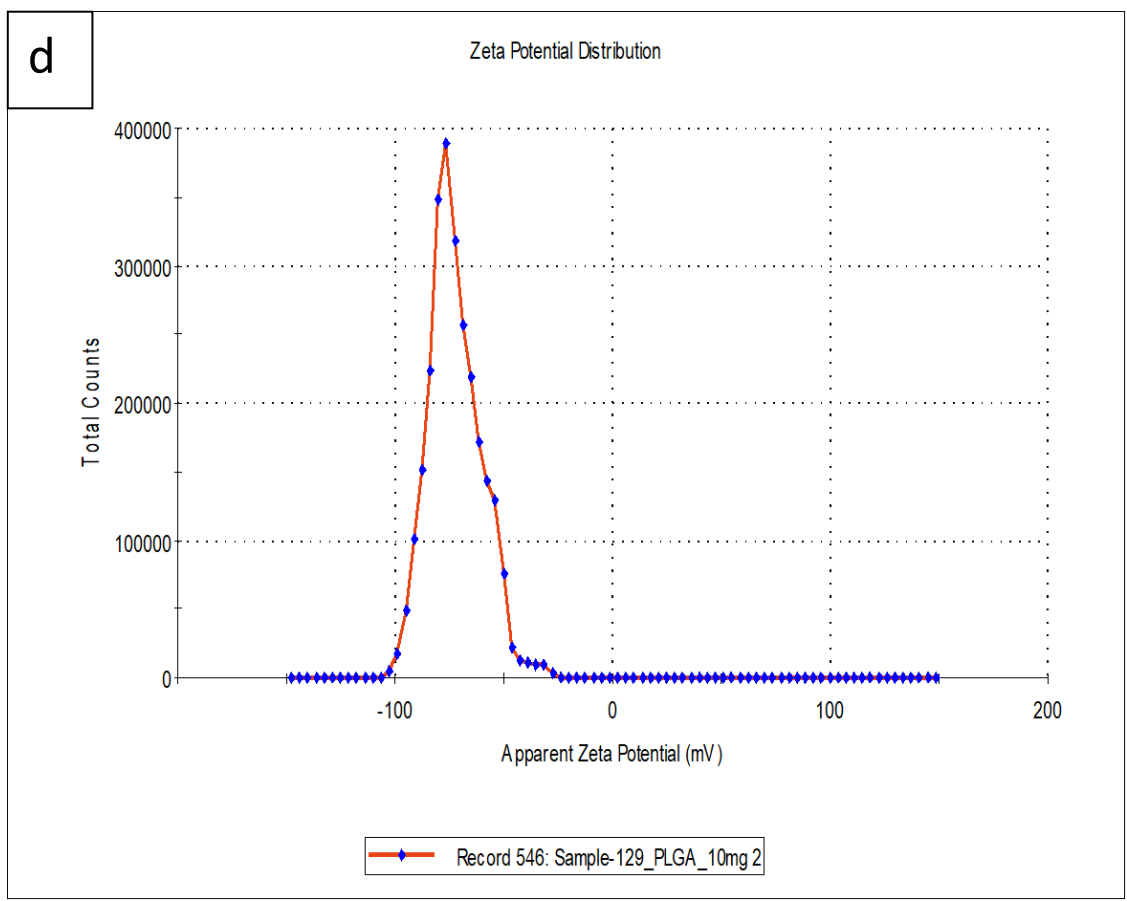

Figure 4.17. Hydrodynamic diameter (a & b) and zeta potential (c & d) distribution of optimized lycopene-PLA-NP.

### Control Release Kinetics of PLA-LNP

Figure 4.18 shows the control release activity of PLA-lycopene nanoparticles, where a burst release was observed at the beginning to the end of day-one when 20.52% was released during the 8 days of study. However, the rate of the release slowed down from the 2nd day and remained constant till the 5th day and slight increase of lycopene on the 6th days by 13.79%. The 2nd burst release was observed after the 6th day where 29.96% of lycopene released from the total load of encapsulated core and continues untilthe end of it. The 8th day observed to be the highest release where day nanoparticles released about 35.73% of total load of lycopene. The 1st burst release happened due to release of lycopene attached on the surface of the nanoparticle; whereas, the 2nd (7th day) and 3rd (8th day) burst release observed was due to molecular degradation of PLA polymer, hence contributing to rapid release of the active compound. Control release kinetics found to be different in literature based on the physicochemical properties of coating materials. The burst release of lycopene was observed when for both polymeric and lipophilic nanoparticles. The nanoliposome, 80% of the total lycopene can be released within 12 hours of its exposure to phosphate buffer solutions (Stojiljkovic et al., 2018). While 50% of the total loaded lycopene released within 24 h when in contact with a release solution (Bano et al., 2020), and hen encapsulated with N-isopropylacrylamide 60% is released within 3 h from solid lipid core of the lycopene nanoemulsion (Nazemiyeh et al., 2016).

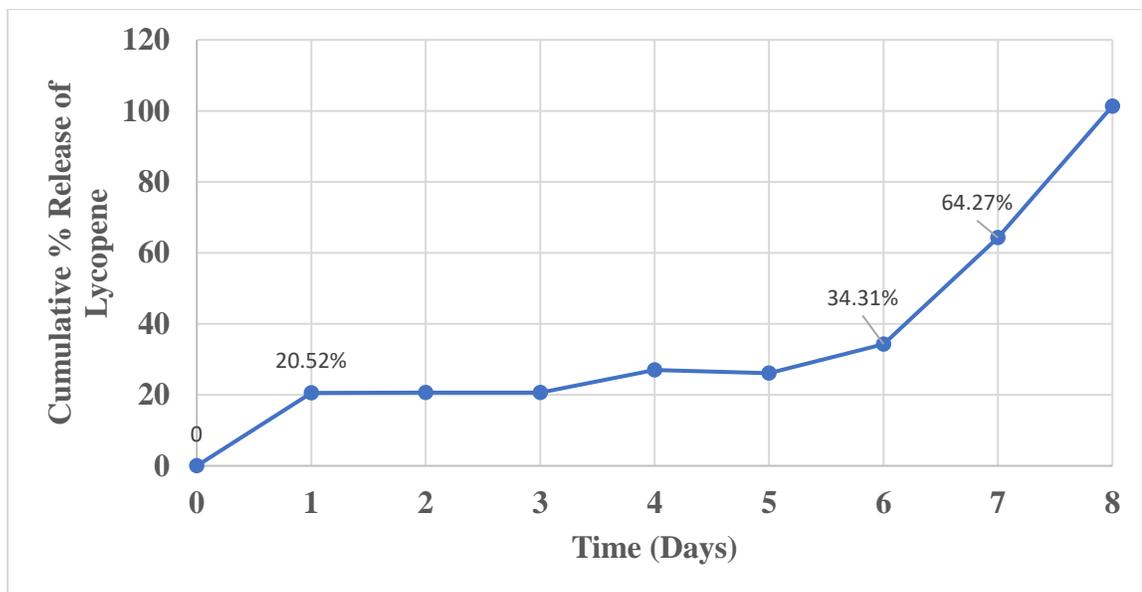

Figure 4.18. Control release activity of optimized PLA-lycopene nanoparticles.

## 4.4. Processing Effect on Physicochemical and Morphological Properties of Lycopene-NP

Heat energy in the form of steam is the most common thermal processing application in the food industry, for the purpose to cook and likewise inactivate enzymes and kill microorganisms to produce a shelf stable quality product. Although, the energy input in conventional pasteurization (CP) may expose the nutrients to extreme conditions for a longer period resulting to undesirable changes in foods particularly the nutrient density (Wang et al., 2018). The high processing temperatures may also cause denaturation of the proteins, development of off-flavor, loss of color, Maillard browning reaction, and destruction of essential vitamins and loss of consistency (Tewari et al., 2008, Gómez et al., 2011).

However, research has proven that microwave pasteurization has lower degradation effect on color compared to infrared and ohmic heating (Vikram et al., (2005)). Similarly, Yousefi et al. (2012) reported that color degradation is more prominent in conventional heat pasteurization compared to microwave pasteurization; hence, their study revealed decreased color with time during conventional pasteurization. Hence, the microwave pasteurization technique may be a potential candidate for pasteurizing fruit juice due to its inherent capacity to induce instantaneous and uniform heating. Microwave nonionizing radiation is localized between the infrared (IR) and radio frequency (RF) waves capable of minimizing nutrient loss in fruit juices during thermal processing (Regier et al., 2017). In general, microwave food processing uses the frequency wave of 2,450 MHz for domestic ovens while 2,450 MHz or 915 MHz is used for industrial ovens (Honda et al., 2014).

PLA Lycopene Nanoemulsion

*Fortification effect on physicochemical and morphological properties of lycopene-PLA nanoemulsion.* Figure-4.19 shows a temperature profile of microwave pasteurization for the 5 and 15º brix sample of the fruit juice model. The profile exhibited the same come-up time (80 s) to get to pasteurization temperature (100°C) for both concentrated model fruit juices (5 and 15º brix). The resident time was 40 s to obtain the desired 5 log reduction of microbial populations (Hashemi et al., 2019).

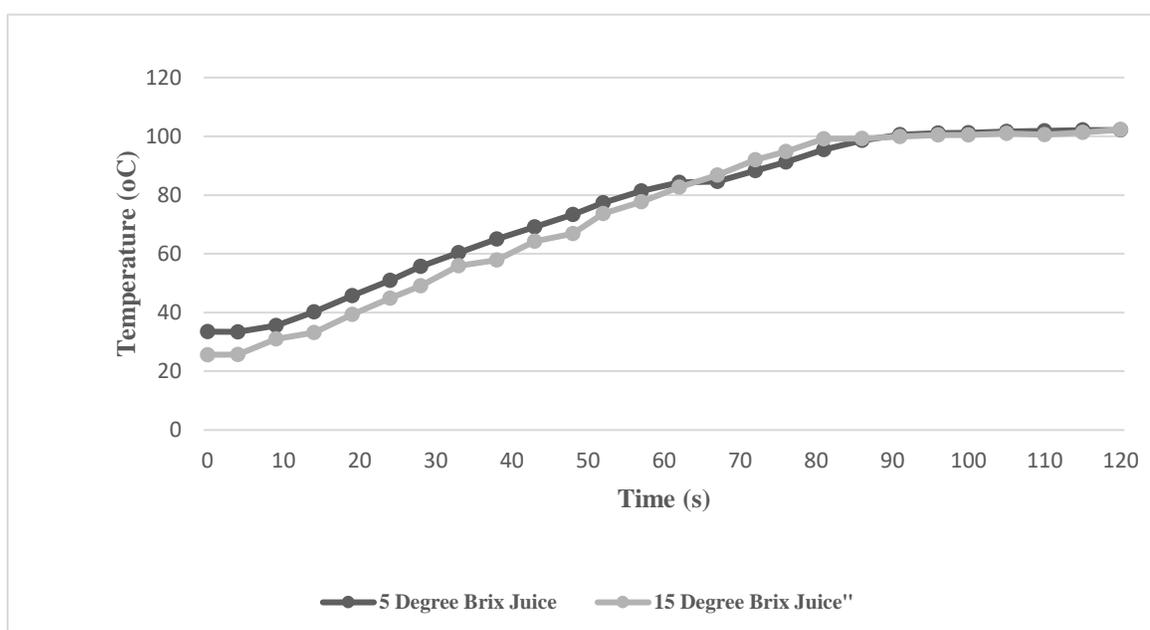

Figure 4.19. Temperature profile of microwave pasteurization for model juice fortified with lycopene nanoparticles.

When encapsulated LNP fortified with different types of fruit juices called fortification. On the other hand, non-encapsulated lycopene blended with juices called non fortified juices. The effect of fortification with encapsulated LNP on the integrity of lycopene nanoparticles in 15° brix fruit juice model significantly (P<0.001) increased hydrodynamic diameter from 293.03 to 480.24 nm as shown in Figure 4.20 (a)) even though no effect (P<0.05) was observed when fortified in decreased juice concentration (5° brix). The sucrose (Table sugar) was used to manufacture model juices usually to provide additional layer of coating on nanoparticles surface which further increases the size of the nanoparticles (Figure 4.20 (a)). On the other hand, it significantly (P<0.001) improved the zeta potential (-15.83 to -25.37 mV) as shown in (Figure 4.20 (c)) when the nanoparticles were fortified in 5° brix fruit juice model The fortification was to have a noticeable reduction effect (p < 0.05) on the electric conductivity (Figure 4.20 (d)) for both 5° and 15° brix juices. However, the reduction (P<0.05) of electric conductivity (Figure 4.20 (d)) of the nanoparticles was even more pronounced when fortified with 15° Brix (more than 8 times) than that of 5° Brix (7 times). The reason behind reduction was due to the poor electrical conductivity of the deionized water and tablel sucrose used to manufacture the fruit juice model (Chereches & Minea, 2019). The sugar provides an additional layer of coating on the nanoparticles, it shows the effect of sugar in the electrical conductivity and zeta potential of the nanoparticles.

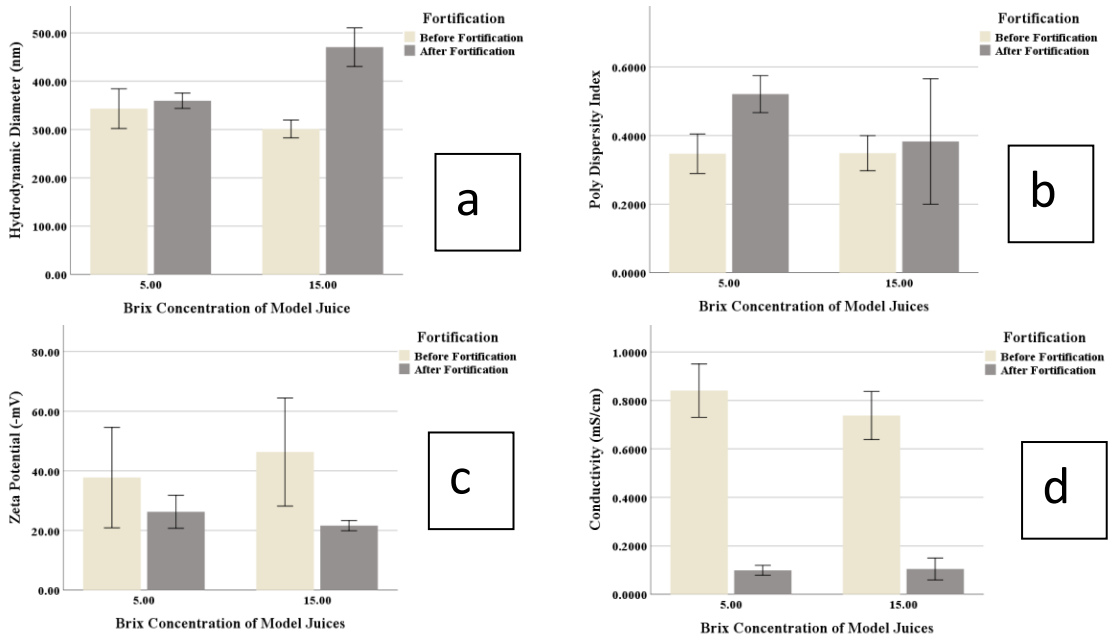

Figure 4.20. Encapsulation effect on HD (a), PDI (b), ZP (c) and conductivity (d) of PLA lycopene nanoemulsion.

The HD distribution curve shows the morphological properties of both fortified and non-fortified PLA-LNP. Figure 4.21 (a & b) shows that fortification in 15$^o$ Brix concentrated juices HD distribution shifted more to the right than that of 5$^o$ Brix juices.

On the other hand, width of distribution also observed to be increased for both 5$^o$ & 15$^o$ Brix juices. The explanation of that increased was the changes of spherical shape to rod shape (Figure 4.21 a & b).

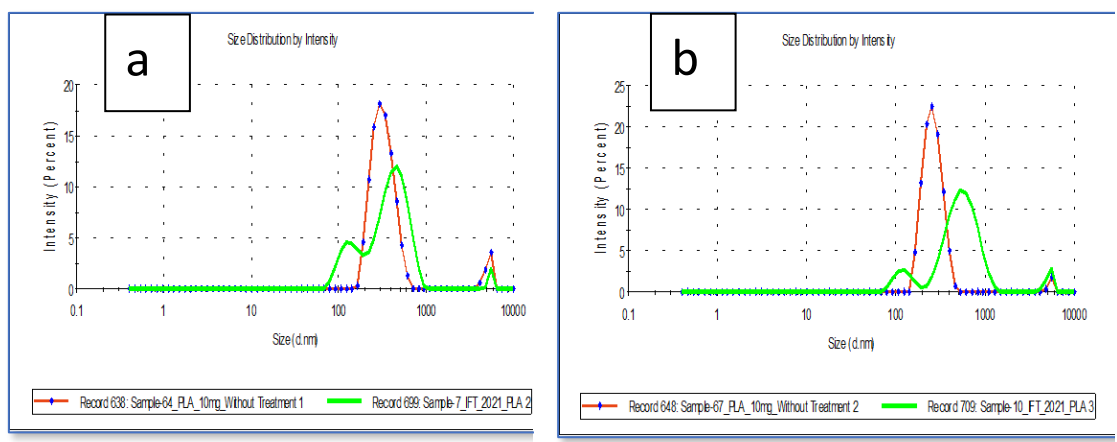

Figure 4.21. Encapsulation effect on HD distribution of lycopene nanoparticles at 5 ºbrix juice (a) and 15ºbrix juice (b). Before fortification with lycopene nanoparticles denoted as without treatments (Record 638: Sample-64 and Record 648: Sample-67) and Record 699: Sample-7 and Record 709: Sample-10 denoted as before fortification with LNP.

*Pasteurization effect on physicochemical and morphological properties of lycopene-PLA nanoemulsion.* The ANOVA results showed that pasteurization and juice concentration had significant interaction (P<0.05) effect on the hydrodynamic diameter, polydispersity index and zeta potential of fortified lycopene nanoparticles. Tukey tests showed that the CP and MP did not have any erosion or aggregation effect (P > 0.05) on size of the nanoparticles (HD) (Figure 4.22 (a)) and stability (Figure 4.22 (c)) of encapsulated lycopene nanoparticles when applied on both juice model system (5 and 15ºbrix). Figure 4.22 (b) showed a non-significant effect (P>0.05) of CTN, CP, and MP on the polydispersity index of lycopene NP. The HD distribution explains the morphological properties of PLA-LNP. As no significant (p > 0.05) changes appeared on the highest peak intensity or width of HD distribution of PLA-LNP so it can be said that spherical shape did not change to rod shape which would significantly affect absorption profile of lycopene nanoparticles.

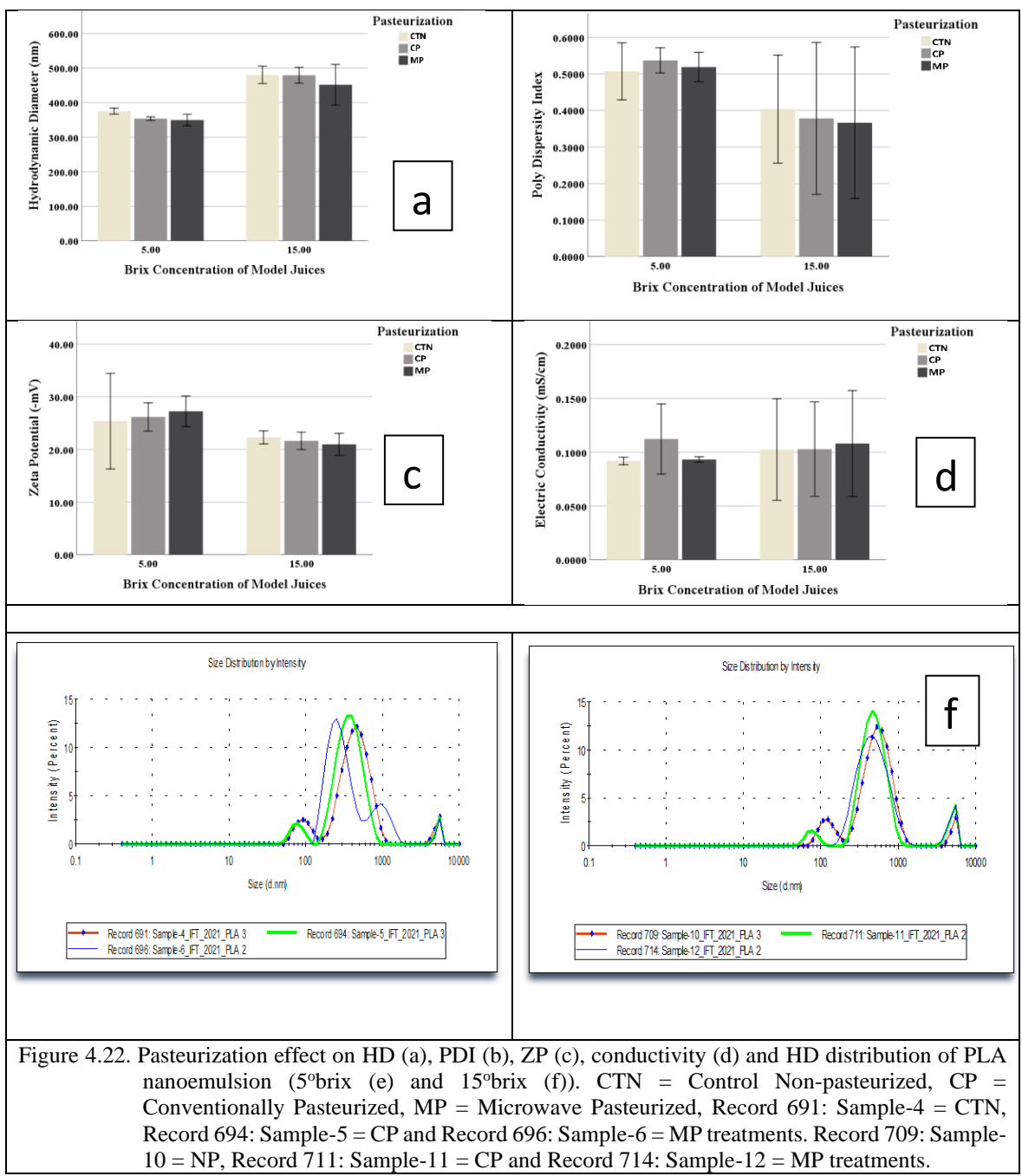

Figure 4.22. Pasteurization effect on HD (a), PDI (b), ZP (c), conductivity (d) and HD distribution of PLA nanoemulsion (5⁰brix (e) and 15⁰brix (f)). CTN = Control Non-pasteurized, CP = Conventionally Pasteurized, MP = Microwave Pasteurized, Record 691: Sample-4 = CTN, Record 694: Sample-5 = CP and Record 696: Sample-6 = MP treatments. Record 709: Sample-10 = NP, Record 711: Sample-11 = CP and Record 714: Sample-12 = MP treatments.

*Pasteurization effect on antioxidant properties of lycopene-PLA nanoemulsion.* The ANOVA results showed that pasteurization, Juice concentration (⁰brix) and fortification had significant interaction effect on the DPPH inhibition activity. No significant ($p > 0.05$) pasteurization treatment effect appeared on DPPH activity of lycopene when it was encapsulated with PLA and fortified in both 5$^o$ and 15$^o$ Brix juice model system. Figure 4.30 revealed that encapsulation provided protection of lycopene from degradation by 70% and more than 60% with CP and MP treatments, respectively. The higher brix (15$^o$ Brix) concentration, no degradation effect (P > 0.05) of lycopene was observed against MP, and CP treatments. The scenario was different in the case of lower juice concentration (5$^o$Brix) as DPPH inhibition was significantly (P<0.05) lower in the CP than the MP treatments. Previous research also supported the idea that conventional heat treatment and microwave pasteurization had significant degradation effect on the carotenoids content (Hashemi et al., 2019). Fratianni et al. (2010) also observed that about 50% of total carotenoids had degraded after 1 min of microwave treatment at 85$^o$C. Similarly, 57% of total carotenoids retained when microwave treatment was applied for 45 s on papaya puree (de Ancos et al., 1999).

Etzbach et al. (2020) observed that higher concentration of orange juices tends to prevent the degradation of total phenolic content. In their research they observed that TPC found to be 71.9 and 73.6% for 0 and 4% concentrated orange juices when thermal treatment (70$^o$C) was applied for 30 s. It is however observed that with low concentration, a much shorter treatment time is required to achieve the same effects as concentrated juices.

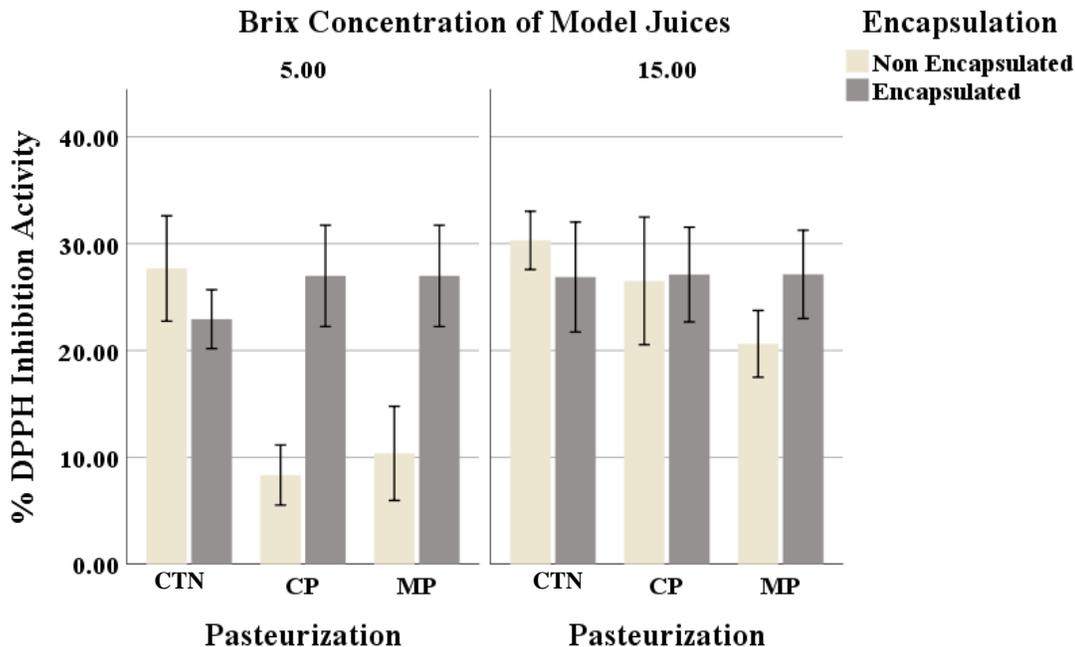

Figure 4.23. Pasteurization and encapsulation effect on antioxidant properties of PLA-lycopene nanoemulsion.

*Effect of pasteurization on color profile of PLA lycopene nanoemulsion.* Figure 4.19 shows the temperature profile of microwave pasteurization for 5° and 15°brix of fruit juice concentrations. The profile exhibited the same come-up time (80 s) to 100°C for both juice concentrations (5° and 15°brix), like the processing time (40 s) to obtain the desired 5 log reduction (Hashemi et al., 2019). The influence of pasteurization (CP, and MP) treatments on the color of fortified or non-fortified fruit juice are shown in Table 4.8. The ANOVA showed a significant ($p < 0.05$) interaction among fortification of encapsulated nanoparticles, pasteurization method and °Brix concentration for both the hue angle (Effect size 15.7%) and $\Delta E$ (Effect size 48.4%). The hue angle means the degree to which a stimulus can be described about how far the color is from the original color described as red, yellow, green, and blue. Moreover, there were significant ($P < 0.05$) differences on "Hue Angle" when non-capsulated lycopene fortified with 5º Brix juices are subjected to MP treatment. The color of the juice changes from yellow (hue angle of 55.93±0.53º) to lime green (88.31 ± 1.80º) due to microwave pasteurization. No significant ($p > 0.05$) changes were observed on hue angle when MP, and CP treatment was applied on the 15º brix juices fortified with non-encapsulated lycopene powder. So, it can be shown that the concentration and the density of the juice provided protection against "Hue angle" at different pasteurization treatments. However, the effect observed was different when fortified with encapsulated lycopene nanoparticles before CP. There were significant differences ($P<0.05$) on the "Hue Angle" of fortified juices when the CP treatment was applied on the different concentrations (15 and 5º Brix). The "Hue Angle" of the juice (5° Brix) spike with encapsulated free lycopene was sensitive to MP treatment due to discoloration, but an inverse effect on the15ºbrix juice.

Table 4.8. Pasteurization effect on color of concentrated (5° and 15°brix) juices fortified with PLA-lycopene nanoparticles.[4]

| Encapsulation | Pasteurization | °Brix | Hue Angle | ΔE |
|---|---|---|---|---|
| Non-capsulated | Non | | $55.93\pm0.53^{a,c,n}$ | 0 |
| | Conventional | 5.00 | $63.84\pm1.50^{a,d,o}$ | $1.51\pm0.19^{b,i*,m*}$ |
| | Microwave | | $88.31\pm1.80^{b*,e*,p}$ | $0.46\pm0.13^{g*,k,n*}$ |
| | Non | | $55.03\pm0.57^{b,c,r}$ | 0 |
| | Conventional | 15.00 | $56.49\pm0.13^{b,d,s}$ | $1.96\pm0.06^{b,j*,q*}$ |
| | Microwave | | $62.22\pm4.05^{b,f*,t}$ | $1.56\pm0.08^{h*,l,r*}$ |
| Capsulated | Non | | $53.06\pm0.38^{g,j,n}$ | 0 |
| | Conventional | 5.00 | $58.27\pm4.28^{g,k,o}$ | $0.13\pm0.06^{c,o*}$ |
| | Microwave | | $53.05\pm0.38^{g,m,q*}$ | $0.14\pm0.05^{c,p*}$ |
| | Non | | $54.08\pm0.58^{h,j,r}$ | 0 |
| | Conventional | 15.00 | $69.01\pm14.86^{i*,l*,s}$ | $0.25\pm0.19^{c,s*}$ |
| | Microwave | | $54.08\pm0.59^{h,m,t}$ | $0.20\pm0.15^{c,t*}$ |

The ΔE of fruit juice was calculated with - Eq. 3.3. Krapfenbauer et al., (2006) which shows a visible color difference varies from 2 to 3.5 and considered slightly noticeable from 0 to 0.5 (Cserhalmi et al., 2006). This study the ΔE values were less than 2 in all cases of the pasteurization treatments. Hence, it is an indication that pasteurization did not have any visual color effect. Zhou et al. (2009) observed the color differences on carrot juices pasteurized with conventional heating. However, thermal treatment made greater color difference (P<0.05) with juices (5° and 15° brix) fortified with unencapsulated lycopene than the encapsulated lycopene nanoparticles. This could be attributed to the polymer coating preventing the color pigments from thermal degradation. On the other hand, greater (P<0.05) effect on color was associated with the CP    fortified juice concentration (15 brix) with nanoparticles to the 5°brix concentration. Moreover, the high solid content retained the heat longer than the low solid content juice due to their respective thermal properties, hence, a higher degradation effect on the color of the juices. Higher color degradation was observed more on the conventional pasteurization than the MP when juice fortified with non-encapsulated lycopene. On the contrary, nanoparticles protect color degradation when pasteurization imposed on both 5° and 15°brix juices. Stinco et al. (2013) also observed that microwave processing specifically changed the color parameters "b" and "L" compared to non-treated samples. By contrast, Cendres et al. (2011) investigation results showed that MP improves color the parameters (b and L). Hashemi et al. (2019) observed the color properties in different hue angles as microwave pasteurization significantly reduced some of the phenolic compounds and β-carotene.

---

[4] Significance measured among the columns. 1st Superscript letter explain significance among different pasteurization treatments; 2nd Superscript letter mean pasteurization effect between two different brix juices and 3rd superscript means pasteurization effect on color profile of the model juices fortified with capsulated or non-capsulated nanoparticles.

Table 4.9 explained that hunter color values "L" and "a" for the juices were significantly (P<0.05) correlated to the DPPH inhibition activity of lycopene nanoparticles. The Pearson correlation indicated  the changes of "L" values are directly proportional  to the change in the DPPH activity of the nanoparticles with a correlation coefficient of 0.56 while the "a" value and the electrical conductivity  are inversely correlated with the DPPH inhibition activity with a correlation coefficient of 0.278  and 0.406, respectively, which explained the  high electric conductivity contributing to  the heat penetration causing the degradation effect of lycopene fortified in the juices. Moreover, the Pearson correlation (Table 4.9) also shows the lightness ("L" value) and redness to greenness ("a" value) had a significant (P<0.05) indirect (r =-0.911) and direct (r = 0.869) effect, respectively, with electrical conductivity. This indicates that higher the electrical conductivity the higher is the heat penetration capacity, thus increase the degradation effect and color change due to browning reaction as a result of caramelization of sucrose. Furthermore, this contributes to the conversion of red to green color of the juice samples. Similarly, Arjmandi et al. (2016) found a significant positive relationship between the amount of lycopene and redness (a) of tomato puree with a Pearson correlation coefficient of 0.758. Vikram et al. (2005) observed microwave pasteurization at 455 W for 180 s had a higher degradation effect on vitamin C compared with conventional heat treatment. However, other authors reported that samples treated at 900 W for 30 sec to 10 min preserve the vitamin C content compared to conventional heat treatment (Igual et al., 2010b; Geczi et al., 2013). Picouet et al. (2009) observed no degradation effect on polyphenol in apples juice with microwave pasteurization at 70 – 80ºC.

Table 4.9. Correlation of physical properties of juices with DPPH inhibition activity of LNP.

| Dependent Variable | Dependent Variable | Pearson Correlation Coefficient | Significance Level (P) |
|---|---|---|---|
| Color "L" | DPPH Inhibition Activity | 0.560 | 0.000 |
| Color "a" | DPPH Inhibition Activity | -0.278 | 0.018 |
| Electric-Conductivity | DPPH Inhibition Activity | -0.406 | 0.000 |
| Electric-Conductivity | Color "L" | -0.911 | 0.000 |
| Electric-Conductivity | Color "a" | 0.869 | 0.000 |

PLGA-Lycopene Nanoemulsions

*Encapsulation effect on physicochemical and morphological properties of lycopene-PLA nanoemulsion.* The ANOVA test (Table 10) shows significant ($p < 0.05$) interactions between the fortification, pasteurization and juice concentration, hence, contributing to the electrical conductivity, zeta potential and DPPH inhibition changes on the lycopene nanoparticles. A bimodal interaction of pasteurization and fortification was observed and found to be significantlychanging the PDI value of the nanoparticles. Additionally, pasteurization, fortification and brix concentration have individual effects on the hydrodynamic diamter of the nanoparticles.

Table 4.10. ANOVA test results to evaluate effect of pasteurization on physicochemica and antioxidant properties of PLGA-LNP.

| Source | Dependent Variable | Type III Sum of Squares | df | Mean Square | F | Sig. |
|---|---|---|---|---|---|---|
| Corrected Model | HD | 1560183.33 | 11 | 141834.848 | 174.469 | 0.000 |
| | PDI | 0.827 | 11 | 0.075 | 11.784 | 0.000 |
| | ZP | 38846.373 | 11 | 3531.488 | 466.183 | 0.000 |
| | Conductivity | 45.837 | 11 | 4.167 | 692.248 | 0.000 |
| | DPPH Inhibition | 1482.419 | 11 | 134.765 | 12.373 | 0.000 |
| Fortification | HD | 1369852.602 | 1 | 1369852.602 | 1685.037 | 0.000 |
| | PDI | 0.547 | 1 | 0.547 | 85.692 | 0.000 |
| | ZP | 38131.139 | 1 | 38131.139 | 5033.592 | 0.000 |
| | Conductivity | 43.657 | 1 | 43.657 | 7252.653 | 0.000 |
| | DPPH Inhibition | 74.330 | 1 | 74.330 | 6.824 | 0.011 |
| Pasteurization | HD | 8823.450 | 2 | 4411.725 | 5.427 | 0.007 |
| | PDI | 0.022 | 2 | 0.011 | 1.758 | 0.181 |
| | ZP | 24.206 | 2 | 12.103 | 1.598 | 0.211 |
| | Conductivity | 1.121 | 2 | 0.560 | 93.076 | 0.000 |
| | DPPH Inhibition | 650.182 | 2 | 325.091 | 29.846 | 0.000 |
| Brix | HD | 41794.542 | 1 | 41794.542 | 51.411 | 0.000 |
| | PDI | 0.022 | 1 | 0.022 | 3.376 | 0.071 |
| | ZP | 368.428 | 1 | 368.428 | 48.635 | 0.000 |
| | Conductivity | 0.043 | 1 | 0.043 | 7.070 | 0.010 |
| | DPPH Inhibition | 436.211 | 1 | 436.211 | 40.048 | 0.000 |

Table 4.10. (Continued).

| | | | | | | |
|---|---|---|---|---|---|---|
| Fortification x Pasteurization | HD | 210.294 | 2 | 105.147 | 0.129 | 0.879 |
| | PDI | 0.067 | 2 | 0.034 | 5.280 | 0.008 |
| | ZP | 29.551 | 2 | 14.776 | 1.951 | 0.151 |
| | Conductivity | 1.124 | 2 | 0.562 | 93.403 | 0.000 |
| | DPPH Inhibition | 342.384 | 2 | 171.192 | 15.717 | 0.000 |
| Fortification x Brix | HD | 26039.290 | 1 | 26039.290 | 32.031 | 0.000 |
| | PDI | 0.050 | 1 | 0.050 | 7.802 | 0.007 |
| | ZP | 1.689 | 1 | 1.689 | 0.223 | 0.639 |
| | Conductivity | 0.059 | 1 | 0.059 | 9.791 | 0.003 |
| | DPPH Inhibition | 294.151 | 1 | 294.151 | 27.006 | 0.000 |
| Pasteurization x Brix | HD | 20452.315 | 2 | 10226.157 | 12.579 | 0.000 |
| | PDI | 0.013 | 2 | 0.006 | 1.013 | 0.369 |
| | ZP | 3.303 | 2 | 1.651 | 0.218 | 0.805 |
| | Conductivity | 0.451 | 2 | 0.225 | 37.431 | 0.000 |
| | DPPH Inhibition | 193.076 | 2 | 96.538 | 8.863 | 0.000 |
| Fortification x Pasteurization x Brix | HD | 3756.306 | 2 | 1878.153 | 2.310 | 0.108 |
| | PDI | 0.019 | 2 | 0.010 | 1.504 | 0.231 |
| | ZP | 150.571 | 2 | 75.286 | 9.938 | 0.000 |
| | Conductivity | 0.470 | 2 | 0.235 | 39.081 | 0.000 |
| | DPPH Inhibition | 90.479 | 2 | 45.239 | 4.153 | 0.020 |
| Corrected Total | HD | 1608960.395 | 71 | | | |
| | PDI | 1.210 | 71 | | | |
| | ZP | 39300.893 | 71 | | | |
| | Conductivity | 46.198 | 71 | | | |
| | DPPH Inhibition | 2135.951 | 71 | | | |

The homogeneity test (Table 4.11) shows microwave pasteurization did not have any effect on nanoparticles size, whereas the conventional pasteurization method had a significant ($P<0.05$) erosion effect on nanoparticles size. It also shows (Table 4.11) encapsulation provided a stronger protection for lycopene encapsulated in PLGA polymer when treated with microwave compared to conventional pasteurization. The homogeneity test also revealed that pasteurization did not have any effect on polydispersity index nor the zeta potential of the nanoparticles, but significantly reduced the DPPH inhibition activity by 5%. Conventional pasteurization decreased the hydrodynamic diameter from 386.22 to 360.86 nm, but stable against microwave pasteurization compared to non-pasteurized nanoparticles.

Table 4.11. Homogeneity test results to identify main effect of pasteurization to physicochemical properties of PLGA-lycopene nanoemulsion.

| Independent Variables | HD (nm) | PDI | ZP (mV) | CD (mS/cm) | Antioxidant Properties (% DPPH Inhibition Activity) |
|---|---|---|---|---|---|
| **Pasteurization** | | | | | |
| CTN | 360.86[a] | 0.46[a] | -40.72[a] | 0.59[a] | 20.63[a] |
| CP | 386.22[b] | 0.49[a] | -39.70[a] | 0.65[b] | 20.99[a] |
| MP | 389.40[b] | 0.51[a] | -39.52[a] | 0.76[c] | 25.57[b] |

Figure 4.31 shows fortification of nanoparticles for both concentrations (5° and 15°Brix concentration) did help to increase the hydrodynamic diameter of the nanoparticles with the additional sugar layer. The Zeta potential and electrical conductivity of the nanoparticles reduced sharply due to the fortification and concentration of the juices. The Zeta potential expressed the surface change of the NPs without considering internal strength of particles. So, it identifies the charges of the sucrose layer in place of the PLGA surface concealed inside the outer layer.

The electrical conductivity of the NPs significantly ($p < 0.05$) reduced from 1/6[th] or lower due to the fortification of juices (5 and 15°brix). Due to the additional coating, the sucrose's electrical conductivity becomes dominant during measurement. The fortification of the juice samples has significantly ($P < 0.05$) reduced the electrical conductivity of nanoparticles, since the surface of the nanoparticles reflects low conductivity of the sugar.

In both concentrated juices, the conductivity was similar as the surface coating materials were sucrose in both cases.

The normalized distribution of the HD was observed to shifted to the right regardless of the concentration, without compromising the width of the hydrodynamic diameter distribution. The similar width of HD distribution tells us fortification did not contribute any changes on the shape of nanoparticles.

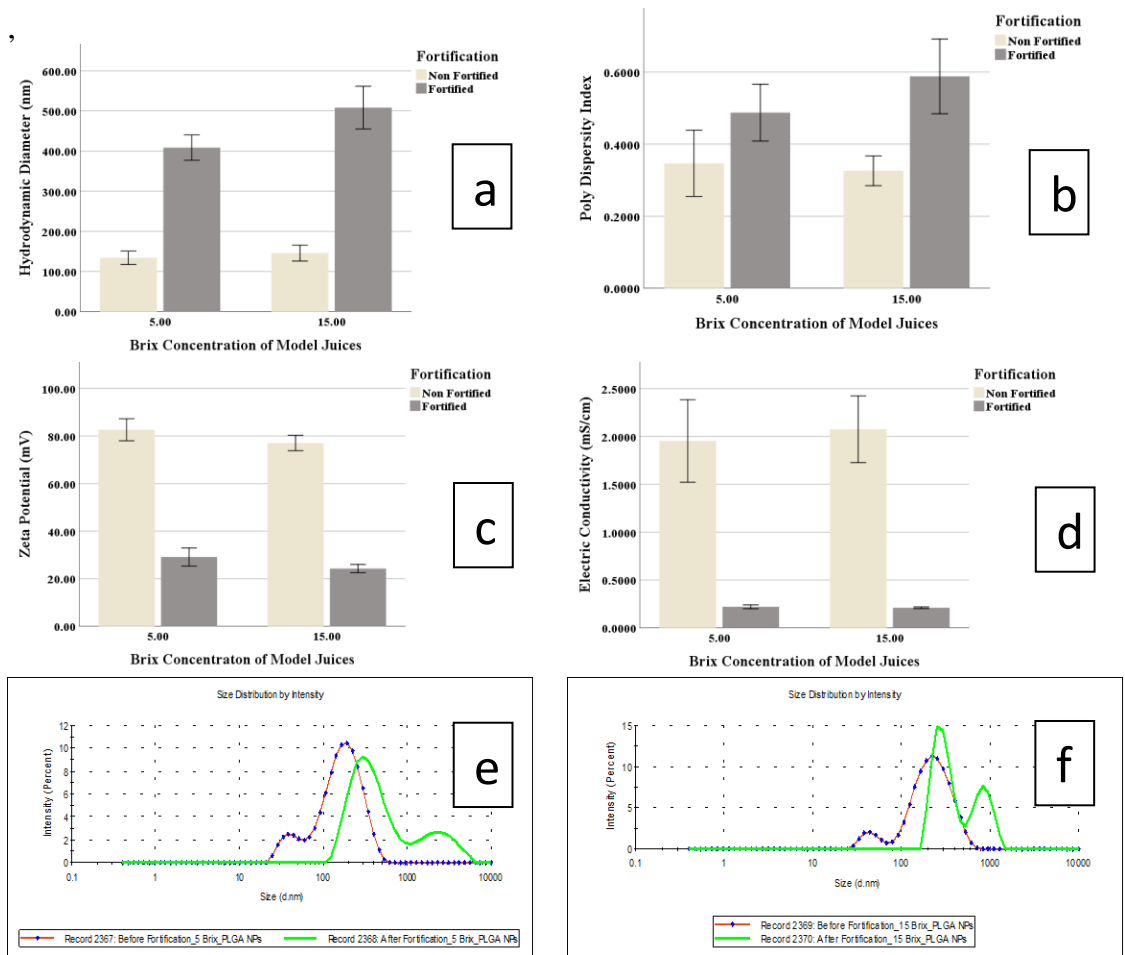

Figure 4.24. Encapsulation (Fortification) effect on HD (a), PDI (b), ZP (c), conductivity (d) and HD distribution of PLGA nanoemulsion (fortified with 5° (e) and 15°brix (f))

*Pasteurization effect on physicochemical and morphological properties of lycopene-PLGA nanoemulsion.* A significant difference ($p < 0.05$) was observed on the polydispersity index of NPs when fortified in 5º Brix juice and pasteurized with microwave. It shows that electromagnetic waves generated from the microwave treatment diminish the homogeneity of the nanoparticles. In 15º brix concentrated juices (Figure 4.25 (b)), fortification reduced the homogeneity (PDI) at different pasteurization treatments (NP, CP and MP) ($P<0.05$). The homogeneity of the PDI of nanoparticles was found to be similar ($P<0.05$) for both types of juices regardless of pasteurization. No significant effect ($p < 0.05$) was found on the size (HD) (Figure 4.25 (a)) stability (ZP) (Figure 4.25 (c)) and conductivity (Figure 4.25 (d)) of the nanoparticles due to pasteurization. It was shown that the stability of the nanoparticles for both concentrations observed to be similar regardless of CP and MP treatments.

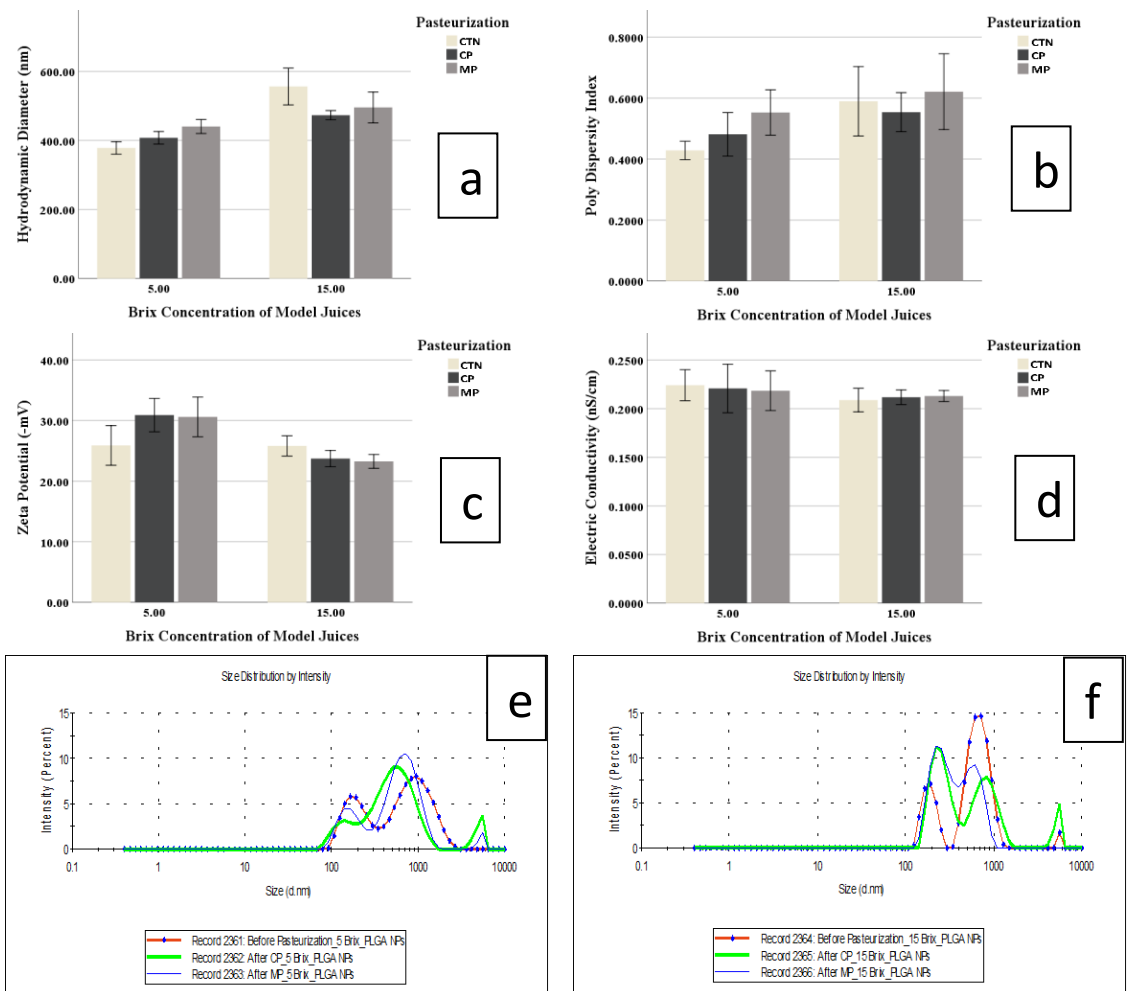

Figure 4.25. Pasteurization effect on HD (a), PDI (b), ZP (c), conductivity (d) and HD distribution of PLGA nanoemulsion both at 5°brix (e) and 15°brix juice model (f).

*Pasteurization effect on antioxidant properties of PLGA-lycopene nanoemulsion.* A significant decrease was observed on the DPPH inhibition activity (Figure 4.26) of LNP when treated with CP or MP. About 60% and more than 50% of the total functional activity of lycopene were degraded for CP and MP treated nanoparticles in 5° Brix juice concentration, respectively. Microwaved (800W) pasteurized samples showed significant ($p < 0.05$) decrease in $\beta$-carotene content after 40s of holding time for papaya puree (de Ancos et al., 1999). de Ancos et al. (1999) also observed a 75% loss of $\beta$-carotene content when 475W of microwave treatment was applied into the papaya puree for 45s. Achir et al. (2016) observed no significant carotenoid degradation in ohmic treatment of grapefruit juice and acerola pulp due to low dissolved oxygen content which contributes to the oxidization of carotenoids during heat treatments. Jaeschke et al. (2016) also observed less degradation effect of microwave pasteurization on $\beta$-carotene than conventional pasteurization. The degradation effect of conventional pasteurization *on* $\beta$-carotene content started after 15 minutes of the treatment.

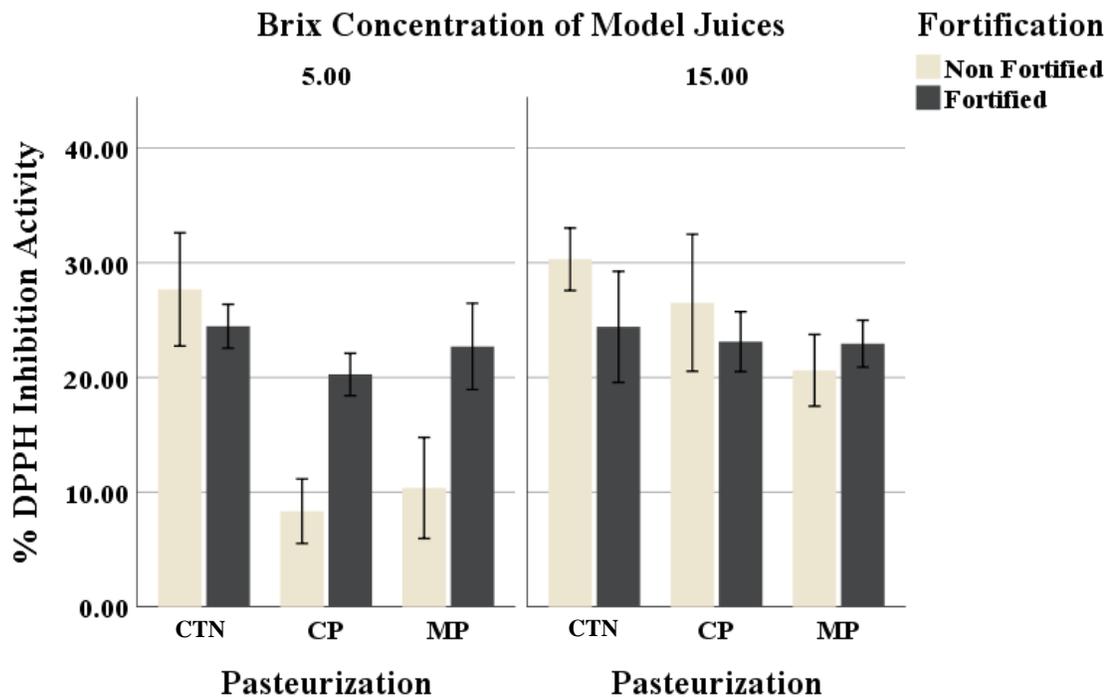

Figure 4.26. Pasteurization and encapsulation (fortification) effect on antioxidant properties of PLGA-lycopene nanoemulsion.

*Pasteurization and encapsulation effect on color profile of PLGA-lycopene nanoemulsion.*

Pasteurization did not have any effect (p > 0.05) on the hunter color value "L" for 15° Brix juices concentration fortified with lycopene nanoparticles. The color "a" value was not affected when pasteurization was applied on both non-encapsulated lycopene fortified juice at different concentrations (15 and 5º Brix). Thus, no pasteurization effect was also observed for color "b" The color of the juices specifically "L" value was significantly reduced for 7.34 ± 0.08 to 5.83 ± 0.14 and 7.34±0.08 to 6.88±0.15 for CP and MP, respectively. The reduction of color "L" value shows the browning reaction higher in the conventional pasterirzation than the microwave pasteurization. Hashemi et al. (2019) showed 30% losses of vitamin C with conventional pasteurization; whereas microwave heating at 400W and 800W contributed losses of about 38 and 42%, respectively. Vitamin C losses were more pronounced at 100 and 200V of ohmic treatment, whereas the losses were 43% and 50% of the total content, respectively (Hashemi et al., 2019).

Table 4.12. Pasteurization effect on color properties of fortified or non-fortified with lycopene nanoparticles.

| Fortification | ºBrix | Pasteurization | L | a | b |
|---|---|---|---|---|---|
| Non-Fortified | 5 | None | 7.34±0.08[a] | 0.3262±0.001[a] | 0.2702±0.001[a] |
| | | Conventional | 5.83±0.14[b] | 0.3188±0.001[ac] | 0.2861±0.003[b] |
| | | Microwave | 6.88±0.15[c] | 0.3052±0.000[bc] | 0.3049±0.001[b] |
| | 15 | None | 7.05±0.03[d] | 0.3345±0.002[d] | 0.2675±0.011[c] |
| | | Conventional | 5.09±0.04[e] | 0.3335±0.000[d] | 0.2782±0.000[d] |
| | | Microwave | 5.49±0.08[e] | 0.3244±0.000[d] | 0.2804±0.000[d] |
| Fortified | 5 | None | 8.31±0.47[f] | 0.3363±0.003[e] | 0.2834±0.002[e] |
| | | Conventional | 7.65±0.22[g] | 0.3373±0.011[e] | 0.2829±0.001[e] |
| | | Microwave | 7.72±0.22[g] | 0.3349±0.003[e] | 0.2829±0.001[e] |
| | 15 | None | 4.71±0.12[h] | 0.3494±0.006[f] | 0.2873±0.002[f] |
| | | Conventional | 4.76±0.09[h] | 0.3389±0.006[g] | 0.2840±0.002[f] |
| | | Microwave | 4.52±0.16[h] | 0.3441±0.004[f] | 0.2853±0.002[f] |

Superscript value explains the significant difference among different pasteurization treatments through the column of the table.

Processing Effect on Rheological Properties of Lycopene Nanoemulsion

Food texture and rheological behavior are significant determinants of consumer acceptance and preference. The rheological behavior may be a potential indicator to predict bio-accessibility and bioavailability of encapsulated lycopene (Lugasi et al., 2003) fortified in fruit juice. Apparently, the rheological characterization of food is also essential for the design of unit operations, process optimization, and quality assurance in food industries (Ibarz, 2003; Rao, 1999). The physicochemical properties of the nanoparticles such as particle size, shape, and surface charge, play a major role in the cellular uptake through the intestine and the Blood Brain Barrier (BBB) (Ciani et al., 2007).   Nanoparticles attach to the cell membrane before absorption which seems to be affected most by the surface charge of the particles (Patila et al., 2007; Chen et al., 2010). However, the cellular surfaces are dominated by negatively charged sulphated proteoglycans molecules which facilitates the absorption of positively charged LNP by binding very tightly on its surface (Bernfild et al., 1999; Mislick et al., 1996). Unlikely, the human lung's fibroblast cells rapidly adsorb negatively charged nanoparticles (Limbach et al., 2005), thus play pivotal roles in cellular proliferation, migration, and motility. However, no existing studies have investigated the effect of conventional and microwave pasteurization effects on the physicochemical and rheological properties of juice fortified with encapsulated lycopene nanoparticles which is the main scope of this study.

PLA-Lycopene Nanoemulsion

The influence of CP and MP treatment on viscosity of juices model system is shown in Figure 4.42 (a), (b) and 4.43 (a) and (b). A Non-Newtonian shear thinning behavior was shown at shear rate ranging from 50 to 230 s$^{-1}$ for both juice concentrations (5 and 10 °Brix). Figure 4.42 (a) & (b) shows that the rheological behavior of the 15 °Brix juices fortified with lycopene nanoparticles; hence, the viscosity of the juice did not change significantly (P<0.05) when treated with CP or MP. Similarly, no change (P<0.05) was observed at lower juice concentration (5° Brix) as shown in Figure 4.43 (a) & (b) for the shear rate ranging from 50 to 230 s$^{-1}$ when pasteurization with conventionally or Microwave.

*Encapsulation and pasteurization effect on viscosity profile of lycopene-PLA nanoemulsion.* The Casson model best explains the rheological behavior of juices before and after fortification with encapsulated lycopene nanoparticles as shown in Table 4.13. The ANOVA test indicated that both fortification & pasteurization, and fortification & juice concentration had significant interaction effects on the plastic viscosity and yield stress of the juices. Meanwhile, when the juices of 5° brix are fortified with PLA-LNP it significantly (P<0.05) reduced its plastic viscosity from 23.97 ± 1.60 to 7.17 ± 0.31 mPas (Figure 4.27) and the yield stress from 10.23 ± 0.30 to 3.34 ± 0.14 mPa.  A similar trend was also observed for juice concentration of 15°brix fortified with PLA-LNP (Figure 4.28) as plastic viscosity reduced from 16.03 ± 3.29 to 2.69 ± 0.08 mPa-s and yield stress from 6.57 ± 1.16 to 2.11 ± 0.86 mPa.

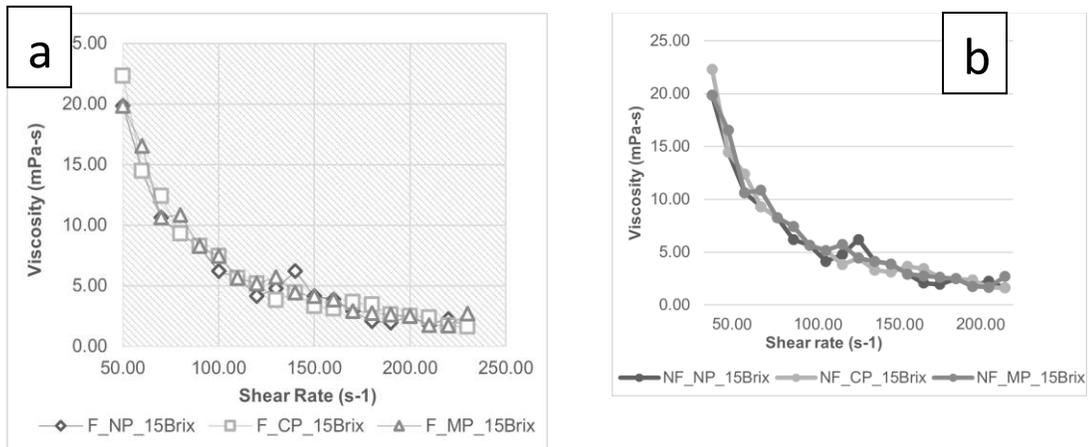

Figure 4.27. Viscosity vs shear rate profile for encapsulated (or fortified) (a) and non-encapsulated (or non-fortified) (b) PLA-lycopene nanoemulsion at 15°Brix concentration[5].

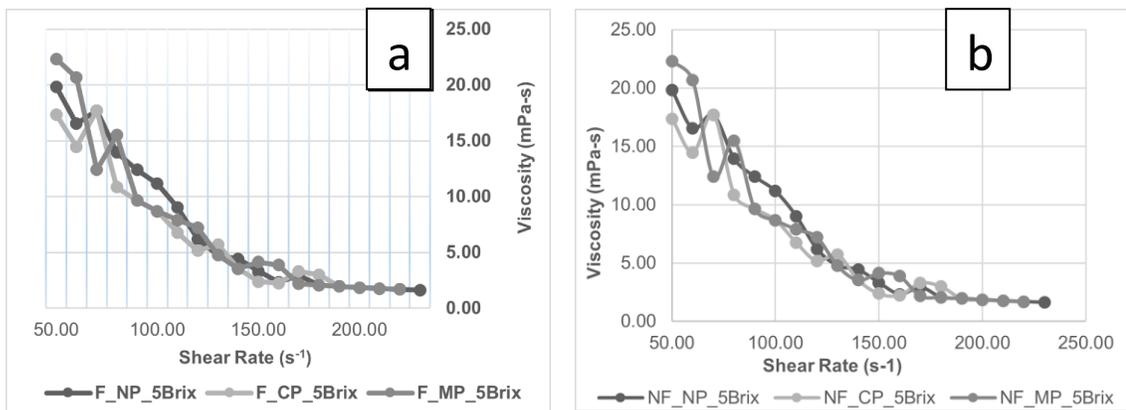

Figure 4.28. Viscosity vs shear rate profile for fortified (a) and non-fortified (b) PLA-lycopene nanoemulsion at 5°Brix concentration.

---

[5] CTN= Control Non-pasteurized, CP= Conventional Pasteurized MP= Microwave Pasteurized, F= After fortification with nanoparticles.

Table 4.13. Rheological model fitness of lycopene nano-emulsions at different
pasteurization treatments[6].

| Encapsulation | Pasteurization | Juice Brix | Casson Model CoF (%) | Power Model CoF (%) |
|---|---|---|---|---|
| Non | Non | 5.00 | 88.93±3.23 | 66.46±9.25 |
| | | 15.00 | 92.33±0.40 | 82.00±3.56 |
| | Conventional | 5.00 | 91.80±1.70 | 78.16±5.36 |
| | | 15.00 | 93.63±1.02 | 84.83±4.65 |
| | Microwave | 5.00 | 90.60±0.00 | 67.50±8.56 |
| | | 15.00 | 91.10±0.00 | 79.70±3.98 |
| Encapsulated | Non | 5.00 | 89.60±4.48 | 77.10±3.85 |
| | | 15.00 | 89.30±2.00 | 76.50±1.00 |
| | Conventional | 5.00 | 87.90±4.39 | 72.70±3.63 |
| | | 15.00 | 90.50±4.52 | 80.70±4.03 |
| | Microwave | 5.00 | 89.60±4.48 | 74.20±3.71 |
| | | 15.00 | 91.10±4.55 | 81.30±4.06 |

*Encapsulation and pasteurization effect on yield stress, plastic viscosity and flow index of PLA-lycopene nanoemulsion.* It can be inferred that the juices fortified with nanoparticles were significantly (P<0.001) lower the plastic viscosity and yield stress (Figure 4.27 (a) & (b)) compared to juices fortified with non-encapsulated lycopene. Based on Figure 4.27 (a) & (b) the higher sugar content in the juice model (15° Brix) potentially prevents the yield stress and plastic viscosity loss due to CP and MP thermal treatment than lower one (5° brix). The flow behavior index shown in Figure 4.46 shows that pasteurization (CP or MP) significantly (P<0.05) influenced the flow behavior index from Newtonian to non-Newtonian behavior. The microwave treatment had a higher effect (P<0.05) on flow behavior index than conventional pasteurization. In contrast, Nindo et al. (2005) did not

---

[6] CoF = Confidence of fitness.

observe any significant concentration effect on flow behavior characteristics on blueberry and rashberry juices.

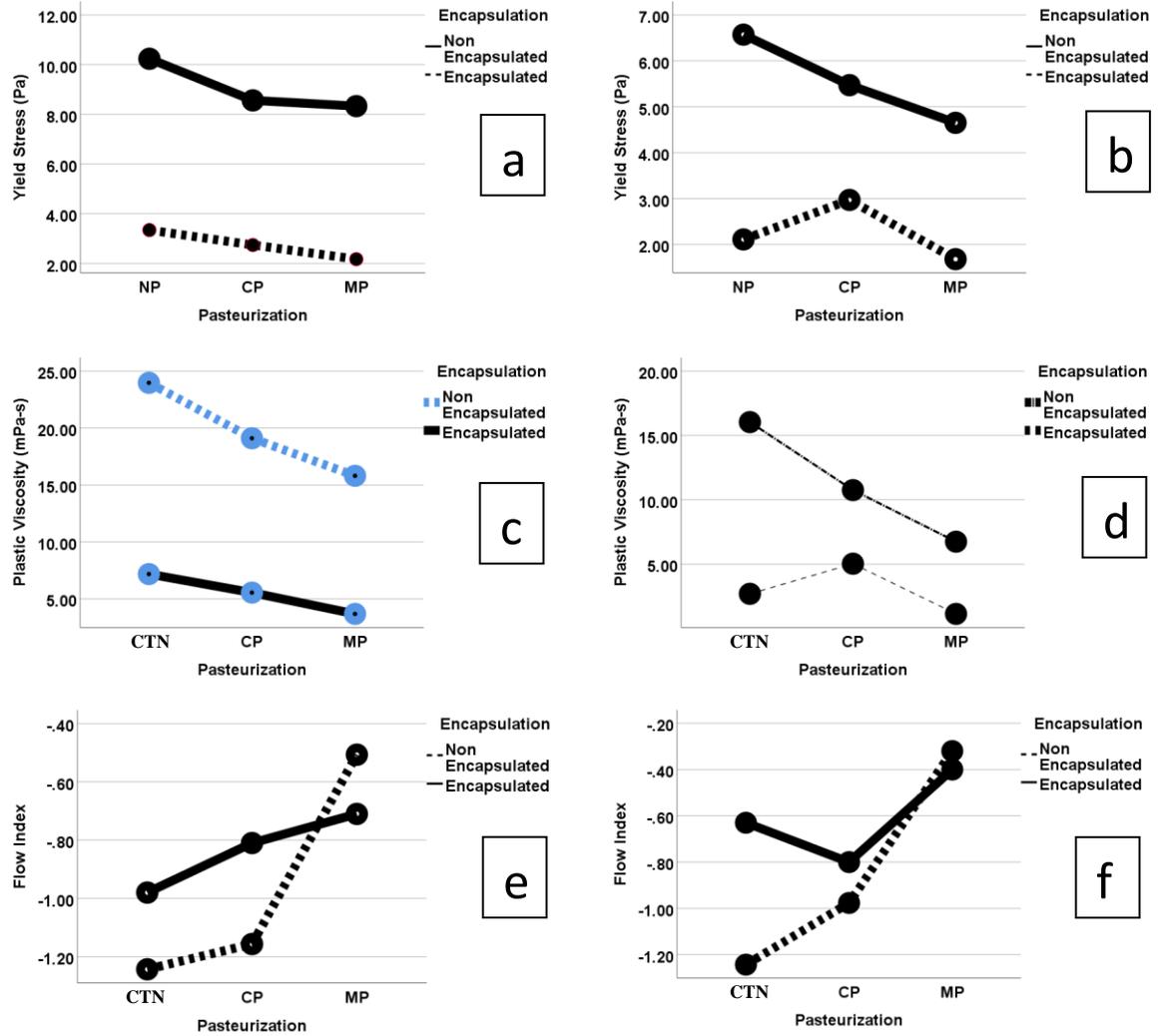

Figure 4.29 Pasteurization effect on yield stress (at 5ºbrix = a and 15ºbrix = b), plastic viscosity (at 5ºbrix = c and 15ºbrix = d) and flow index (at 5ºbrix = e and 15ºbrix = f) of different concentrated juice model due to fortification with PLA-LNP[7].

<hr />

[7] CTR= Control; CP= Conventional Pasteurized; MP= Microwave Pasteurized.

PLGA-Lycopene Nanoemulsion

*Encapsulation and pasteurization effect on viscosity profile of PLGA-lycopene nanoemulsion.* The Casson model best described the rheological behavior of both fortified and non-fortified juice concentrations ($5^{\circ}$ or $15^{\circ}$Brix). The confidence of fitness range of Casson model for non-fortified juice samples with LNP were between $91.30 \pm 1.39$ and $93.47 \pm 1.44$, compared to $92.16 \pm 1.99$ and $93.74 \pm 0.97$ when LNP fortified in juice concentration ($5^{\circ}$Brix). Confidence of fitness was a little bit higher ($94.47 \pm 0.40$ and $94.70 \pm 0.17$) for non-fortified juices ($15$ $^{\circ}$Brix) and ($94.56 \pm 0.94$ and $94.76 \pm 0.79$) after fortification with LNP. The interaction effect for fortification, concentration of the juices and pasteurization treatment significantly influenced the confidence of fitness of the Casson and Bingham model. However, the Power-Law model only deviated significantly due to bimodal interaction of fortification and pasteurization treatment effect.

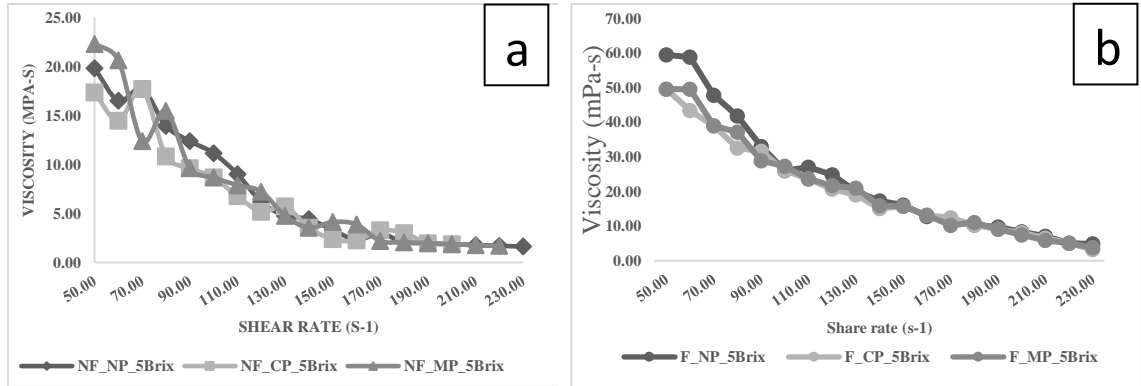

Figure 4.30 Viscosity vs shear rate profile for fortified (b) and non-fortified (a) PLA-lycopene nanoemulsion at 5ºBrix concentration[8].

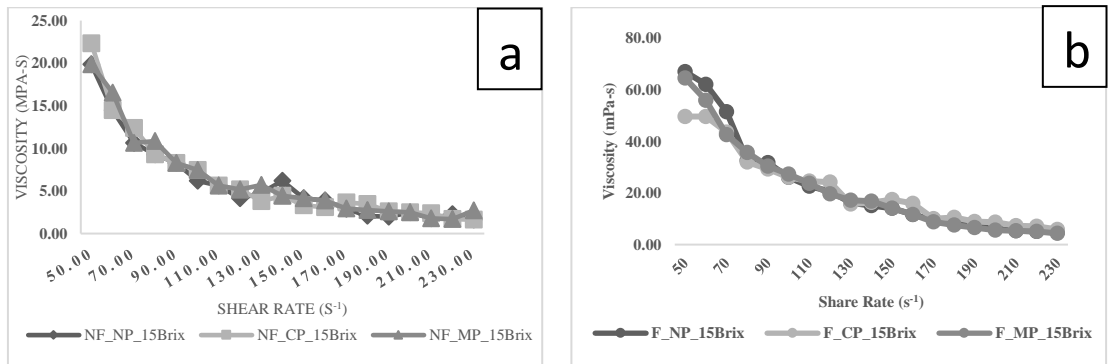

Figure 4.31 Viscosity vs shear rate profile for fortified (b) and non-fortified (a) PLA-lycopene nanoemulsion at 15ºBrix concentration[9].

Table 4.14 Rheological model fitness of lycopene nano-emulsions at different pasteurization treatments[10].

| Fortification | Brix | Pasteurization | Bingham Model | Casson Model | Power-Law Model |
|---|---|---|---|---|---|
| | | | Mean±SD | Mean±SD | Mean±SD |
| Non-Fortified | 5°Brix | NP | 85.73±2.89 | 91.30±1.39 | 77.10±2.42 |
| | | CP | 87.90±0.87 | 92.53±0.98 | 80.40±2.42 |
| | | MP | 89.47±2.66 | 93.47±1.44 | 82.40±3.29 |
| | 15°Brix | NP | 90.50±0.52 | 94.47±0.40 | 84.63±2.02 |
| | | CP | 91.40±1.04 | 94.70±0.17 | 85.93±0.92 |
| | | MP | 88.90±1.38 | 94.60±0.00 | 86.93±0.98 |
| Fortified | 5°Brix | NP | 90.02±1.39 | 93.74±0.97 | 83.27±2.73 |
| | | CP | 86.17±2.13 | 92.29±0.97 | 81.79±2.25 |
| | | MP | 86.52±3.37 | 92.16±1.99 | 79.63±5.87 |
| | 15°Brix | NP | 88.43±1.33 | 94.68±0.56 | 85.88±2.94 |
| | | CP | 89.17±1.32 | 94.56±0.94 | 84.87±1.89 |
| | | MP | 90.40±0.31 | 94.76±0.79 | 85.26±1.47 |

*Encapsulation and pasteurization effect on yield stress, plastic viscosity and flow index of PLGA-lycopene nanoemulsion.* The encapsulation of lycopene and further fortification with model juices did not have any effect on the plastic viscosity when the 5° Brix concentrated (Figure 4.32) juices were fortified with PLGA-LNP. But significant improvement had been observed when it was fortified with 15°brix. The improvement of plastic viscosity also shows in different pasteurization treatment for 15° Brix juices when fortified with non-encapsulated lycopene.

---

[10] CTR= Control Non-pasteurized; CP= Conventional Pasteurized MP= Microwave Pasteurized.

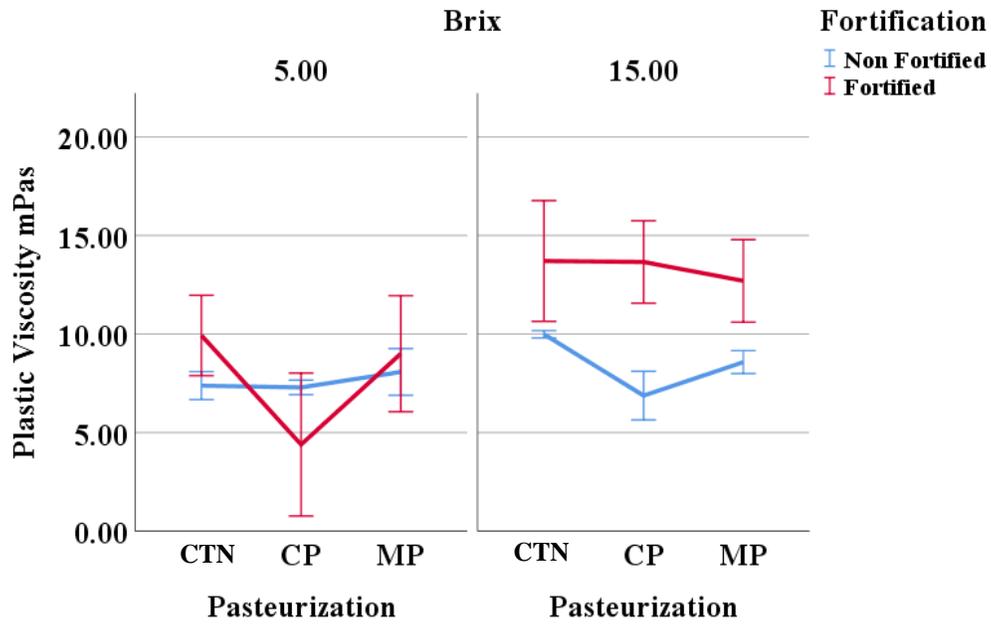

Figure 4.32. Pasteurization effect on plastic viscosity of PLGA-lycopene nanoemulsion[11].

A significant improvement on the yield stress was observed on the samples except juices of 5°brix concentration (Figure 4.33) that were conventionally pasteurized. However, fortification had significant ($P<0.05$) effect on yield stress of juice model of 15ºbrix concentration irrespective of pasteurization treatments.

---

[11] CTR= Control Non-pasteurized; CP= Conventional Pasteurized MP= Microwave Pasteurized.

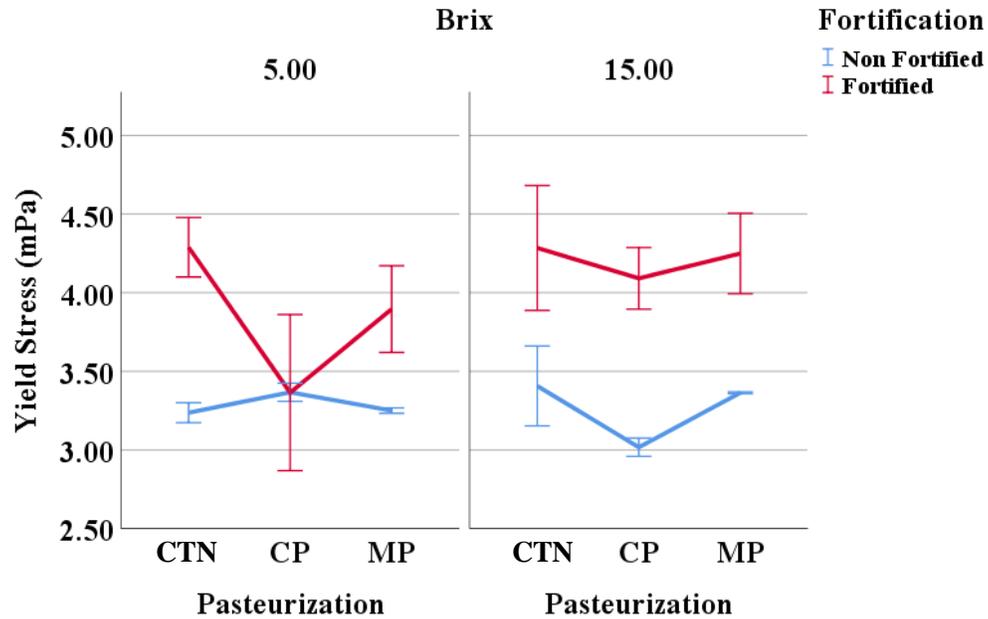

Figure 4.33. Pasteurization effect on yield stress of different PLGA-lycopene nanoemulsion[12].

Consistency index shows that no pasteurization effect appeared when PLGA-lycopene nanoparticles fortified with both 5 and 15°Brix concentrated (Figure 4.34) juice model. Significant reduction of consistency index observed when PLGA lycopene nanoparticles fortified with 15°brix juices.

---

[12] CTN= Control Non-pasteurized; CP= Conventional Pasteurized MP= Microwave Pasteurized.

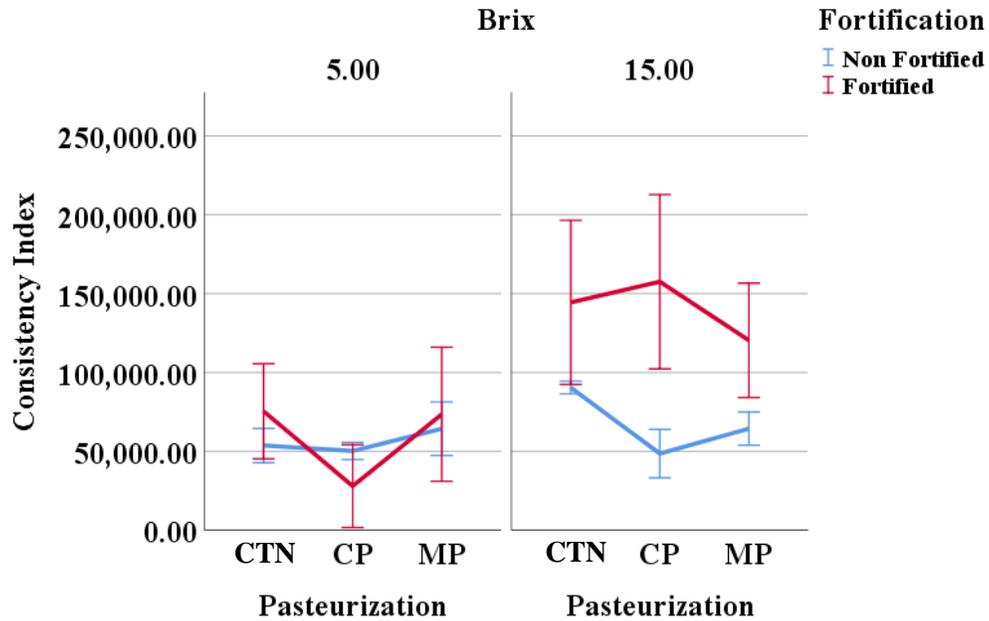

Figure 4.34. Pasteurization effect on consistency index of different PLGA-lycopene nanoemulsion.[13]

For both 5° brix juice model, significant (P<0.05) change was observed on flow index when conventional pasteurization was applied on it after fortification with encapsulated lycopene nanoparticles. Similarly, for 15° brix juice model, significant (P<0.05) change was observed before fortification with encapsulated lycopene nanoparticles.

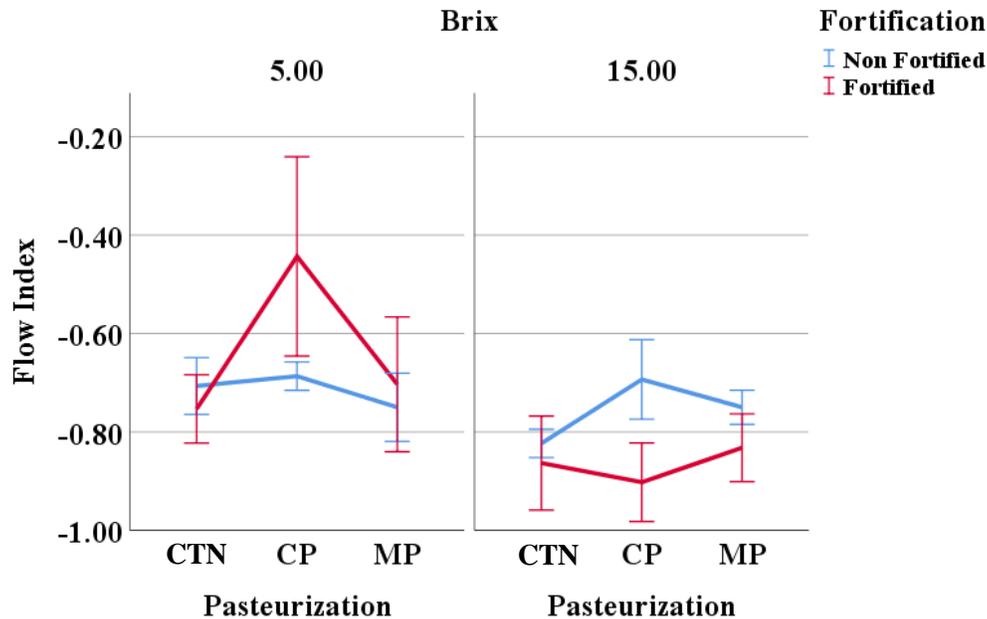

Figure 4.35. Pasteurization effect on flow index of different PLGA-lycopene nanoemulsion[14].

The Color values "a" and "b" have been positively or negatively correlated ($P <$ 0.05) to the DPPH Inhibition activity with the Pearson correlation coefficient of 0.406 and -0.57, respectively. Hence, increasing the red color intensity of the juices shows increased lycopene content, thus show increase DPPH Inhibition activity. The plastic viscosity, yield stress, consistency index and behavioral flow index is the determinant of rheological properties of the juices fortified with LNP. Based on the results shown on Table 4.15, a highly significant relationship ($P < 0.01$) was appeared between the rheological properties and color of the treated juices samples. The plastic viscosity had a weak corelation ($P<0.05$) with the DPPH Inhibition activity. The juices containing the encapsulated and non-encapsulated lycopene nanoparticles contributed to the red color of the juices which is indicative of the "a" color value. As rheological properties correlated with color "a" value and color "a" was also correlated with functional properties of encapsulated lycopene so, suitability of pasteurization can also be verified, considering minimum nutrient loss, by evaluating rheological properties of juices.

---

[14] CTR= Control Non-pasteurized; CP= Conventional Pasteurized MP= Microwave Pasteurized.

Table 4.15. Correlation study for rheological and physicochemical properties of PLGA-lycopene nanoemulsion.

| Variable-1 | Variable-2 | Pearson Correlation Coefficient | Significance Level |
|---|---|---|---|
| HD | Zeta Potential | 0.935 | 0.000 |
| % DPPH Inhibition activity | Plastic Viscosity | 0.233 | 0.049 |
| % DPPH Inhibition activity | Color "a" | 0.406 | 0.000 |
| % DPPH Inhibition activity | Color "b" | -0.570 | 0.000 |
| Plastic Viscosity | Color "L" | 0.413 | 0.000 |
| Plastic Viscosity | Color "a" | 0.317 | 0.007 |
| Yield Stress | Color "L" | -0.380 | 0.001 |
| Yield Stress | Color "a" | 0.352 | 0.002 |
| Consistency Index | Color "L" | -0.542 | 0.000 |
| Consistency Index | Color "a" | 0.292 | 0.012 |
| Flow Index | Color "L" | 0.503 | 0.000 |

## 4.6  Development of Simulation Model for Human GIT

Studying digestive processes has made the bridge between food science and biomedical engineering (Bornhorst et al., 2016). However, much advancement has been accomplished to determine the nutrient reaction kinetics in the human GIT, but the flow dynamics and mixing mechanisms of food in every major digestion compartment is still not understood. Moreover, food digestion involves many interlinked biological and engineering phenomena through enzymatic reactions, fluid flow dynamic a, particle diffusion, and mass transfer. Industrial unit operations can help build a simulation system to understand the complex mechanisms of action of foods during digestion. The unit operations can be verified by a mathematical model to establish the complex mechanism of action of foods in human GIT, hence, solving the simulation challenges of human digestive systems (Bornhorst et al., 2016). The digestion of food can be divided into four processing components namely the mouth, the stomach, the small intestine, and the large intestine; hence, the whole digestive system can be segmented in such a way that mastication is  simulated by crushing or grinding of solid and liquid to form dough , bolus transport through the esophagus, and chyme transport from the stomach to the intestine by automated time controlled peristaltic pump, enzymatic hydrolysis with a bioreactor to simulate  fermentation and absorption going on in different digestion chamber (mouth, stomach, and intestine) (Bornhorst et al., 2016). The key variable of the unit operation are key variables and characteristic parameters to be used to model the simulate human digestion system.

The optimization of the in-vitro system is only possible if the factors controlling the bioavailability of bioactives are well understood and process simulate through mathematical modeling. The bioavailability is a balance between clearance or metabolism and the rates of absorption, and transportation can only be simulated through automatic computerized in-vitro systems. Thus, the food digestibility can be assessed in a relatively simple way, and more complex models can be chosen if the question is more about the rate of appearance of specific nutrients or bio-actives in the bloodstream (Egger et al., 2019).

The simulation model presented by most researchers mimic the mouth, stomach, small & large intestine (Lamond et al., 2019). Hence, none of the researchers have tried to simulate intestinal absorption of nanoparticles in their respective systems, though Barrett (2014) reported, 80% absorption of nanoparticles through the duodenum by diffusion. The study was conducted on a four-chamber simulation model. The three chambers were developed to simulate the digestion in the mouth, stomach, and intestine, respectively. The fourth chamber simulates the bio-accessibility of nanoparticles. Hence, the automatic programmable in-vitro system is named as AK AutoGIT (Anwar-Kassama Auto Gastrointestinal Tract).

<div align="center">Control of Chyme and Enzymes Transfer</div>

The peristaltic doser or transferring pump provides a platform for single or multiple scheduling options to dispense specific dosing volume. The dosing pump has an option to provide weekly programing with specified time scheduling to ensure that the dosing and transferring of chyme and enzymes are fully controlled automatically through a preset programming. The programming could be used to automatically control the rate of dosing, minimum or maximum dosing volume, as well as individual dosing or reverse dosing.

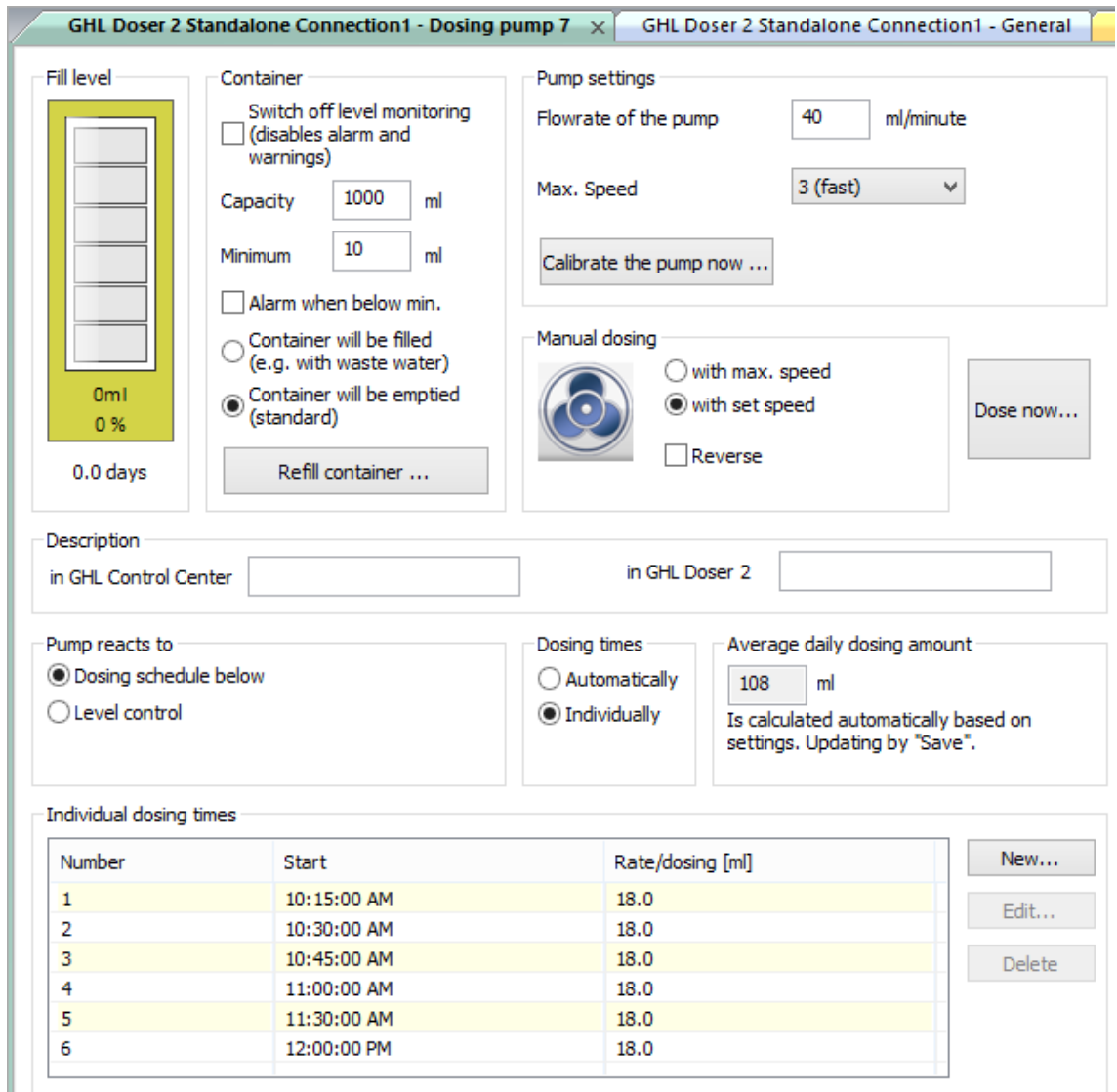

Figure 4.36. GHL Doser 2.1 software for controlling every dosing or transfer to the in-vitro system.

The pH Control of the Digestive System

The pH control panel can monitor and control pH in an automatic fashion through a preset programming. The pH can be set at a distinct point by manipulating the required field. The "ON Time" (Figure 4.37) button in the app display identifies the dosing time, where the "Off Time" provides the time to adjust the pH after a single dosing. On the other hand, the "High alarm" point prevents the dosing of the base solution while the "low alarm" stops the acid dosing. The two different modes are available to be selected named as "control" and "monitor" by two sequential execution points. Preset on the in "control" mode, automatic dosing continues based on the "high/low alarm" setup point; whereas the "monitoring" mode performs only the monitoring of the pH changes over time without any dosing event as shown in Figure 4.37. The "required pH" point is used to identify the desired point of control at which system is runned.

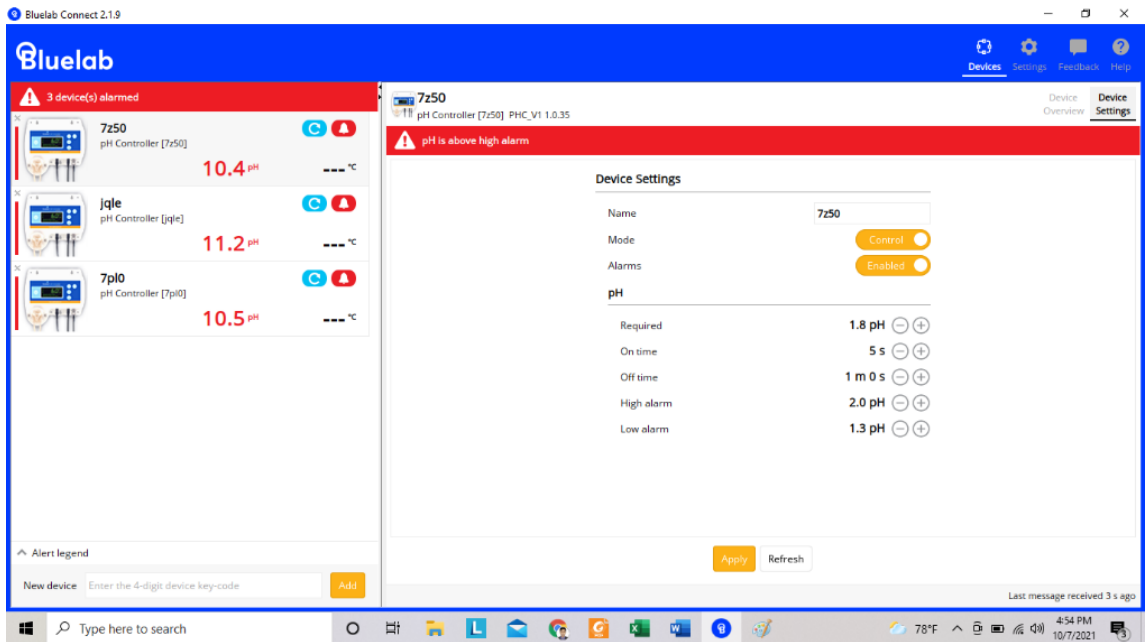

Figure 4.37. Controlling of pH at three different chambers of in-vitro digestion by pH controller app.

Control of Automatic Stirring for Each Digestion Chamber

The automatic stirrers were used to simulate food break down inside mouth, stomach, and intestinal chamber as shown in Figure 4.38. The stirrers were scheduled through programmable WiFi connection platform. In each stirrer, the rpm, stirring time and duration can be programmed before running the digestion system.

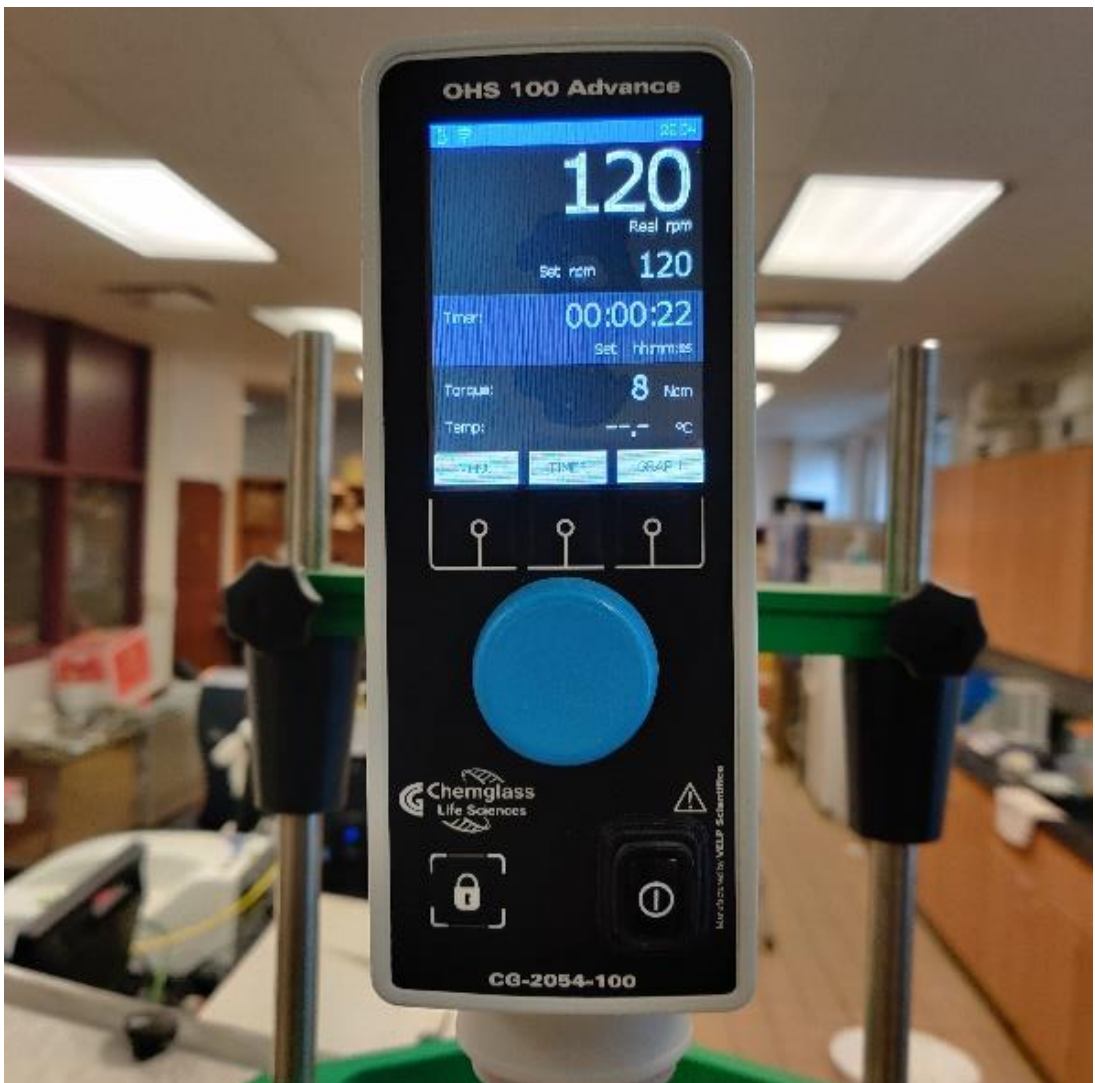

Figure 4.38. Automatic programmable stirrer for Anwar-Kassama Automatic Gastrointestinal Tract (A-K AutoGIT).

Function of Complete Automatic In-Vitro Digestive System

*Simulation of mouth digestion.* A double-jacketed bioreactor was used to simulate the mouth A transfer pump (pump 1) with four sub-pumps, two of which were used to control all the fluid transfer from food and saliva source to the mouth. A sub-pump1 was designed to transfer fruit juices from the food source to the mouth. The sub-pump2 was assigned to transfer saliva from saliva source to mouth. The automatic stirrer-1 was used to stir liquid inside mouth to simulate digestion with the saliva.

*Simulation of stomach digestion.* A double-jacketed bioreactor was also used to as a stomach simulator just like the mouth discusssed above. The peristaltic pump 2, contained 4 sub-pumps, was used to transfer all the fluids from acid, base and gastric juice container to the stomach, hence mimicking the gastric emptying by transferring fluids from stomach to intestine for the final digestion. That peristaltic pump is also used for collecting samples from the different points of stomach digestion system. The two pH meters were set to monitor and control the pH with regards to the preset ranges and the stirrer 2 simulates the digestion inside stomach.

The gastric emptying follows a mathematically modelled by the following equation 4.4 (Elashoff et al., 1982):

$$y(t) = 2^{-\left(\frac{t}{T/2}\right)^{\beta}} \qquad (4.4)$$

where $y(t)$ refers to the fraction of the chyme remaining in the stomach at time ($t$); $\beta$ the coefficient describing the shape of the curve; and T/2 is the gastric emptying half time calculated using Equation 4.5 (Hunt & Stubbs, 1975).

$$T/2 = V_o \times (0.1797 - 0.1670^{-K}) \tag{4.5}$$

The gatric empying half-time (T/2) depends on $V_0$ which is the volume of food going through digestion, and $K$ is the caloric density of the foods.

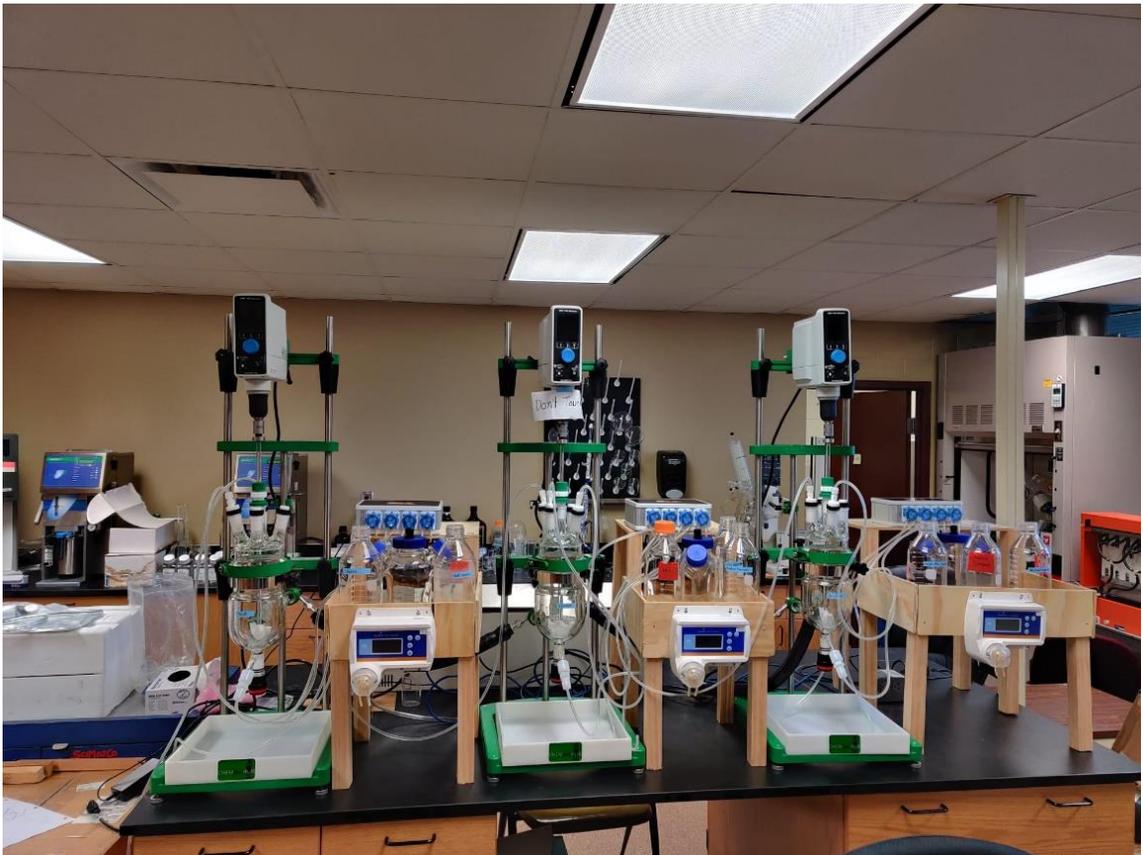

Figure 4.39. The AK-AutoGIT automatic programmable silumator of the human digestive system.

*Simulation of intestinal digestion.* The third double-jacketed bioreactor was used as the intestinal simulator like the other two chambers. It also includes pump 3 that controls the fluid transfer from the intestinal juices, and pH base solution to the stomach and the intestine to sample collection source. The pH meter 3 was used to monitor and control the pH throughout the whole period of the intestinal digestion.

## Technical Glitches Associated to the Automatic In-Vitro System

Although, the model system used for this study was designed for fully automatic digestion simulation, there have been few associated glitches involved. First, the control through "Internet cloud" can be problematic in case of interrupted internet connectivity which sometimes may interrupt transferring the programmed commands to run the digestion automatically. On the other hand, connecting to the Center App" may be problematic when the windows operating system of the computer (PC) goes into the sleep mode.

## 4.7.    In-Vitro Bio-Accessibility and Release Kinetics of Lycopene Nanoparticles

### Human Digestion

Digestion begins in the mouth, where it breaks the complex compounds into monomers. Chewing food has distinct functions such as breakdown of particles, lubrication, mixing, to form a bolus and subsequent transport to the stomach. However, for liquid foods, the mixing of the food particles with saliva is accomplished by the combined action of the tongue and palate (Hutchings & Lillford, 1988; Sharma, 1973). Similarly, paste mixers and kneading machines can be used to mix Newtonian, shear thickening, or shear-thinning food items (McCabe et al., 2005) to simulate the mixing of liquid food items into the human GIT. Meanwhile, biopolymers convert to simpler molecules into the human GIT having the food particles mixed with salivary enzymes can be put into simulation by bioreactors with automatic stirrers (De Almeida et al., 2008).

In the stomach, the boluses are mixed with the gastric secretions (electrolytes, enzymes, mucus, and HCl (Guyton & Hall, 2006) to breakdown boluses by the peristaltic contractions of the gastric wall which is comparable to peristaltic pumping accompanying mixing. Nevertheless, the bioreactor can provide a similar anaerobic condition for acid and enzymatic hydrolysis of the stomach digestion as mixing in the GIT. Some important characteristic and the most challenging parameter to model is the mixing rate in the stomach, the pH distributions, gastric emptying rates, and the bolus breakdown. Gastric mixing should be verified with the gastric index developed by Bornhorst et al. (2014) as mechanical properties to evaluate gastric emptying rate which ultimately affects the degree of mixing. The residence time is one of the characteristic parameters used to describe reactor operation. Gastric emptying defines the time it takes to empty half of the stomach contents (Marciani et al., 2001; Marciani et al., 2013; Whitehead et al., 1998). When bolous has completed its digestion in presence of stomach enzymes and chemicals it called chyme. The flow rate of chyme from stomach to intestine is called gastric emptying rate.

The small intestine receives chyme from the stomach through the pylorus with a pore size of 1 µm. Functions of the small intestine is harmonized with unit operations of a bioreactor, membrane separator, and peristaltic pumps. The chyme is mixed with the intestinal secretions by peristaltic contraction movement of the small intestinal wall. The chyme mixes with the pancreatic secretions in the duodenum where pH is maintained at 7.0 to activate the enzymes such as lipase, phospholipase, α-amylase, trypsin, and others (Barrett, 2014; Bornhorst & Singh, 2013). The mixing and flow of the chyme are related to industrial processes at mixing tanks and peristaltic pumping through the pipe fittings. Also, the enzymatic hydrolysis of chyme is simulated as a unit process with a bioreactor (Riedlberger & Weuster-Botz, 2012).

Eighty percent (80%) of the nutrients are absorbed through the small intestinal lining. Water diffuses through the mucus wall to the lumen, and further transported through the epithelium's cell membrane to the bloodstream (Barrett, 2014). This separation process is analogous to membrane separation processes, such as reverse osmosis system. The Mass transfer through the membrane controls by the fluid flow and the chemical and physical interaction of the nutrients with the membranous intestinal wall with an average pore size of 200 nm.

*Physicochemical properties of lycopene nanoparticles.* The hydrodynamic diameter of the PLA-LNP appears to be significantly (P<0.05) reduced when compared the pre and post in-vitro digestion irrespective of the pasteurization treatments (CNP, CP and MP).

Table 4.16. ANOVA test results to evaluate the effect of pasteurization and digestion time on hydrodynamic diameter of PLA-LNP.

| Source | Sum of Squares | df | Mean Square | F | Sig. |
|---|---|---|---|---|---|
| Corrected Model | 23533.98 | 8 | 2941.75 | 21.17 | 0.000 |
| Intercept | 2718296.41 | 1 | 2718296.41 | 19564.44 | 0.000 |
| Pasteurization | 5140.73 | 2 | 2570.37 | 18.50 | 0.000 |
| Digestion Time | 18330.71 | 2 | 9165.35 | 65.97 | 0.000 |
| Pasteurization X Digestion Time | 1542.45 | 4 | 385.61 | 2.77 | 0.029 |
| Total | 4367405.35 | 162 | | | |

It is visualized that the non-pasteurized PLA-LNP had higher erosion due to digestion as HD reduced from 164.37±12.39 (BD_CTN_0) to 126.54±14.37 (BD_CTN_240) compared to CP and MP treated PLA-LNP. Hydrodynamic Diameter reduced from 170.11±4.69 (BD_CP_0) to 145.37±6.77 (BD_CP_240) and 169.67±14.35 (BD_MP_0) to 146.78±9.8 (BD_MP_240) for CP and MP treated PLA-LNP, respectively. It means that both CP (erosion was 14.5%) and MP (erosion was 13.5%) protecting the degradation of PLA-LNP compared to CTN treated PLA-LNP where erosion appeared to be 23%. It can be concluded that pasteurization provided some additional protection against erosion of LNP, or release and degradation of lycopene thus helped to improve bio-accessibility of lycopene encapsulated in it.

Table 4.17. Homogeneity test results to evaluate the in-vitro digestion effect on hydrodynamic diameter of PLA-LNP[15].

| Treatments | N | Subset 1 | 2 | 3 | 4 |
|---|---|---|---|---|---|
| AD_CTN_240 | 9 | 126.55 | | | |
| AD_CP_240 | 9 | | 145.37 | | |
| AD_MP_240 | 9 | | 146.78 | | |
| AD_CTN_180 | 9 | | 155.79 | 155.79 | |
| BD_CTN_0 | 36 | | | 164.37 | 164.37 |
| BD_MP_0 | 36 | | | 169.67 | 169.67 |
| BD_CP_0 | 36 | | | 170.11 | 170.11 |
| AD_MP_180 | 9 | | | | 172.98 |
| AD_CP_180 | 9 | | | | 176.23 |
| Sig. | | 1.000 | 0.433 | 0.080 | 0.257 |

The hydrodynamic diameter (HD) (Figures 4.57 and 4.58) of the non-pasteurized PLGA-LNP increase ($P<0.05$) from $114.80 \pm 7.76$ to $244.77 \pm 0.40$ and $169.60 \pm 3.20$ nm at 180 and 240 minutes of in-vitro digestion, respectively. The reason behind this increment could be attributed to the adsorption of water on the surface of the PLGA-LNP during hydrolysis of nanoparticles containing chyme in the presence of lipase, amylase and protease. Similarly, the HD of the CP treated PLGA-LNP increased ($P<0.05$) from $126.41 \pm 3.31$ to $157.73 \pm 0.13$ and $160.87 \pm 1.39$; whereas, the MP treated PLGA-NPs, t from $116.42 \pm 8.45$ to $156.50 \pm 2.87$ and $152.30 \pm 0.98$ after 180 and 240 minutes of in-vitro digestion, respectively. The HD of the MP treated PLGA-LNP was found to be the lowest after in-vitro digestion compared to the CP or CNP treated PLGA-LNP.

---

[15] Value at different subset identify significant differences. AD = After digestion, BD = Before digestion, NP = non-pasteurized, CP = Conventional pasteurization, MP = Microwave pasteurization, 0 = At the start of digestion, 180 = 180 minutes, 240 = 24 minutes.

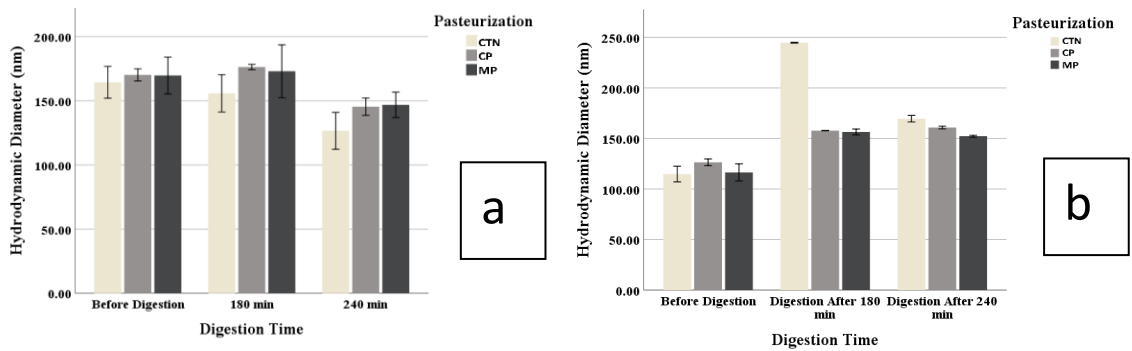

Figure 4.40. Pasteurization and digestion effect on hydrodynamic diameter profile of PLA (a) and PLGA (b) lycopene nanoemulsion.

Table 4.18. Homogeneity test results to evaluate the effect of in-vitro digestion on hydrodynamic diameter of PLGA-LNP[16].

| Treatments | N | Subset | | | | | |
|---|---|---|---|---|---|---|---|
| | | 1 | 2 | 3 | 4 | 5 | 6 |
| BD_CTN_0 | 36 | 114.80 | | | | | |
| BD_MP_0 | 36 | 116.42 | | | | | |
| BD_CP_0 | 36 | | 126.41 | | | | |
| AD_MP_240 | 9 | | | 152.30 | | | |
| AD_MP_180 | 9 | | | 156.50 | 156.50 | | |
| AD_CP_180 | 9 | | | 157.73 | 157.73 | | |
| AD_CP_240 | 9 | | | | 160.87 | | |
| AD_CTN_240 | 9 | | | | | 169.60 | |
| AD_CTN_180 | 9 | | | | | | 244.77 |
| Sig. | | 0.494 | 1.000 | 0.060 | 0.160 | 1.000 | 1.000 |

The ANOVA test shows that the interaction of pasteurization and digestion time had negative effect ($p < 0.05$) on the zeta potential values or stability of the PLA-LNP. The zeta potential of the PLA-LNP reduced from -18.71±2.40 (BD_CTN_0) to -2.03±1.13 (BD_CTN_180) and -3.87±0.37 (BD_CTN_0) at 180 and 240 minutes of digestion, respectively. With regards to the CP treatment, the reduction was from -23.42±2.10 (BD_CP_0) to -3.77±0.43 (BD_CP_180) and -3.48±0.20 (BD_CP_240) for 180 and 240 minutes of digestion, respectively. Similarly, PLA-LNP are sensitive against MP treatment as the stability of the nanoparticles reduced from -21.86 ± 2.71 (BD_MP_0) to -5.42 ± 2.09 (BD_MP_180) and -7.26±1.8=72 (BD_MP_240) at 180 and 240 minutes of digestion, respectively. Microwave treatment made some conformational changes of PLA-LNP which prevent the degradation against enzymatic action of lipase and amylase. It is hence shown that the MP treated PLA-LNP were observed to be more stable than that of CTN or CP treated PLA-LNP.

---

[16] Value at different subset identify significant differences. AD = After digestion, BD = Before digestion, CTN = Control Non-pasteurized, CP = Conventional pasteurization, MP = Microwave pasteurization, 0 = At the start of digestion, 180 = 180 minutes, 240 = 240 minutes.

Table 4.19. ANOVA test results to evaluate the effect of pasteurization and digestion on for zeta potential value of PLA-LNP.

| Source | Type III Sum of Squares | df | Mean Square | F | Sig. |
|---|---|---|---|---|---|
| Corrected Model | 10996.20 | 8 | 1374.52 | 306.14 | 0.000 |
| Intercept | 10756.88 | 1 | 10756.88 | 2395.81 | 0.000 |
| Pasteurization | 200.10 | 2 | 100.05 | 22.28 | 0.000 |
| Digestion Time | 10459.76 | 2 | 5229.88 | 1164.82 | 0.000 |
| Pasteurization X Digestion Time | 145.92 | 4 | 36.48 | 8.12 | 0.000 |
| Error | 686.95 | 153 | 4.49 | | |
| Total | 51388.11 | 162 | | | |
| Corrected Total | 11683.150 | 161 | | | |

Table 4.20. Homogeneity test results to evaluate the effect of pasteurization and in-vitro digestion on zeta potential value of PLA-LNP[17].

| Treatments | N | Subset 1 | 2 | 3 | 4 | 5 |
|---|---|---|---|---|---|---|
| BD_CP_0 | 36 | -23.42 | | | | |
| BD_MP_0 | 36 | -21.86 | | | | |
| BD_CTN_0 | 36 | | -18.72 | | | |
| AD_MP_240 | 9 | | | -7.26 | | |
| AD_MP_180 | 9 | | | -5.42 | -5.42 | |
| AD_CTN_240 | 9 | | | | -3.87 | -3.87 |
| AD_CP_240 | 9 | | | | -3.77 | -3.77 |
| AD_CP_180 | 9 | | | | -3.48 | -3.48 |
| AD_CTN_180 | 9 | | | | | -2.03 |
| Sig. | | 0.677 | 1.000 | 0.454 | 0.382 | 0.464 |

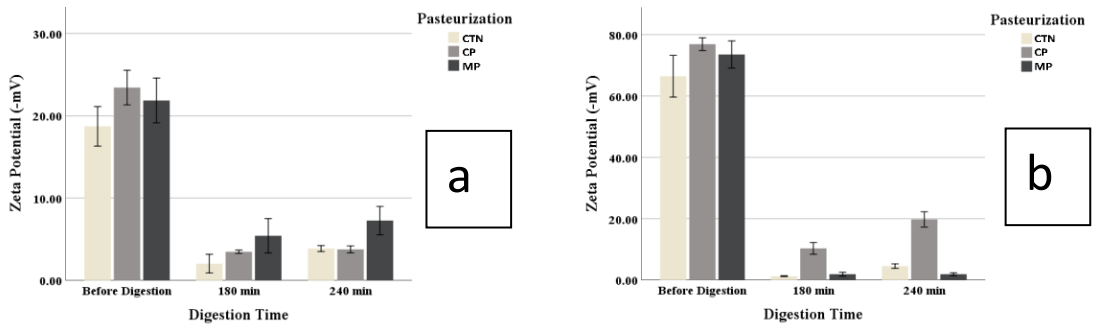

Figure 4.41. Pasteurization and digestion effect on zeta potential profile of PLA (a) and PLGA-LNP (b).

Table 4.21. Homogeneity test results to evaluate th in-vitro digestion and pasteurization effect on zeta potential of PLGA-LNP[18].

| Treatments | N | Subset 1 | 2 | 3 | 4 | 5 | 6 |
|---|---|---|---|---|---|---|---|
| AD_CTN_180 | 9 | 1.26 | | | | | |
| AD_MP_180 | 9 | 1.90 | | | | | |
| AD_MP_240 | 9 | 1.97 | | | | | |
| AD_CTN_240 | 9 | 4.54 | | | | | |
| AD_CP_180 | 9 | | 10.33 | | | | |
| AD_CP_240 | 9 | | | 19.77 | | | |
| BD_CTN_0 | 36 | | | | 66.42 | | |
| BD_MP_0 | 36 | | | | | 73.50 | |
| BD_CP_0 | 36 | | | | | | 76.87 |
| Sig. | | 0.204 | 1.000 | 1.000 | 1.000 | 1.000 | 1.000 |

Zeta potential value (Figure 4.59) shows the stability of lycopene NPs in which the higher the value the greater the stability. At the beginning of the digestion the stability of PLGA-NPs was observed to be very high (<-70 mV); whereas post in-vitro digestion was -4.54±0.71, -19.77±2.48 and -1.91±0.46 for NP, CP and MP treated PLGA-NPs, respectively. This means PLGA-NPs were sensitive to the enzymatic action of amylase in intestine.

Interaction effect of pasteurization and digestion time found to be significant (P<0.05) on polydispersity index of the PLA-LNP; whereas the homogeneity test shows that the conventional and microwave pasteurization contributed to the uniformity of the nanoparticles. PDI found to be lower before digestion and increases with the progress of the digestion time. The CP treatment effect on the PDI where it increased from 0.253 ± 0.02 to 0.651 ± 0.01 after the enzymatic action of lipase, amylase and preotease. However, due to MP treatment the increment of PDI value occurred from 0.239 ± 0.06 to 0.645 ± 0.05. The invitro loss of homogeneity was found to be highest with the CTN PLA-LNP (0.258 ± 0.13 to 0.775 ± 0.18).

---

[18] AD = After digestion, BD = Before digestion, CTN = non-pasteurized, CP = Conventional pasteurization, MP = Microwave pasteurization, 0 = At the start of digestion, 180 = 180 minutes, 240 = 24 minutes.

Table 4.22. ANOVA test results to evaluate the effect of in-vitro digestion and pasteurization on polydispersity index of PLA-LNP.

| Source | Type III Sum of Squares | df | Mean Square | F | Sig. |
|---|---|---|---|---|---|
| Corrected Model | 6.167 | 8 | 0.77 | 83.19 | 0.000 |
| Intercept | 28.862 | 1 | 28.86 | 3114.40 | 0.000 |
| Pasteurization | 0.223 | 2 | 0.11 | 12.03 | 0.000 |
| Digestion Time | 5.859 | 2 | 2.93 | 316.11 | 0.000 |
| Pasteurization X Digestion Time | 0.211 | 4 | 0.05 | 5.70 | 0.000 |
| Error | 1.418 | 153 | 0.01 | | |
| Total | 31.414 | 162 | | | |
| Corrected Total | 7.585 | 161 | | | |

Table 4.23. Homogeneity test results to evaluate the effect of digestion and pasteurization on PDI of PLA-LNP[19].

| Treatments | N | Subset 1 | 2 | 3 | 4 |
|---|---|---|---|---|---|
| BD_MP_0 | 36 | 0.239 | | | |
| BD_CP_0 | 36 | 0.253 | | | |
| BD_CTN_0 | 36 | 0.258 | | | |
| AD_CP_180 | 9 | | 0.488 | | |
| AD_MP_240 | 9 | | | 0.648 | |
| AD_MP_180 | 9 | | | 0.649 | |
| AD_CP_240 | 9 | | | 0.651 | |
| AD_CTN_180 | 9 | | | 0.691 | 0.691 |
| AD_CTN_240 | 9 | | | | 0.775 |
| Sig. | | 1.000 | 1.000 | 0.976 | 0.456 |

The CNP and CP treated PLGA-LNP, PDI increased ($p < 0.05$) from $0.22 \pm 0.03$ and $0.25 \pm 0.01$ to $0.36 \pm 0.05$ and $0.26 \pm 0.002$, respectively after the completion of the 240 minutes in-vitro digestion where for MP treated PLGA-LNP, PDI became unchanged ($p < 0.05$) until the end of digestion.

Table 4.24. Homogeneity test results to evaluate the effect of in-vitro digestion pasteurization on polydispersity index of PLGA-LNP[20].

| Treatments | N | Subset 1 | 2 | 3 | 4 | 5 |
|---|---|---|---|---|---|---|
| AD_CTN_180 | 9 | 0.20 | | | | |
| BD_CTN_0 | 36 | 0.22 | 0.22 | | | |
| BD_MP_0 | 36 | | 0.23 | 0.23 | | |
| AD_MP_180 | 9 | | 0.23 | 0.23 | | |
| AD_MP_240 | 9 | | 0.23 | 0.23 | | |
| AD_CP_180 | 9 | | | 0.25 | 0.25 | |
| BD_CP_0 | 36 | | | 0.25 | 0.25 | |
| AD_CP_240 | 9 | | | | 0.27 | |
| AD_CTN_240 | 9 | | | | | 0.37 |
| Sig. | | 0.10 | 0.53 | 0.29 | 0.24 | 1.00 |

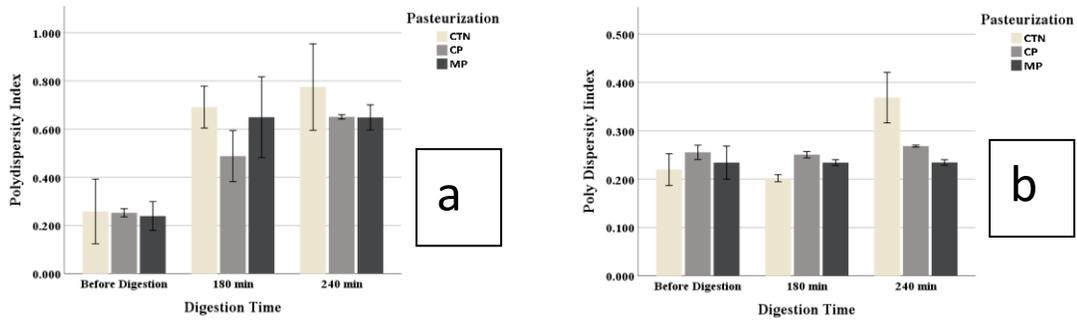

Figure 4.42. Pasteurization and digestion effect on polydispersity index profile of PLA (a) and PLGA-LNP (b).

Table 4.25. ANOVA test to evaluate the effect of pasteurization and in-vitro digestion effect on bio-accessibility of PLA-LNP.

| Source | Type III Sum of Squares | df | Mean Square | F | Sig. |
|---|---|---|---|---|---|
| Corrected Model | 578.940 | 8 | 72.368 | 7.75 | 0.000 |
| Intercept | 58220.44 | 1 | 58220.44 | 6231.83 | 0.000 |
| Pasteurization | 395.00 | 2 | 197.50 | 21.14 | 0.000 |
| Digestion Time | 62.87 | 2 | 31.44 | 3.36 | 0.037 |
| Pasteurization X Digestion Time | 249.64 | 4 | 62.41 | 6.68 | 0.000 |
| Total | 90372.21 | 162 | | | |

Table 4.26. Homogeneity test to evaluate the effect of pasteurization and in-vitro digestion on bio-accessibility of PLA-LNP[21].

| Treatments | N | Subset 1 | 2 | 3 | 4 | 5 |
|---|---|---|---|---|---|---|
| AD_CP_240 | 9 | 18.65 | | | | |
| AD_MP_180 | 9 | 20.38 | 20.38 | | | |
| AD_MP_240 | 9 | 21.39 | 21.39 | 21.39 | | |
| BD_MP_0 | 36 | | 22.94 | 22.94 | 22.94 | |
| BD_CP_0 | 36 | | 23.12 | 23.12 | 23.12 | |
| BD_CTN_0 | 36 | | | 24.41 | 24.41 | 24.41 |
| AD_CP_180 | 9 | | | 24.77 | 24.77 | 24.77 |
| AD_CTN_240 | 9 | | | | 26.12 | 26.12 |
| AD_CTN_180 | 9 | | | | | 27.17 |
| Sig. | | 0.415 | 0.414 | 0.154 | 0.218 | 0.405 |

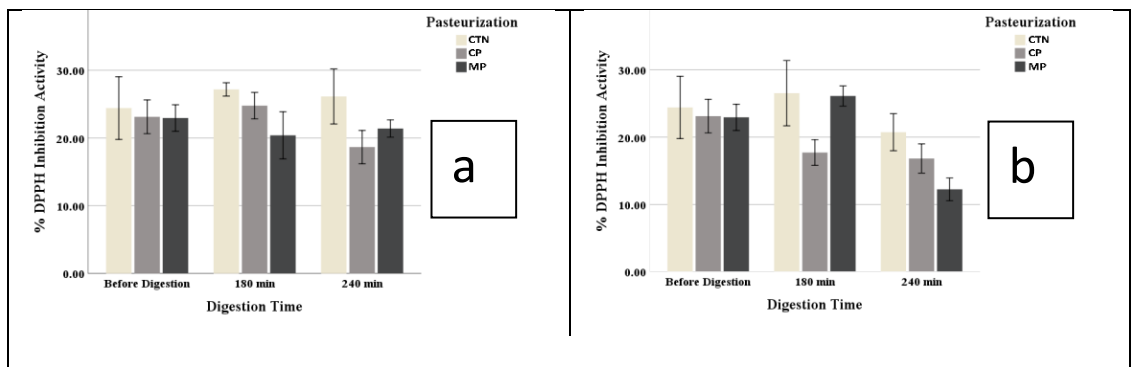

Figure 4.43. Digestion effect on bio-accessibility profile of pasteurized PLA (a) and PLGA-LNP (b).

The ANOVA shows a significant interaction between pasteurization and digestion time on the DPPH inhibition activity of encapsulated nanoparticles. The percent DPPH inhibition activity of the conventionally pasteurized (CP) PLA-LNP reduced from 23.12 ± 2.49 (BD_CP_0) to 18.65±2.46% (AD_CP_240) due to consecutive action of amylase and lipase. No significant difference was found for NP and MP treated PLA-LNP until the end of in-vitro digestion (Figure 4.43 (a)). It was also observed that the CP had degradation effect on functional activity of PLA-LNP but not MP.

[21] Value at different subset identify significant differences. AD = After digestion, BD = Before digestion, CTN = Control Non-pasteurized, CP = Conventional pasteurization, MP = Microwave pasteurization, 0 = At the start of digestion, 180 = 180 minutes, 240 = 24 minutes.

No digestion effect observed on functional properties of lycopene (DPPH inhibition activity) on nonpasteurized PLGA-LNP. On the other hand, functional properties of lycopene reduced from 23.12±2.49 to 16.83±2.16 and from 22.94±1.95 to 12.25±1.69 when CP and MP were applied on PLGA-LNP, respectively. It shows that MP (47%) had greater degradation effect on PLGA-LNP compared to the CP (27%) (Figure 4.43 (b)).

Recently researcher have observed the effect of different methods of pasteurization, such as low temperature (Plaza et al., 2011), high pressure (Vervoot et al., 2011; Garcia et al., 2001; Esteve et al., 2009), pulse electric field (Plaza et al., 2011; Sanchez-Mareno et al., 2005; Vervoot et al., 2011; Esteve et al., 2009), and conventional pasteurization (Sanchez-Mareno et al., 2005) and observed thathigher carotenoid content in orange juice when treated with high pressure pasteurization system, as could be attributed to the release of carotenoids from pectin bonding of fruit cells due to the extreme pressure applied on it (Plaza et al., 2011; Sanchez-Mareno et al., 2003; Sanchez-Mareno et al., 2005). On the contrary, a lower retention of carotenoids was observed by Brasili et al. (2017), Lee et al. (2003), and Zulueta et al. (2010) when thermal treatment applied on food composites. Etzbach et al. (2020) evaluated that conventionally pasteurized juice and the hot filled juice showed a significantly lower total carotenoid content compared to the low pasteurized juice. High temperatures during processing, such as conventional pasteurization and hot filling might cause the instability of the polyene chain of the carotenoids, resulting in the carotenoid degradation through isomerization, oxidation, and cleavage which appeared with low bio-accessibility of carotenoids.

The thermal treatment of fruit juices samples helps release the carotenoids from the fruit matrix, but when incorporated in the lipid droplets into the gastric juice, micellarization in the intestine is prevented due to large lipid droplets formulation, hence the carotenoid binds with the lipids to form the carotenoids-lipid complex (Etzbach et al., 2020). The bile salt secreted from the gallbladder into the small intestine works as the surfactant ultimately plays a vital role for carotenoids to be incorporated into micelles.

*Morphological and chemical characteristics of PLA and PLGA-LNP.* The HD distribution (Figure 4.44, 4.45 & 4.46) of the nanoparticles provides an important direction for bio-accessibility and bioavailability of nanoparticles. No significant change was observed due to the CP and MP treatments where for NP PLA-LNP, size distribution shifted to the right. It means that non-pasteurized PLA-LNP adsorbed water swell the structure during in-vitro digestion. Addition of water on the surface of the nanoparticles is the first step of hydrolysis which made LNP bigger than the original.

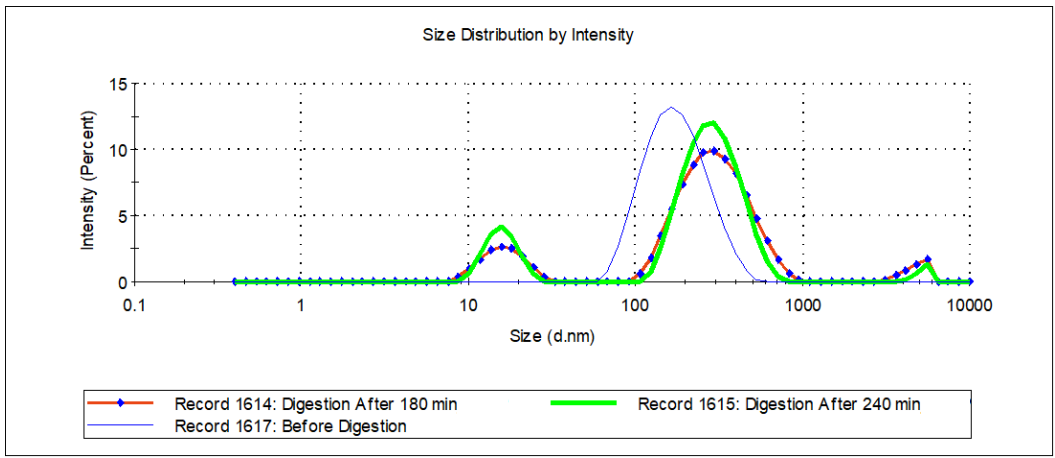

Figure 4.44. In-vitro human digestion effect on HD distribution of non-pasteurized PLA-LNP.

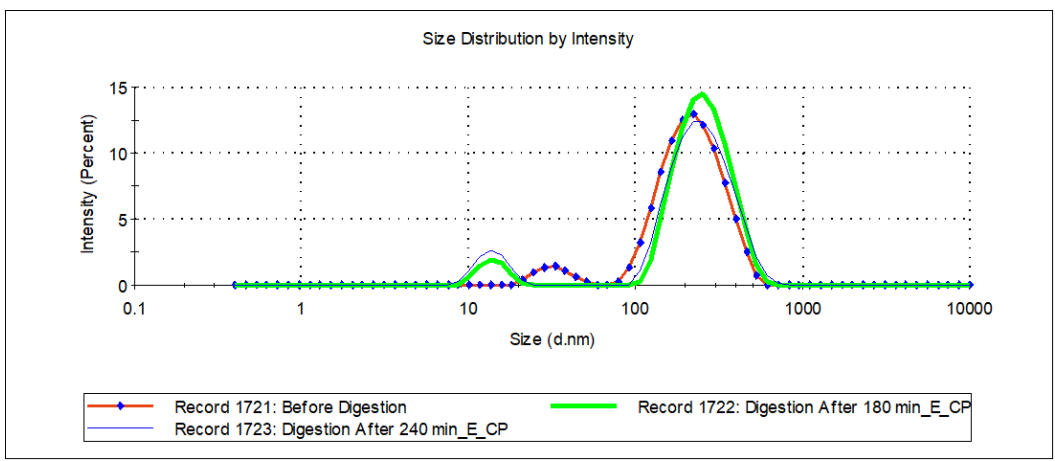

Figure 4.45. In-vitro human digestion effect on HD distribution of conventionally pasteurized PLA-LNP.

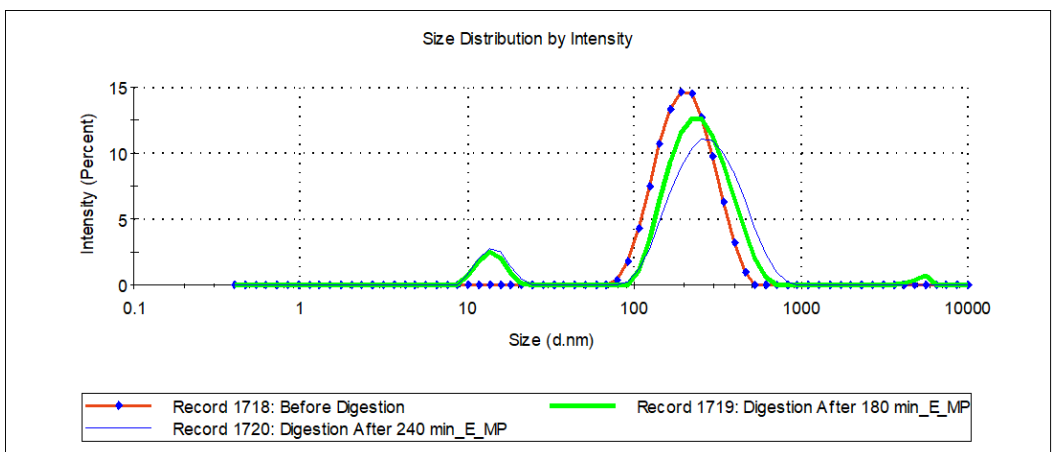

Figure 4.46. In-vitro human digestion effect on HD distribution of microwave pasteurized PLA-LNP.

In Figure 4.47, 4.48 and 4.49. HD distribution shifted to the right sharply for MP treated PLGA-LNP compared with CTN treated one. With regards to CP treated PLGA-LNP, no sharp shifting appeared on HD distribution, but width of the nanoparticles increased both for microwave and conventionally pasteurized PLA-LNP. This increment of distribution width explained that the particles' shape was transferred from spherical to rod shape which will slow down the absorption capacity of the lycopene nanoparticles.

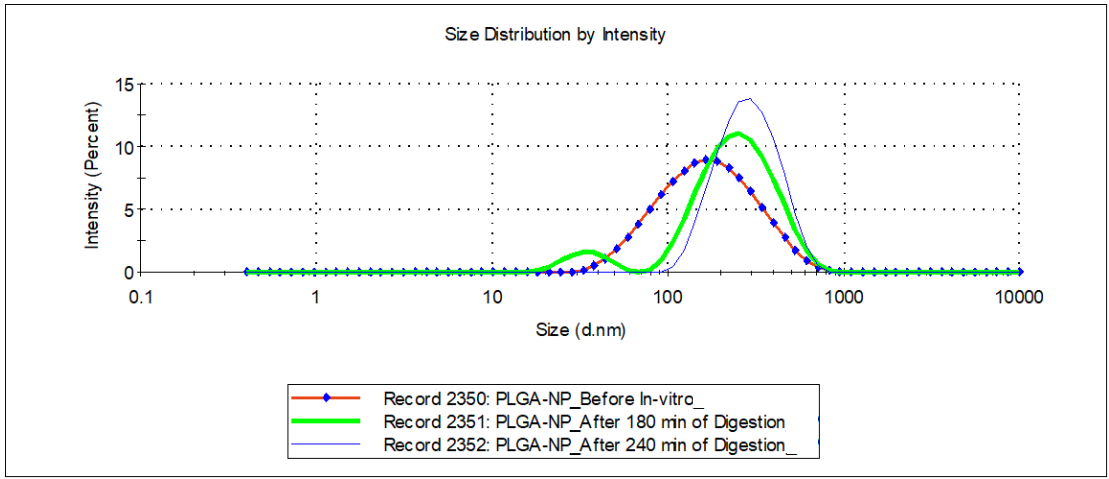

Figure 4.47. Effect of in-vitro digestion on size distribution of non-pasteurized PLGA-LNP.

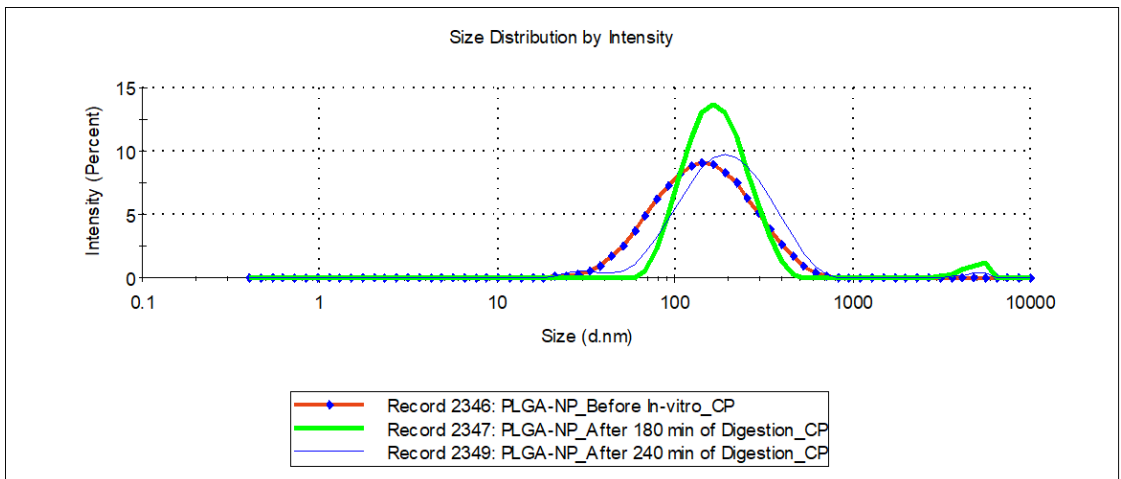

Figure 4.48. Effect of in-vitro digestion on size distribution of conventionally pasteurized PLGA-LNP.

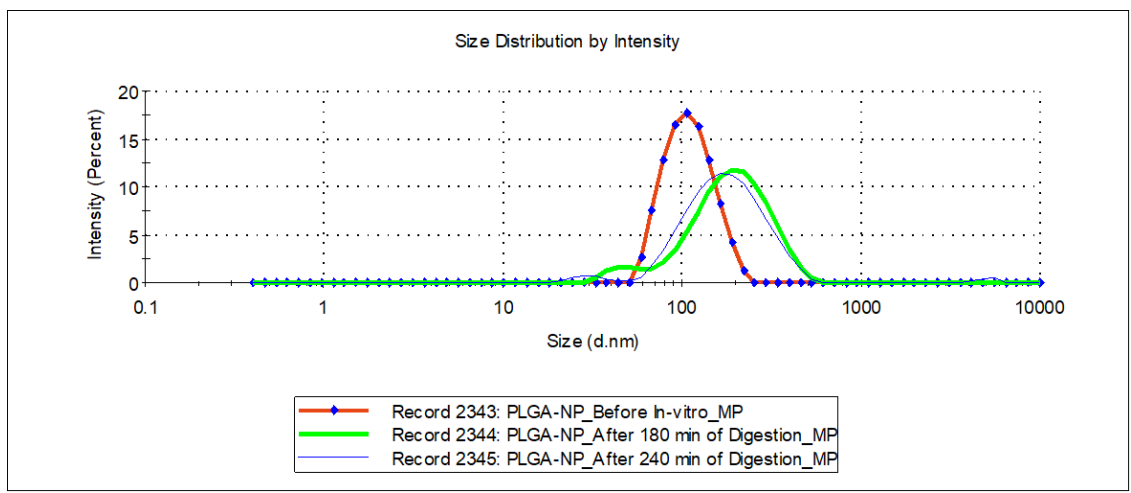

Figure 4.49. Effect of in-vitro digestion on size distribution of microwave pasteurized PLGA-LNP.

*Chemical Properties of PLA and PLGA-LNP.* Figures 4.50 and 4.51 can explain the evidence of lycopene presence after the in-vitro digestion of both PLA and PLGA-lycopene LNP. Figure 4.50 shows that standard lycopene had a % transmittance peak at 2178.82, 2020.40 and 1978.38 cm$^{-1}$ wave number which also observed after the digestion of non-pasteurized PLA-LNP at 2158.26, 2030.98 and 1967.91 cm$^{-1}$ wavenumber. For CP treated PLA-LNP, transmittance peak observed at 2158.66, 2031.56 and 1976.89 cm$^{-1}$ wavenumber where MP treated PLA-LNP the transmittance peak appeared at 2158.35, 2031.60 and 1975.41 cm$^{-1}$ wavenumber. All the consecutive peaks were very close to each other depicting the same characteristic properties of lycopene. For PLGA-LNP (Figure 4.51), peak observed at 2170.46, 2035.12 and 1966.05 cm$^{-1}$ wavenumber for non-pasteurized condition where for CP treatment the transmittance peak appeared at 2176.60, 2055.92 and 1968.33 cm$^{-1}$ wavenumber. Transmittance peak observed at 2170.01, 2034.93 and 1966.34 for MP treated PLGA-LNP after the completion of in-vitro digestion.

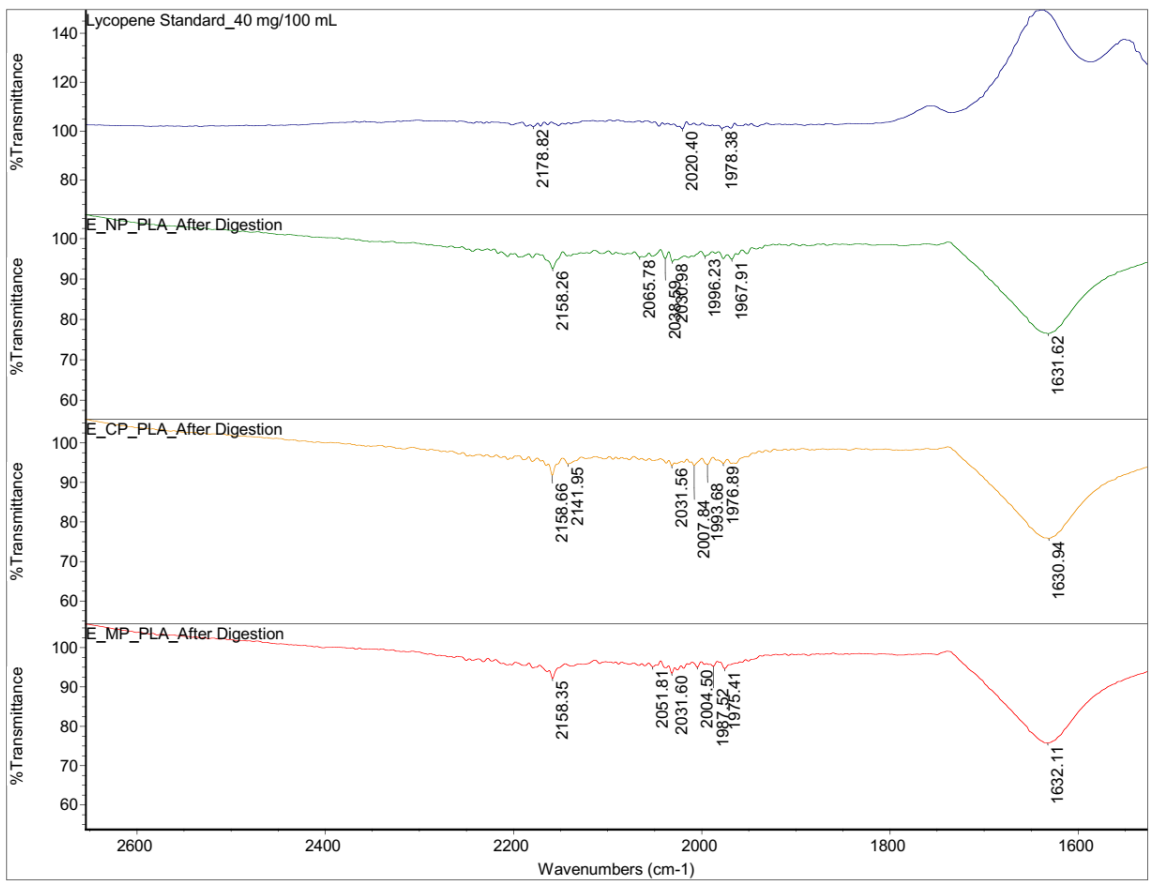

Figure 4.50. Pasteurization effect on chemical structure of PLA-LNP measured by FT-IR spectroscopy.

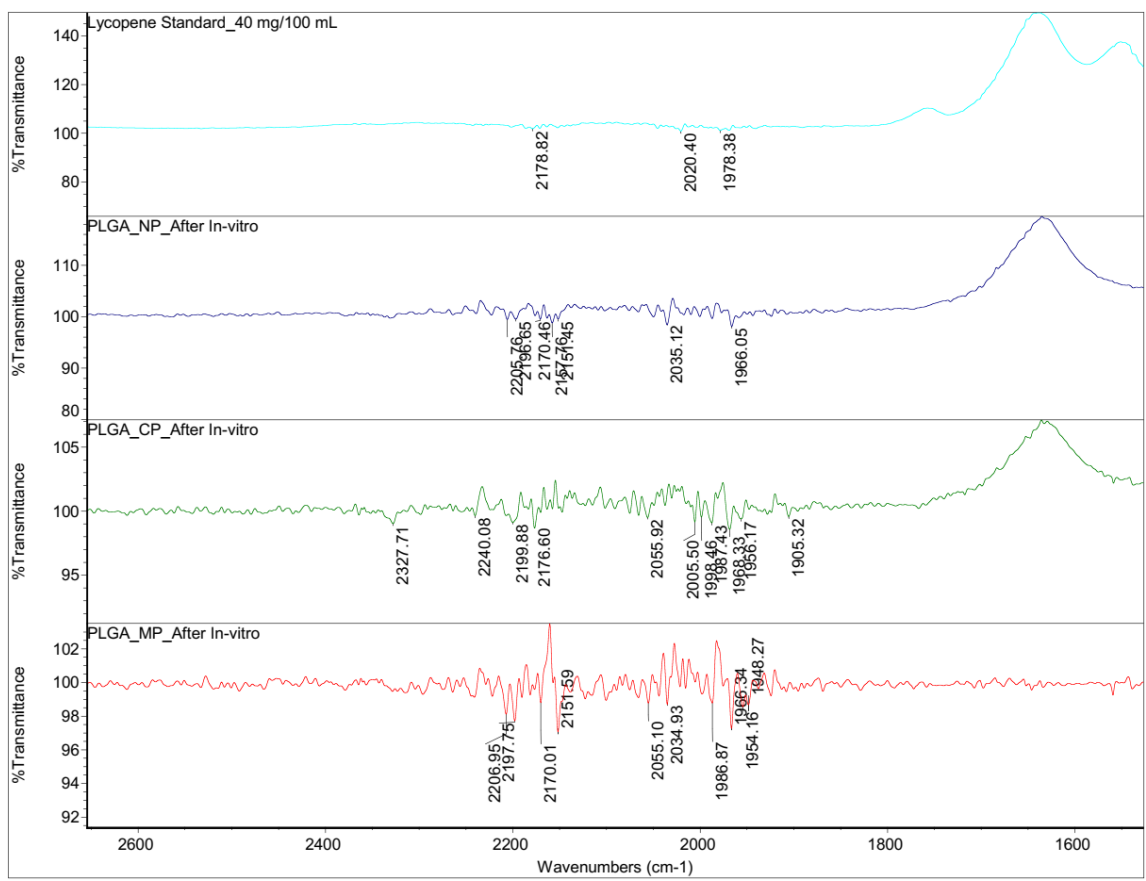

Figure 4.51. Pasteurization effect on chemical structure of PLGA-LNP measured by FT-IR spectroscopy.

*In-vitro release kinetics.* The ANOVA test shows that interaction of pasteurization, digestion time and fortification have significant effect ($p < 0.05$) on the DPPH inhibition activity of the lycopene nanoparticles. However, the homogeneity test depicts that conventional pasteurization had less degradation effect than microwave pasteurization. No significant change was observed on the DPPH inhibition activity of the nanoparticles until 120 minutes of in-vitro human digestion when non-encapsulated lycopene fortified in the juices which was not significant during the 120 to 240 minutes of in-vitro digestion. It was also observed that the HCl and gastric lipase had greater degradation effect than the intestinal lipase secreted from the pancreas. However, the encapsulated lycopene nanoparticles appeared to have a mix effect due to conventional and microwave pasteurization. The reduction of the DPPH activity was observed at 120 mininutes of digestion for microwave treated PLA-LNP which was appeared to be unchanged until the end of 240 min.  The conventionally pasteurized nanoparticles start to degrade after 180 minutes of in-vitro human digestion. It was also observed conventionally pasteurized NP were more stable than microwave pasteurized lycopene nanoparticles.

Table 4.27. ANOVA test results to evaluate the effect of pasteurization and in-vitro digestion on control release kinetics of PLA and PLGA-LNP.

| Source | Type III Sum of Squares | df | Mean Square | F | Sig. |
|---|---|---|---|---|---|
| Corrected Model | 3007.15 | 80 | 37.59 | 4.00 | 0.000 |
| Intercept | 153975.71 | 1 | 153975.71 | 16391.16 | 0.000 |
| Pasteurization | 34.97 | 2 | 17.48 | 1.86 | 0.159 |
| Digestion | 1105.78 | 8 | 138.22 | 14.71 | 0.000 |
| Encapsulation | 88.37 | 2 | 44.18 | 4.70 | 0.010 |
| Pasteurization X Digestion | 729.42 | 16 | 45.59 | 4.85 | 0.000 |
| Pasteurization X Encapsulation | 69.65 | 4 | 17.41 | 1.85 | 0.121 |
| Digestion X Encapsulation | 333.94 | 16 | 20.87 | 2.22 | 0.006 |
| Pasteurization X Digestion X Encapsulation | 645.039 | 32 | 20.157 | 2.146 | .001 |
| Total | 158504.667 | 243 | | | |

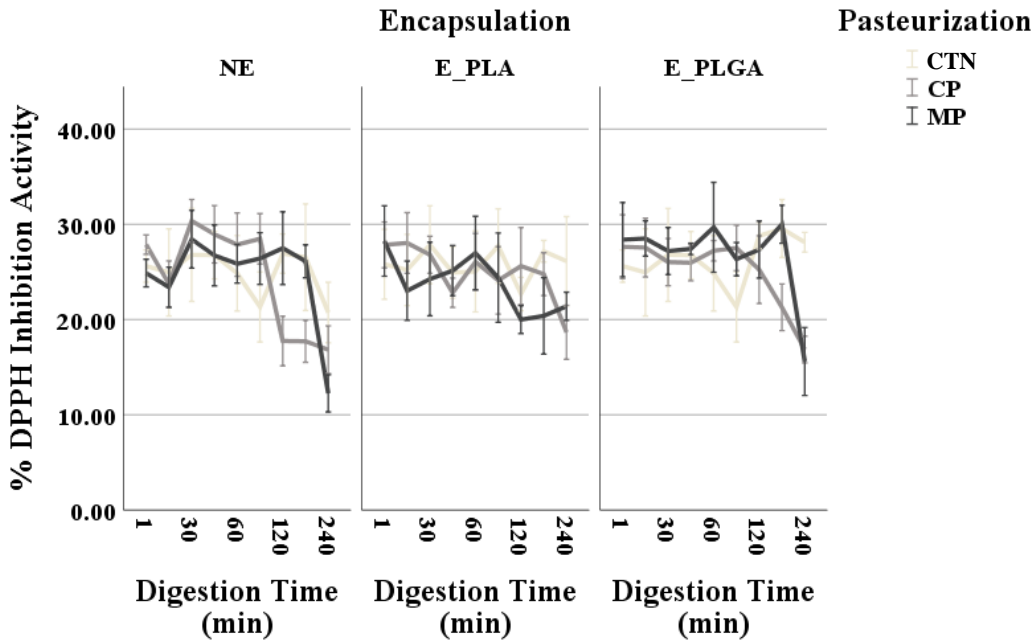

Figure 4.52. Pasteurization and in-vitro digestion effect on bio-accessibility profile of PLA and PLGA-LNP[22].

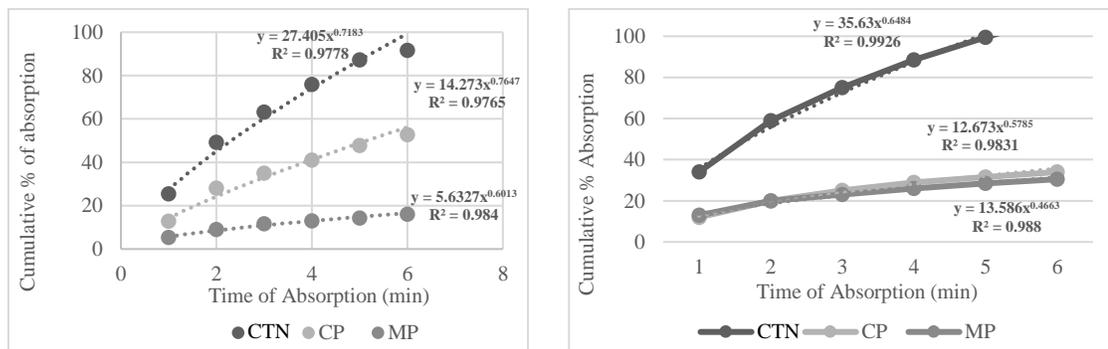

Figure 4.53. Pasteurization effect on PLA (a) and PLGA-LNP (b) absorption profile through the in-vitro intestinal lining simulator

---

Figure 4.53 (a) shows the pasteurization effect on PLA-LNP absorption through the membrane filter with a pore size of 200 nm, built to simulate intestinal absorption. The cumulative percentage shows that PLA-LNP releases lycopene at slowest rate when it was treated with MP compared with CP and CTN PLA-LNP. After observing the digestion effect on the hydrodynamic diameter of different, pasteurized PLA-LNP, it can be concluded that the MP was the highest compared to the CP and CTN which made the absorption profile the slowest with MP PLGA (Figure 4.53 (b)) and PLA-LNP. The Korsmeyer-Peppas Model Fitness Test shows the LNP treated with different techniques had the same $R^2$ value (98%). The exponential model shows that all the PLA-LNP follow a non-Fickian diffusion when passed through the in-vitro intestinal absorption simulator, except the PLGA-lycopene nanoparticles (Fickian type of diffusion) pasteurized with microwave.

## Rumen Digestion

Nanotechnology has emerged as a novel field with countless applications and although its use in animal nutrition is still scarce (Hill et al., 2017; Swain et al., 2015). Nanocarriers in drug delivery have been used in human medicine (Sun et al., 2014; Kandasamy & Maity, 2015; Swierczewska et al., 2016; Ashraf et al., 2016). Lipid or polymeric nanotechnology has received a considerable focus due to its safety and ease of production and scalability (Albuquerque et al., 2020; Ekambaram et al., 2012; Mashaghi et al., 2013; Weber et al., 2013; Carron et al., 2015). The application of lipid or polymeric nanoparticles may increase its efficiency through two main properties: smaller particles clear the rumen more quickly and display improved absorption at the intestinal level (Owens, 1987; Lescoat & Sauvant, 1995; Tarr & Yalkowsky, 1989).

*Physicochemical properties.* The ANOVA test shows the effect of the dependent and independent variables. The interaction effect of pasteurization, digestion time and polymer types only appeared to affect the hydrodynamic diameter of the nanoparticles. On the other hand, a bimodal interaction of digestion time and polymer types affects the changes of hydrodynamic diameter, PDI and zeta potential value of the LNP.

Table 4.28. ANOVA test to evaluate the effect of digestion, polymer types and pasteurization on physicochemical and bio-accessibility of LNP.

| Source | Dependent Variable | Type III Sum of Squares | df | Mean Square | F | Sig. |
|---|---|---|---|---|---|---|
| Corrected Model | HD | 38138.364 [a] | 7 | 5448.338 | 217.637 | 0.000 |
| | PDI | 0.762 [b] | 7 | 0.109 | 131.043 | 0.000 |
| | ZP | 29091.790 [c] | 7 | 4155.970 | 39.212 | 0.000 |
| | DPPH | 420.787 [d] | 7 | 60.112 | 3.889 | 0.004 |
| Pasteurization | HD | 496.755 | 1 | 496.755 | 19.843 | 0.000 |
| | PDI | .006 | 1 | 0.006 | 6.995 | 0.013 |
| | ZP | 22.423 | 1 | 22.423 | 0.212 | 0.649 |
| | DPPH | 38.209 | 1 | 38.209 | 2.472 | 0.127 |
| Digestion Time | HD | 2.163 | 1 | 2.163 | 0.086 | 0.771 |
| | PDI | 0.649 | 1 | 0.649 | 781.113 | 0.000 |
| | ZP | 9997.451 | 1 | 9997.451 | 94.326 | 0.000 |
| | DPPH | 192.245 | 1 | 192.245 | 12.436 | 0.001 |
| Polymer Type | HD | 217.431 | 1 | 217.431 | 8.685 | 0.006 |
| | PDI | 0.055 | 1 | 0.055 | 66.507 | 0.000 |
| | ZP | 6223.932 | 1 | 6223.932 | 58.723 | 0.000 |
| | DPPH | 119.273 | 1 | 119.273 | 7.716 | 0.010 |

Table 4.28 (Continued).

| | | | | | | |
|---|---|---|---|---|---|---|
| Pasteurization X Digestion Time | HD | 1045.159 | 1 | 1045.159 | 41.749 | 0.000 |
| | PDI | 0.001 | 1 | 0.001 | 1.514 | 0.229 |
| | ZP | 0.378 | 1 | 0.378 | 0.004 | 0.953 |
| | DPPH | 47.352 | 1 | 47.352 | 3.063 | 0.091 |
| Pasteurization X Polymer Type | HD | 4892.583 | 1 | 4892.583 | 195.437 | 0.000 |
| | PDI | 0.003 | 1 | 0.003 | 3.369 | 0.077 |
| | ZP | 39.043 | 1 | 39.043 | 0.368 | 0.549 |
| | DPPH | 8.940 | 1 | 8.940 | 0.578 | 0.453 |
| Digestion Time X Polymer Type | HD | 26631.243 | 1 | 26631.243 | 1063.800 | 0.000 |
| | PDI | 0.017 | 1 | 0.017 | 20.330 | 0.000 |
| | ZP | 6459.677 | 1 | 6459.677 | 60.947 | 0.000 |
| | DPPH | 0.783 | 1 | 0.783 | 0.051 | 0.824 |
| Pasteurization X Digestion Time X Polymer Type | HD | 6443.395 | 1 | 6443.395 | 257.385 | 0.000 |
| | PDI | 0.001 | 1 | 0.001 | 1.494 | 0.232 |
| | ZP | 13.537 | 1 | 13.537 | 0.128 | 0.723 |
| | DPPH | 19.251 | 1 | 19.251 | 1.245 | 0.274 |
| Total | HD | 841279.042 | 36 | | | |
| | PDI | 3.820 | 36 | | | |
| | ZP | 73379.504 | 36 | | | |
| | DPPH | 20356.705 | 36 | | | |
| Corrected Total | HD | 38839.318 | 35 | | | |
| | PDI | 0.786 | 35 | | | |
| | ZP | 32059.456 | 35 | | | |
| | DPPH | 853.629 | 35 | | | |

Hydrodynamic diameter (Figure4.76) of PLA and PLGA-LNP have changed significantly ($p < 0.05$) after 120 minutes of enzymatic digestion in rumen fluid at in-vitro conditions. The HD of the PLA-LNP, was observed to be significantly reduced (P<0.05) from 171.77±7.74 to 154.40±4.17 nm when the condition was without pasteurization as shown in Figure4.76, but the PLGA swelling was observed due to the addition of extra water during enzymatic hydrolysis and diameter increases from 122.93±3.02 to 164.20±1.85 nm.

PLGA is more hygroscopic compared to PLA which enhance the swelling of PLGA LNP and further size increment. Similarly, with the MP treatment, the diameter of the PLGA LNP also increased (P<0.05) from 122.83±1.61 to 198.00±11.09 nm while the PLA LNP decreased (P<0.05) from 178.97±3.27 to 81.98±0.85 nm.

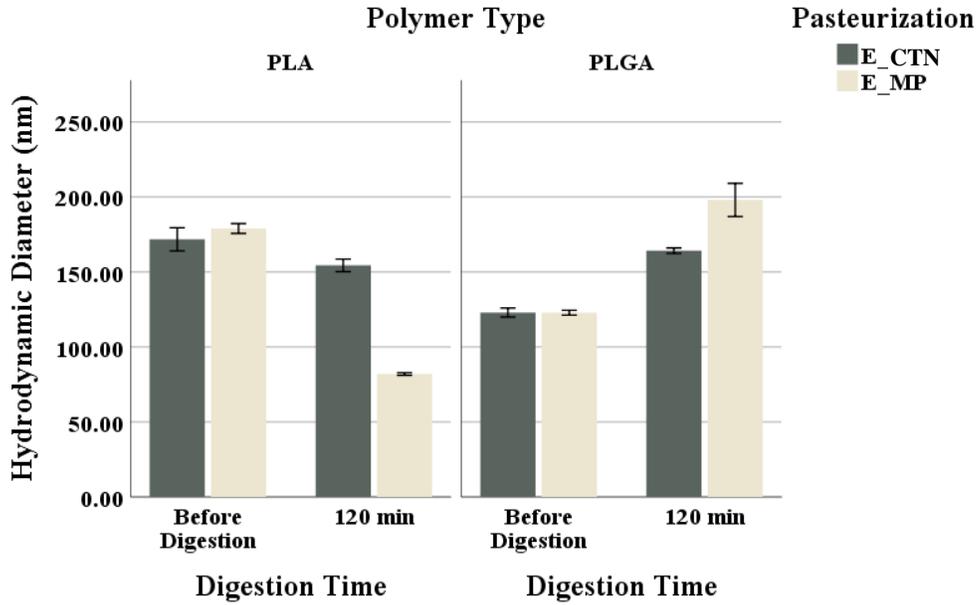

Figure 4.54. Rumen digestion effect on hydrodynamic diameter profile for pasteurized PLA and PLGA-LNP[23].

Pasteurization did not have any effect (P<0.05) on PDI value (Figure 4.55) of LNP due to the rumen fluid digestion for either PLA-LNP or PLGA-LNP. PDI increased from 0.141±0.014 to 0.497±0.009 and 0.264±0.008 to 0.503 ±0.082 for PLA-LNP and PLGA-LNP, respectively. The CTN samples MP show a PDI shifting from 0.120±0.024 to 0.426±0.008 and 0.256±0.009 to 0.495±0.049 for PLA and PLGA LNP, respectively, after

---

[23] E_CTN = Encapsulated but control, E_MP = Encapsulated but microwave pasteurized.

120 minutes of rumen fluid fermentation. Hence, the size distribution of the PLGA and PLA-LNP shifted from a more homogeneous to heterogeneous due to enzymatic actions on inducing surface erosion of the LNP.

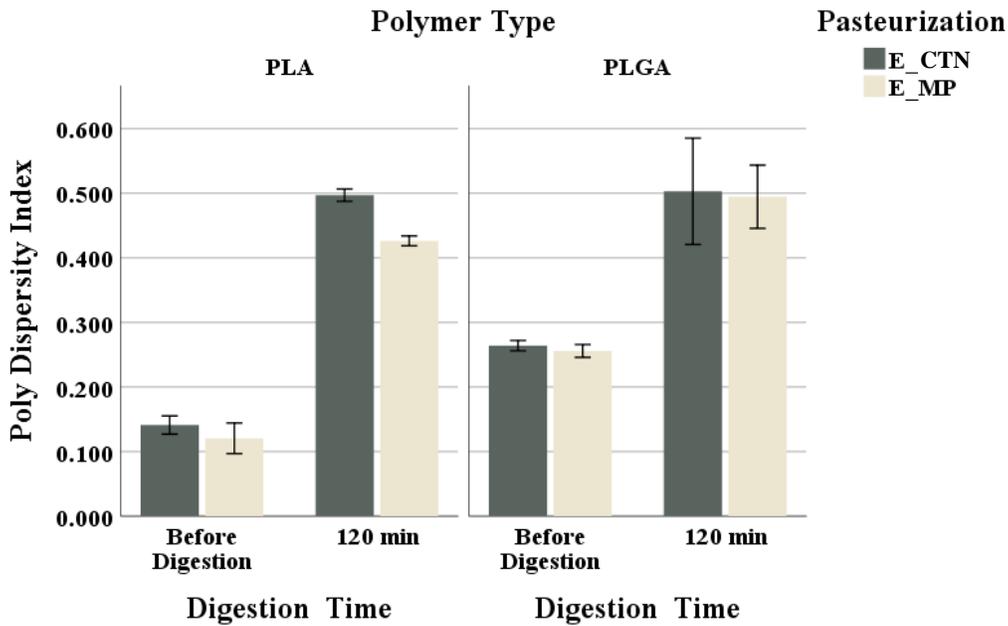

Figure 4.55. Rumen digestion effect on polydispersity index profile for pasteurized PLA and PLGA-LNP[24].

Zeta potential shows the stability of the nanoparticles under rumen fluid digestion, the higher the values, higher was the stability of the LNP. The non-pasteurized PLGA nanoparticles (Figure 4.56) zeta potential value observed a sharp reduction from -75.22±2.12 to 12.53±0.51 mV; whereas the PLA shows almost 50% reduction from -18.00 ± 1.57 to -9.55 ± 0.67 mV after the completion of rumen digestion. Similarly, samples that were MP, the ZP values were reduced from -72.42±24.18 to -7.57±0.99 mV and from -

17.02±0.82 to -11.60±1.21 mV for PLGA and PLA-LNP, respectively. Hence, no significant changed (p < 0.05) appeared between zeta potential value for PLGA and PLA-LNP after 120 minutes of fermentation with rumen fluid.  The stability of PLA and PLGA-LNP was found to be similar after rumen digestion though the initial stability was far better in PLGA-LNP compared to PLA-LNP. Hence, it was shown   that the PLGA-LNP was more susceptible to rumen digestion than the PLA-LNP.

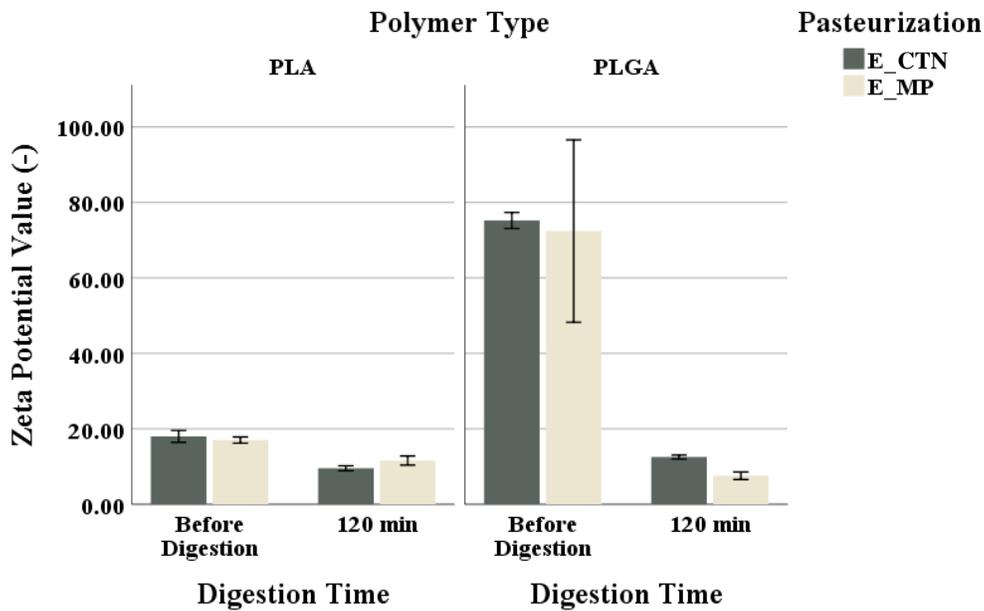

Figure 4.56. Rumen digestion effect on zeta potential profile for pasteurized PLA and PLGA-LNP[25].

*Bio-accessibility of PLA and PLGA-LNP.* Pasteurization (Figure 4.57) had significant effect (P<0.05) on the non-encapsulated lycopene emulsion when the functional activity was reduced from 21.5±0.12 to 14.42±0.50. But no significant (P>0.05) effect was observed between the bio-accessibility of the LNP (10.32±3.41) and MP (11.8±2.92) treated non-encapsulated lycopene emulsion. The Rumen digestion time did not have any difference (P>0.05) effect on the bio-accessibility of PLGA-LNP or PLA-LNP except when PLGA-lycopene nano emulsion digested in rumen fluid before pasteurization. Hence, the encapsulation prevents the functional properties of lycopene nanoparticles. The bio-accessibility was found to be improved by more than 50% due to encapsulation compared to the non-encapsulated lycopene after rumen digestion. No significant difference (p < 0.05) was observed between bio-accessibility of PLA-LNP and PLGA-LNP after 120 minutes rumen digestion. The formulation of the PLA-LNP with tween-80 made the nanoparticles more stable (Albuquerque et al., 2020) compared with PLGA-LNP though, the initial stability of PLGA-LNP was observed to be far too higher than that the PLA-LNP. This could be attributed to the fact that tween-80 contains oleic acid, a long chain fatty acid which was not degraded by rumen inoculum and provide extra protection for PLA-LNP.

---

[25] E_CTN = Encapsulated but Control Non-pasteurized, E_MP = Encapsulated but microwave pasteurized.

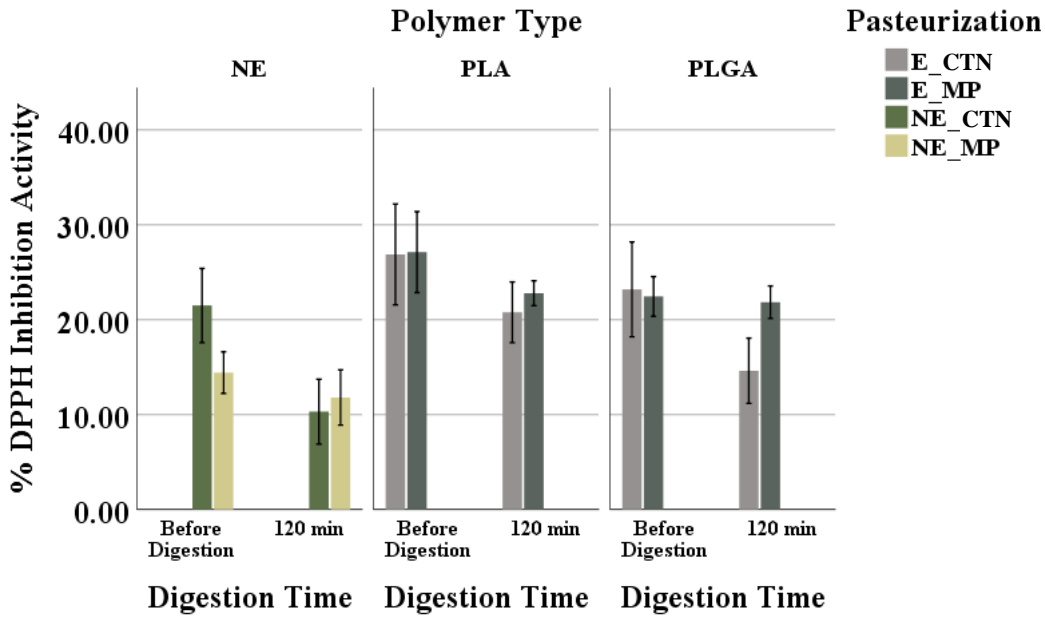

Figure 4.57. Rumen digestion effect on bio-accessibility profile for pasteurized PLA and PLGA-LNP[26].

*Morphological and chemical properties of PLA and PLGA-LNP.* The size distribution shows that NP PLGA-LNP shifted to the right after 120 minutes of rumen fluid digestion (Figure 4.58). A similar effect was also observed when the MP was applied on PLGA-NP; however, the PLA-LNP size distribution flatten and shifted to the right when CNP and MP were applied. The fluctuation of the size distribution up and down was due to the adsorption of water from the rumen fluid which caused surface erosion of polymer for PLA-LNP and PLGA-LNP during hydrolytic reactions.

---

[26] E_CTN = Encapsulated but Control Non-pasteurized, E_MP = Encapsulated but microwave pasteurized, NE_CTN = Non-encapsulated but Control Non-pasteurized, NE_MP = Non-encapsulated but microwave pasteurized.

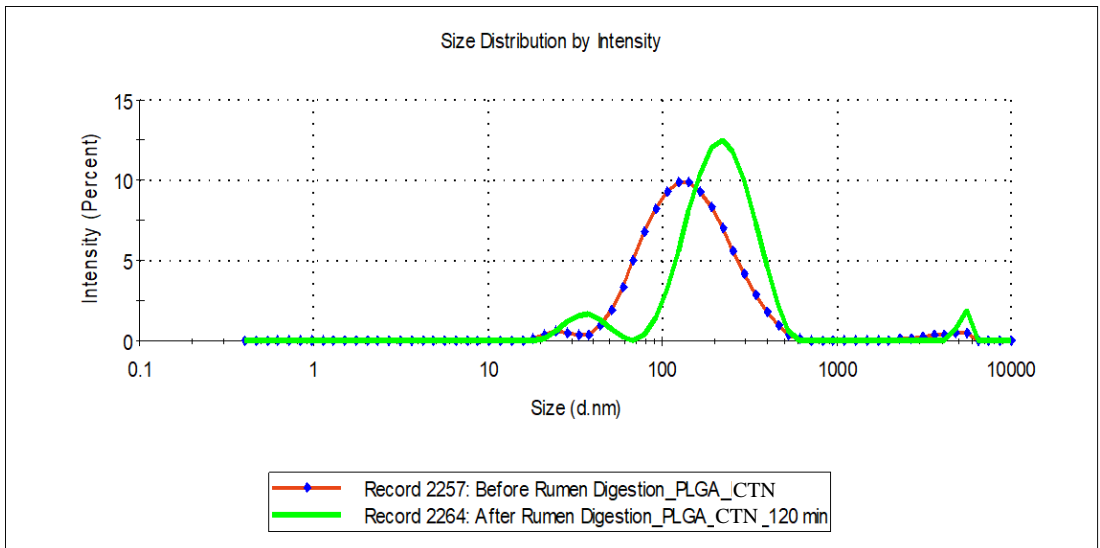

Figure 4.58. Effect of in-vitro rumen digestion on size distribution of non-pasteurized PLGA-LNP.

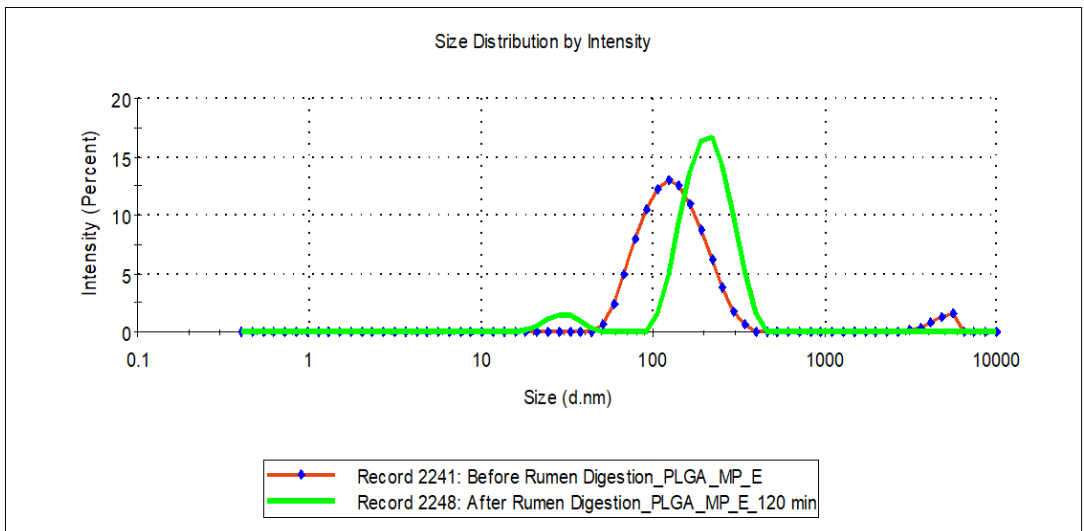

Figure 4.59. Effect of in-vitro rumen digestion on size distribution of microwave pasteurized PLGA-LNP.

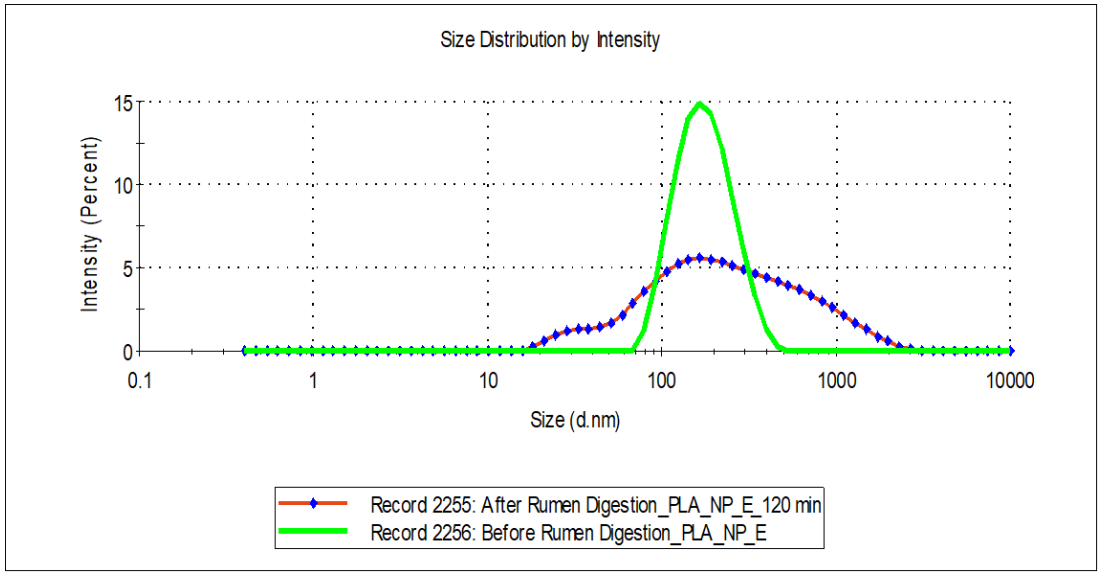

Figure 4.60. Effect of in-vitro rumen digestion on size distribution of non-pasteurized PLA-LNP.

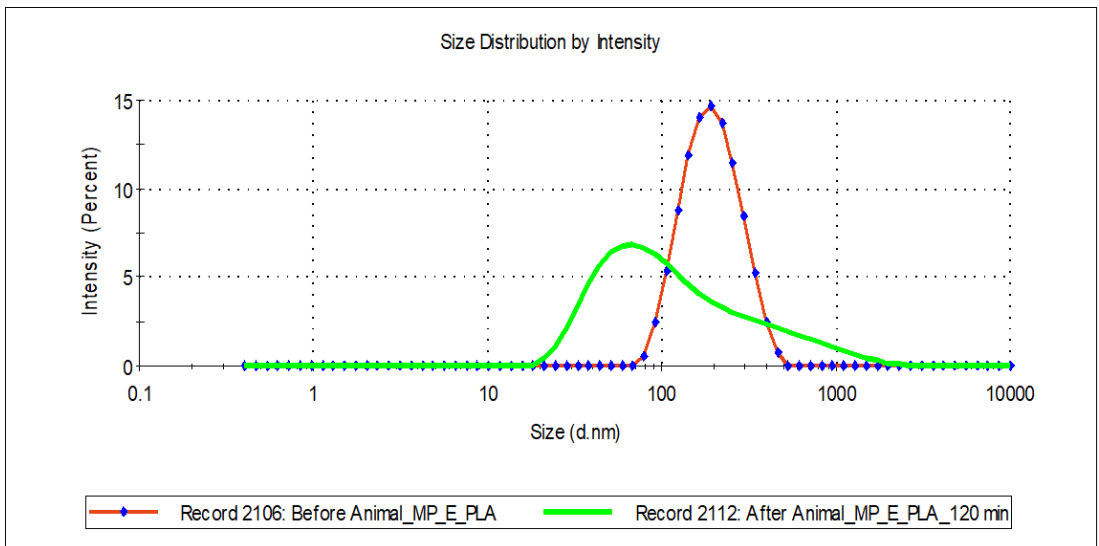

Figure 4.61. Effect of in-vitro rumen digestion on size distribution of microwave pasteurized PLA-LNP.

*In-vitro release kinetics.* The Tukey test for the digestion time and bio-accessibility shows

significant degradation after 30 minutes of rumen fermentation which was reached at 68%

after 120 minutes of fermentation. On the other hand, the Tukey HSD test for bio-accessibility on the polymer type shows that PLA preserved higher ($p < 0.05$) functional activity (24.32%) of lycopene than that of PLGA (22.16%). The in-vitro release kinetics shows a clear picture that explains the effect of pasteurization, polymer and fermentation on preservation of functional activity of lycopene (Figure 4.62). Figure 4.62 shows that due to MP treatment functional activity of PLA-LNP was reduced from 27.12±4.27% to 22.79±1.31% after 120 minutes of rumen digestion. On the contrary, the MP did not have any effect ($p > 0.05$) on the lycopene when synthesized in PLGA-LNP as the bio-accessibility changes were not significant (P<0.05). However, both LNP showed high sensitivity to rumen fermentation which leads to the loss of the functional activity of the lycopene. The degradation was higher for PLGA-LNP, from 23.19±5.00% to 15.58±2.52%, compared to PLA-LNP from 26.87±5.31% to 20.78±3.19%. Hence, it was shown that MP samples made conformational changes on the PLA and PLGA LNP and further providing the additional protection for the encapsulated lycopene.

Table 4.29. Homogeneity test results to evaluate the effect of rumen digestion on bio-accessibility of lycopene nanoparticles (PLA/PLGA).

| Digestion Time | N | Subset | | | |
| | | 1 | 2 | 3 | 4 |
|---|---|---|---|---|---|
| 120 min | 18 | 17.48 | | | |
| 90 min | 18 | 18.02 | | | |
| 60 min | 18 | 18.97 | 18.97 | | |
| 45 min | 18 | | 21.86 | 21.86 | |
| 30 min | 18 | | | 22.89 | 22.89 |
| 15 min | 18 | | | 24.77 | 24.77 |
| Before Digestion | 36 | | | | 25.61 |
| Significance | | 0.864 | 0.192 | 0.184 | 0.255 |

Table 4.30. Homogeneity test results to evaluate the effect of rumen digestion on bio-
accessibility of lycopene nanoparticles.

| Type Polymer | N | Subset | | |
| | | 1 | 2 | 3 |
|---|---|---|---|---|
| NE | 48 | 19.23 | | |
| PLGA | 48 | | 22.16 | |
| PLA | 48 | | | 24.31 |
| Sig. | | 1.000 | 1.000 | 1.000 |

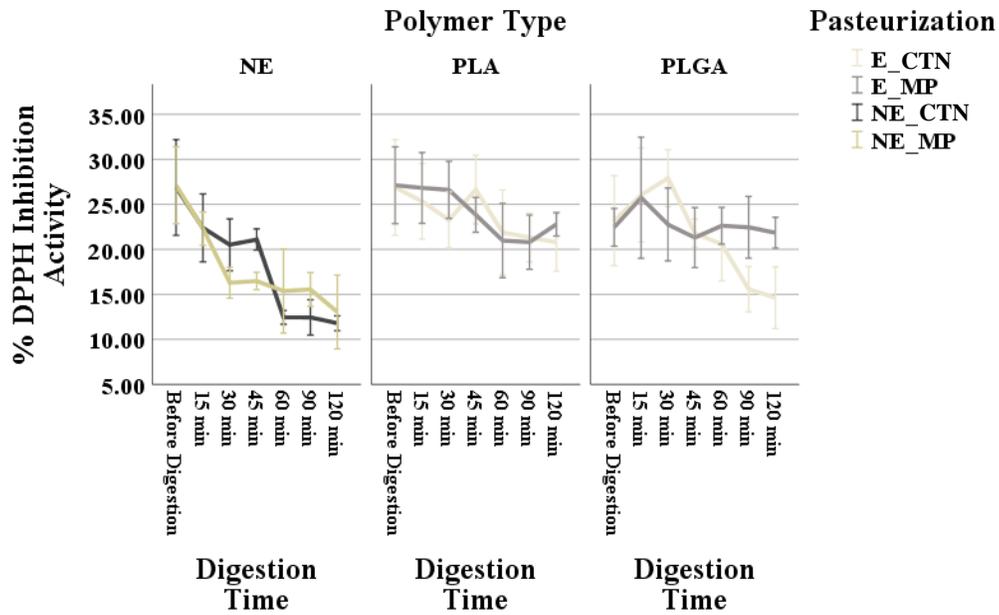

Figure 4.62. Pasteurization, polymer and rumen digestion time effect on
release/degradation kinetics of free or encapsulated lycopene. E_CTN
= Encapsulated but Control Non-pasteurized, E_MP = Encapsulated
but microwave pasteurized, NE_CTN = Non-encapsulated and
Control Non-pasteurized, NE_MP = Non-encapsulated and
microwave pasteurized and NE = Non-encapsulated.

## CHAPTER 5

## CONCLUSION AND RECOMMENDATIONS

### 5.1.    Conclusion

The automatic in-vitro digestion simulation was developed to evaluate the bioaccessibility study of loaded encapsulated lycopene nanoparticles. Lycopene was encapsulated in two polymeric (PLGA and PLA) nanoparticles. Sonication time, surfactant, and polymer concentration were evaluated the inherent factors as well that influence the formulation and encapsulation of lycopene nanoparticles. In addition, the process parameters were optimized and their impact on the physicochemical properties and effects on the bio-accessibility of lycopene nanoparticles in the GIT were evaluated. The sonication time strongly ($p < 0.05$) influenced the nano-emulsion stability with the zeta potential value and encapsulation efficiency (63.8 and 85.7%) when PLGA nanoparticles were used to synthesize the lycopene nanoparticles.  The surfactant (-20.3%) and polymer (49.9%) concentration strongly influenced (P<0.05) the hydrodynamic diameter and electrical conductivity of the PLGA-lycopene nanoparticles, respectively. The bio-accessibility of lycopene, due to encapsulation, was observed to be $90.87 \pm 4.00\%$ which is 3 - 9 times greater than the non-encapsulated lycopene when ingested orally.

The food model (fruit juices) was developed and fortified with lycopene nanoparticles. The effect of encapsulation was evaluated; hence, it was found out that the apparent viscosity, flow behavior index, and yield stress had significantly changed (P<0.05). However, the pasteurization process (CP and MP) was not significantly affected (P>0.05). The high sugar content in the 15°Brix juice concentration introduced additional coating to the encapsulated lycopene nanoparticle which further preserved its functional integrity against the pasteurization process compared to the 5°Brix juices concentration.

The in-vitro study shows the hydrodynamic diameter (HD) of the non-pasteurized PLGA-LNP increased (P < 0.05) from 114.80±7.76 to 169.60±3.20 nm after in-vitro digestion. The reason behind this increment was the result of water adsorption on the surface of the PLGA-LNP which is a principal step for hydrolysis. Similarly, The HD increased for the conventionally pasteurized PLGA-LNP significantly (P < 0.05) from 126.41±3.31 to 160.87±1.39 nm whereas the microwaved pasteurized also increased from 116.42 ± 8.45 to 152.30 ± 0.98 nm after the in-vitro digestion. The HD of the MP PLGA-LNP was found to be the lowest after in-vitro digestion compared to CP or non-pasteurized PLGA-LNP. The effect of MP treatment probably caused configurational structural changes which accelerated PLGA erosion that also led to further lycopene release and degradation.

The research results exhibited that higher zeta potential is to be synonymous to stable nano-emulsion. The zeta potential value for non-pasteurized was observed to reduce from -18.71 ± 2.40 to -3.87 ± 0.37 after in-vitro digestion. Similarly, a reduction from -23.42 ± 2.10 to -3.48 ± 0.20 for CP PLA-LNP in nano-emulsion was also observed. Unlike the MP treated PLA-LNP nano-emulsion, the degradation was reduced as a result of stability from -21.86 ± 2.71 to -7.26 ± 1.8=72 at the end of intestinal digestion. It can be concluded from the results indicated in the above discussion that PLA-LNP nano-emulsion MP was observed to be more stable than that of non-pasteurized or CP PLA-LNP nano-emulsion. The PLGA-LNP nano-emulsion was observed to be stable in the beginning of digestion because of the very high zeta potential ($\leq$-70 mV) while the values dropped to 4.54 ± 0.71, 19.77 ± 2.48, and 1.91 ± 0.46 after in-vitro digestion when the following pasteurized methods (non-pasteurized CP and MP) were applied on PLGA-LNP, respectively. Hence, PLGA-NP nano-emulsion was more stable against CP compared with non-pasteurized or MP.

The absorption profile of PLA-LNP nano-emulsion pasteurized with microwave had the slowest absorption rate compared to those treated conventionally and non-pasteurized. The digestion effect on the average hydrodynamic diameter of different pasteurized nano-emulsions with PLA-LNP pasteurized with microwave was higher which kept particle movement slower as compared to the other two treatments (non-pasteurized or CP). The Korsmeyer-Peppas Model Fitness Test was used to evaluate the data from nano-emulsions with different LNP that were treated with different pasteurization techniques. The goodness of fit was the same with the coefficient of determinant (98%) which followed the non-Fickian type of diffusion when it passed through the in-vitro intestinal absorption simulator except when PLGA-lycopene nanoparticles treated with microwave pasteurization is used due to their low polymeric degradation rate as compared to other treated LNP.

## 5.2.    Recommendations

The research conducted in this dissertation effort to develop two novel polymeric encapsulation procedure was to improve lycopene bio-accessibility in the GIT. Due to the lack of access to adequate instrumentation, the ability to conduct certain relevant experiment limited research efforts. The following recommendations henceforth need to be pursued to complement this research project:

- DPPH activity was used in the study to estimate the bioaccessibility; hence, an HPLC method if applied will provide a more accurate estimation of bioaccessibility.

- Most of the size characterization was based on hydrodynamic diameter which is usually larger than the actual size; hence, an SEM and TEM analysis is required to provide a more accurate characterization of the size and structure.

- The in-vitro digestion was applied on this dissertation study; therefore, efforts of this research study realized the limitation of the in-vitro model. However, further research efforts with in-vivo animal (mouse) models are required to establish a correlation between the in-vitro to in-vivo.

- Toxicity studies should be done for both PLGA and PLA nanoparticles to validate safety measure before using in functional food products.

- The novel nanoparticles should be applied against cancer cells and other chronic diseases treatments.

VITA

Mohammad Anwar Ul Alam, the son of the Late Mohammad Shamsul Alam and Shahida Alam, was born in Dhaka, Bangladesh. He completed his bachelor's degree program in 2009 and his master's degree program in 2010, both from the University of Dhaka located in Dhaka, Bangladesh under the Institute of Nutrition and Food Science Program with concentrations in nutrition and food science. He successfully defended his dissertation in the Department of Food and Animal Sciences with concentrations in food safety and processing and graduated in August 2022 with a Doctoral of Philosophy degree in Food Science from Alabama A&M University located in Normal, Alabama.